\documentclass[a4paper,11pt]{article}
\pdfoutput=1
\usepackage{jcappub}
\usepackage{caption}
\usepackage{subcaption}
\usepackage{graphicx}
\usepackage{hyperref}
\usepackage{amsmath}
\usepackage{mathtools}
\usepackage{upgreek}
\usepackage{tablefootnote}
\usepackage{soul}
\usepackage{color}
\usepackage{verbatim}
\usepackage{float}
\usepackage{bm}
\usepackage{placeins}
\usepackage{enumitem}
\usepackage[usenames,dvipsnames,svgnames,table]{xcolor}
\usepackage{simplewick}
\usepackage{simpler-wick}
\usepackage{amssymb}
\usepackage{epstopdf}
\graphicspath{ {./Figures/} }
\usepackage{dcolumn}
\usepackage{bm}
\usepackage{mhchem}
\usepackage[normalem]{ulem}
\usepackage{multirow}
\usepackage{booktabs}
\usepackage{tikz}
\usepackage[compat=1.1.0]{tikz-feynman}
\usepackage{contour}
\usetikzlibrary{positioning}
\usetikzlibrary{shapes.geometric}

\newcommand{\bsL}{\boldsymbol{L}}
\newcommand{\bsl}{\boldsymbol{\ell}}
\newcommand{\bsk}{\boldsymbol{k}}

\newcommand{\bsq}{\boldsymbol{q}}
\newcommand{\bsp}{\boldsymbol{p}}

\begin{document}
\title{The reconstructed CMB lensing bispectrum}

\author[a,b]{Alba Kalaja,}
\author[a]{Giorgio Orlando,}
\author[c,d]{Aleksandr Bowkis,}
\author[c,d,f]{Anthony Challinor,}
\author[a]{P. Daniel Meerburg,}
\author[e]{Toshiya Namikawa}

\affiliation[a]{Van Swinderen Institute for Particle Physics and Gravity, University of
Groningen, Nijenborgh 4, 9747 AG Groningen, The Netherlands}
\affiliation[b]{Center for Computational Astrophysics, Flatiron Institute, 162 5th Avenue, New York, NY 10010, USA}
\affiliation[c]{Institute of Astronomy, Madingley Road, Cambridge CB3 0HA, UK}
\affiliation[d]{Kavli Institute for Cosmology, Cambridge, Madingley Road, Cambridge
CB3 0HA, UK}
\affiliation[f]{DAMTP, Centre for Mathematical Sciences, Wilberforce Road, Cambridge
CB3 0WA, UK}
\affiliation[e]{Kavli Institute for the Physics and Mathematics of the Universe (WPI),
UTIAS, The University of Tokyo, Kashiwa, Chiba 277-8583, Japan}

\emailAdd{a.kalaja@rug.nl}
\emailAdd{g.orlando@rug.nl}

\abstract{Weak gravitational lensing by the intervening large-scale structure (LSS) of the Universe is the leading non-linear effect on the anisotropies of the cosmic microwave background (CMB). The integrated line-of-sight mass that causes the distortion -- known as lensing convergence -- can be reconstructed from the lensed temperature and polarization anisotropies via estimators quadratic in the CMB modes, and its power spectrum has been measured from multiple CMB experiments. Sourced by the non-linear evolution of structure, the bispectrum of the lensing convergence provides additional information on late-time cosmological evolution complementary to the power spectrum. However, when trying to estimate the summary statistics of the reconstructed lensing convergence, a number of noise-biases are introduced, as previous studies have shown for the power spectrum. Here, we explore for the first time the noise-biases in measuring the bispectrum of the reconstructed lensing convergence.
We compute the leading noise-biases in the flat-sky limit and compare our analytical results against simulations, finding excellent agreement. Our results are critical for future attempts to reconstruct the lensing convergence bispectrum with real CMB data.}
\maketitle

\section{Introduction}
The photons of the cosmic microwave background (CMB) are deflected by the large-scale structure (LSS) of the universe as they travel from the last scattering surface ($z\sim 1100$) towards our detectors. This effect, called weak gravitational lensing of the CMB~\cite{lewis:weak_lensing_review}, traces the distribution of matter over a wide range of redshifts ($0.1<z<5$), making it a powerful probe of late-time cosmological evolution. 

Lensing alters the CMB power spectra and introduces non-Gaussianities in the distribution of the anisotropies. It also generates a $B$-mode polarization signal that represents the main source of confusion for the signal from primordial gravitational waves generated during inflation (see, e.g.,~\cite{kamionkowski:bmodes_review} and references therein). From this perspective, lensing can be a nuisance that needs to be removed (by delensing), e.g., to extract the primordial $B$-mode signal~\cite{knox:limit_detectability,seljak:lensing_contaminant_gravity}. 
On the other hand, the induced statistics of lensed CMB modes can be used to reconstruct the line-of-sight integrated gravitational potential, also known as the lensing potential ($\phi$), whose gradient determines the lensing deflections and whose second derivatives determine the lensing shear and convergence (denoted with $\kappa$).

The first measurement of the CMB lensing signal was achieved in cross-correlation with radio galaxy counts~\cite{smith:first_lensing_detection}, whereas the detection with CMB data only was done for the first time by ACT~\cite{das:detection_lens_act,das:detection_lens_act2}, followed by \textit{Planck}~\cite{planck13:gravitational_lensing}, POLARBEAR~\cite{ade:polarbear_lens_detection}, the South Pole Telescope (SPT)~\cite{spt:lensing_measurement,story:spt_lens_detection2} and BICEP/Keck \cite{bk_lens_detection}. The most recent measurement by \textit{Planck}~\cite{planck18:gravitational_lensing} detected the lensing power spectrum at $40\,\sigma$ significance.

Given that CMB lensing is most sensitive to structure growth at redshift $z\sim 2$, its power spectrum is particularly useful to constrain the content of the Universe around that redshift, such as the dark energy equation-of-state~\cite{act11:dark_energy_lensing,spt:lensing_measurement,planck13:gravitational_lensing} and the sum of the neutrino masses~\cite{kaplinghat:nu_from_CMB,battye:neutrino_mass_cmb_lensing}. In addition, the cross-correlation between CMB lensing and other tracers of large-scale structure (LSS), e.g., galaxy surveys, can help circumvent cosmic variance~\cite{seljak:png_without_cosmic_variance,tommaso:cmb_lensing_png} and break degeneracies between cosmological parameters, for example, galaxy bias and growth in the LSS~\cite{modi:cmb_lens_cleft,banerjee:neutrino_darkenergy,schmittfull:cross_corr_cmb_galaxy}.

Upcoming experiments including Simons Observatory (SO)~\cite{so:forecast} and CMB-S4~\cite{cmbs4:science_book} will provide measurements of the CMB sky with unprecedented precision. In this context, CMB lensing plays a major role in refining our knowledge of the late-time Universe over a wide range of redshifts, both as a cosmological probe on its own as well as in combination with correlated LSS tracers.
Thus far, cosmological applications of CMB lensing have relied on the assumption that the lensing potential is Gaussian, namely that the underlying matter distribution is well described by linear theory. 
However, at late times, nonlinear clustering in the matter density fluctuations generates large non-Gaussianities, which in turn induce a small, but non-zero, bispectrum in the CMB lensing convergence field. While this effect is negligible for the sensitivity reached by, e.g., \textit{Planck}, future CMB experiments should be sensitive to the non-Gaussian convergence field \cite{namikawa:lensing_bispec,pratten:post_born_lensing,Liu:2016nfs}. Interestingly, it has been shown that ongoing Stage-III experiments, like the Atacama Cosmology Telescope (ACT)~\cite{act:forecast} and Simons Array~\cite{simonsarray:experiment}, may already be able to measure the lensing convergence bispectrum arising from the non-linear growth of the LSS -- the next higher-order correlation function -- and that in the future Stage-IV experiments, like CMB-S4, have the potential to detect it at $50\,\sigma$ statistical significance~\cite{namikawa:lensing_bispec}.

A simple and efficient (albeit somewhat suboptimal) method to reconstruct the lensing potential employs estimators that are quadratic in the CMB anisotropies. Here, we work with the widely used Hu-Okamoto (HO) quadratic estimator~\cite{hu:mass_reconstruction_CMB_polarization,okamoto:cmb_lens_reconstruction_fullsky}. Because of the nature of the quadratic estimator, additional terms called noise-biases arise when estimating the $n$-point functions of the lensing potential, and are comparable to or even larger than the signal in the recent power spectrum measurements.
These noise-biases have been worked out in detail for the power spectrum in Refs.~\cite{kesden:lensing_reconstruction,hanson:lens_rec} and methods developed for their practical subtraction in survey data~\cite{namikawa:bias_hardened,story:spt_lens_detection2}.
However, when including non-Gaussian effects in the convergence field, new reconstruction bias terms contribute to the measured lensing power spectrum and must be taken into account in future analysis (see, e.g., Ref.~\cite{bohm:cmb_bias_bispectrum,Bohm:2018omn}). This non-Gaussian bias is also important for the cross-correlation between CMB lensing and other LSS tracers \cite{Fabbian:2019tik}. 

Besides these additional biases in the measurement of the lensing power spectrum, non-Gaussianity of the lensing convergence field encodes cosmological information that could be used to tighten the bounds on neutrino masses and dark energy models. Fisher forecasts show that the expected constraints on the dark energy equation-of-state and sum of neutrino masses from the lensing bispectrum\footnote{Note that from here on we will use the terminology lensing bispectrum, for which we intend the bispectrum of the lensing convergence field, not to be confused with the CMB bispectrum induced by lensing, such as ISW-lensing~\cite{sachs:sachs_wolfe,goldberg:sachs_wolfe}.} alone are comparable to those obtained from the power spectrum~\cite{namikawa:lensing_bispec}. Additionally, the combination of the power spectrum and the bispectrum suggests the potential to improve constraints by around $30\,\%$, making the lensing bispectrum an appealing cosmological tool comparable and complementary to the lensing power spectrum. 

In this work, we aim to characterize for the first time the leading-order noise-biases\footnote{The largest noise-biases for the lensing bispectrum are Gaussian, therefore they do not introduce a bias in the bispectrum signal. However, we keep this nomenclature to highlight the analogy with the power spectrum reconstruction.} in the reconstructed lensing convergence bispectrum. We limit our analysis to temperature modes only, but the extension to polarization is straightforward.
The procedure to compute these biases follows the same perturbative methods that were applied to obtain the noise-biases for the power spectrum.
However, the bispectrum case can be significantly more challenging compared to the power spectrum, as the bispectrum of the estimated convergence field depends on the 6-pt correlation function of the lensed CMB modes. While in principle the calculations can still be carried out using Wick's theorem for the products of (Gaussian) unlensed CMB fields and low-order cumulant expansions for the (non-Gaussian) lensing convergence fields that appear in the perturbative expansion of the lensed CMB $n$-pt functions, keeping track of all terms in the expansion can become increasingly difficult. Therefore we will additionally use and adapt a more convenient way to organize these calculations based on Feynman diagrams~\cite{jenkins:feynman_diagrams1,jenkins:feynman_diagrams2}, which allows one to write down the noise-biases without resorting to lengthy calculations. The noise-bias term can be identified by combining the proper propagators, similarly to how these diagrams are used in quantum field theory. 
In this way, we will be able to characterize the noise-bias terms $N_B^{(0)}$, $N_B^{(1)}$, $N_B^{(3/2)}$ in the flat-sky limit (the superscripts on the bispectrum noise-biases denote the leading-order dependence on the lensing potential power spectrum).
Finally, in order to validate our analytical and numerical computations of the noise-biases, we will compare the results of the largest noise-biases, $N_B^{(0)}$ and $N_B^{(1)}$, with simulations of the lensed temperature CMB maps. 

This paper is organized as follows. In Sec.~\ref{sec:defQE}, we introduce the HO quadratic estimator for the lensing convergence field and establish our notation. In Sec.~\ref{sec:noisebiases_comp}, we discuss the origin of noise-biases that arise in the bispectrum of the reconstructed convergence, starting from the cumulant expansion of the lensed CMB 6-pt function. We also present the direct (``Wick expansion'') calculation of these biases along with the diagrammatic approach. 
In Sec.~\ref{sec:comp:N}, we show our numerical results and in Sec.~\ref{sec:validation} we validate $N_B^{(0)}$ and $N_B^{(1)}$ using simulations of the lensed temperature CMB maps. We draw our conclusions and outline perspectives for future work in Sec.~\ref{sec:conclusions}. Finally, in App.~\ref{appen:N32} we give the full technical details of the computation of the $N_B^{(3/2)}$ bias, and in App.~\ref{app:paired_sims} we describe an alternative set of simulations to test our theoretical findings. 

\paragraph*{\textbf{Notation and conventions}} Throughout the paper, we adopt the flat-sky approximation. We denote 2D Fourier wavenumbers by $\bsl$ for CMB temperature fields and $\bsL$ for the lensing potential/convergence. In general, the modulus of a vector is defined as, e.g., $\ell \equiv |\bsl|$. 

The power spectrum and the bispectrum of the relevant observables are defined, respectively, as
\begin{equation}\label{eq:def_power_spectrum}
    \langle X(\bsl)X(\bsl')\rangle = (2\pi)^2\, \delta^{(2)}(\bsl+\bsl') C_\ell^{XX},
\end{equation}
and
\begin{equation}\label{eq:def_bispectrum}
    \langle X(\bsl_1) \,X(\bsl_2) \, X(\bsl_3) \rangle= (2 \pi)^2 \delta^{(2)}(\bsl_1 + \bsl_2 + \bsl_3) \, B_X(\ell_1 ,\ell_2 , \ell_3) \, ,
\end{equation}
where $X$ can be either $T$, $\phi$ or $\kappa$. We further use the notation $n_C$ to denote connected $n$-pt functions.
We use the Fourier convention (in 2D)
\begin{align}\label{eq:fourier_transform}
    \mathcal{F}[Y(\hat{\boldsymbol{n}})](\bsl) &=\int \mathrm{d}^2\hat{\boldsymbol{n}} \, Y(\hat{\boldsymbol{n}})\, e^{- i \bsl \cdot \hat{\boldsymbol{n}}},\\ 
    \mathcal{F}^{-1}[Y(\bsl)](\hat{\boldsymbol{n}}) &= \int \frac{\mathrm{d^2}\bsl}{(2\pi)^2} \, Y(\bsl)\, e^{i \bsl \cdot \hat{\boldsymbol{n}}}.
\end{align}

\noindent For brevity, we introduce the compact notations $\boldsymbol{\ell}_{ij\dots n} =  \boldsymbol{\ell}_i+\boldsymbol{\ell}_j +\dots+\boldsymbol{\ell}_n$, $\bar{\bsL}_{ij} = \bsL_i-\bsl_j$ (where either $i=j$ or $i\neq j$) and 
\begin{equation*}
    \int_{\bsl_i,\dots,\,\bsl_n} = \int \frac{\mathrm{d^2}\bsl_i}{(2\pi)^2}\ldots  \int \frac{\mathrm{d^2}\bsl_n}{(2\pi)^2}\,.
\end{equation*}

\noindent Additionally, the lensed temperature fields $\widetilde{T}(\bsL_i-\bsl_j)$ and $\widetilde{T}(\bsl_i)$ are sometimes represented as $\widetilde{T}_{ij}$ and $\widetilde{T}_i$, respectively. 
In our numerical computations, we consider a flat $\Lambda$CDM cosmology, with cosmological parameters in accordance with the latest \textit{Planck} results~\cite{planck2018:cosmological_parameters}, summarized in Tab.~\ref{tab:cosmology}. 
\begin{table}[t]
    \centering
    \begin{tabular}{l  l  l}
     \toprule
     & $\boldsymbol{\Lambda}$\textbf{CDM} \textbf{parameters} &\\
     \midrule
        $H_0 = 67.32\,\text{km\,s}^{-1}\,\text{Mpc}^{-1}$ &  $\Omega_\mathrm{b} h^2 = 0.02238$ & $\sum m_\nu = 0.06\,\text{eV}$\\
        $\Omega_K = 0$ &  $\Omega_\mathrm{c} h^2 = 0.12010$  & $\tau = 0.0543$\\
         $n_s = 0.9660$& $A_\mathrm{s} = 2.1005\times 10^{-9}$  & $r = 0$\\
        \bottomrule
    \end{tabular}
    \caption{Best-fit Planck parameters (specifically, Tab.~1 of Ref.~\cite{planck2018:cosmological_parameters} with $TT,TE,EE\text{+low$E$+lensing+BAO}$) used in our numerical computations.}
    \label{tab:cosmology}
\end{table}

\section{The Hu \& Okamoto quadratic estimator} \label{sec:defQE}

In this section, we briefly review the effect of weak gravitational lensing on the CMB temperature anisotropies, and introduce the HO quadratic estimator~\cite{hu:mass_reconstruction_CMB_polarization} in the flat-sky approximation. 

The effect of lensing is usually expressed as a remapping of the primary CMB anisotropies
\begin{equation}
\label{eq:lens_remap}
    \widetilde{T}(\hat{\boldsymbol{n}}) = T(\hat{\boldsymbol{n}}') = T(\hat{\boldsymbol{n}}+\nabla\phi(\hat{\boldsymbol{n}}))\, ,
\end{equation}
where $\hat{\boldsymbol{n}}$ is the line-of-sight direction on the sky, $T$ refers to the \textit{unlensed} temperature field and $\widetilde{T}$ indicates the \textit{lensed} counterpart. We shall adopt this notation from here on. 

The lens remapping in Eq.~\eqref{eq:lens_remap} is given by the deflection angle $\boldsymbol{d}(\hat{\boldsymbol{n}})=\nabla\phi(\hat{\boldsymbol{n}})$, where the lensing potential $\phi(\hat{\boldsymbol{n}})$ is determined by the projection of the gravitational potential along the line-of-sight:
\begin{equation}\label{eq:def_phi}
     \phi(\hat{\boldsymbol{n}}) = -2 \, \int_{0}^{\chi_*} \mathrm{d}\chi \, \frac{\chi_* - \chi}{\chi_*\chi} \,\Psi (\chi\hat{\boldsymbol{n}};\eta_0-\chi)\,.
\end{equation}
Here, $\chi$ denotes the comoving distance, $\eta$ the conformal time (with present-day value $\eta_0$) and $\Psi$ is the (Weyl) gravitational potential. The definition in Eq.~\eqref{eq:def_phi} assumes the Born approximation, whereby the integration is carried out along the undeflected photon path, which is valid when the deflection angles are small compared to the scales on which the lenses vary.\footnote{Post-Born corrections to lensing are, however, an important source of the convergence bispectrum along with the contribution from non-Gaussian matter clustering~\cite{pratten:post_born_lensing}. Post-Born effects also introduce a curl-mode to the lensing deflections, so they are no longer describable by the lensing potential alone. We ignore the small effect of curl-modes in the deflections throughout this paper and leave their analysis to future work. }

In the following we will also use the lensing convergence $\kappa(\hat{\boldsymbol{n}})$, which captures the isotropic distortion due to weak gravitational lensing and is related to the lensing potential via
\begin{equation} \label{eq:def_kappa}
    \kappa(\hat{\boldsymbol{n}})= -\frac{1}{2}\nabla^2\phi(\hat{\boldsymbol{n}}) \, .
\end{equation}

We can expand the lensed temperature modes Eq.~\eqref{eq:lens_remap} (up to second order in $\phi$) as
\begin{equation} \label{eq:lens_expanded}
\widetilde{T}(\hat{\boldsymbol{n}}) =  T(\hat{\boldsymbol{n}}) +  \nabla T(\hat{\boldsymbol{n}})  \cdot \nabla \phi(\hat{\boldsymbol{n}}) + \frac{1}{2}  \nabla_i \nabla_j T(\hat{\boldsymbol{n}}) \,\, \nabla^i \phi(\hat{\boldsymbol{n}}) \, \nabla^j \phi(\hat{\boldsymbol{n}}) \, + \mathcal O(\phi^3) \, .
\end{equation}
Replacing the Fourier transform of the CMB temperature anisotropies and the lensing potential as defined in Eq.~\eqref{eq:fourier_transform}, we find
\begin{equation}
        \widetilde{T}(\bsl) =  T(\bsl) + \delta T(\bsl) + \delta^2 T(\bsl) + \mathcal O(\phi^3) \,, 
\label{eq:expansion_fourier_theta}
\end{equation}
where the $\mathcal O(\phi)$ and $\mathcal O(\phi^2)$ terms are respectively given by
\begin{align}
    \delta T(\bsl) &=  -\int_{\bsp}  \, \Big[\bsp\cdot (\bsl-\bsp)\Big] \,  \phi(\bsl-\bsp) \, T(\bsp) \, ,\label{eq:delta_T}\\
    \delta^2 T(\bsl) &= - \frac{1}{2} \int_{\bsp} \int_{\bsq}  \,
 \Big[\bsp\!\cdot\!(\bsp +\bsq-\bsl)\Big]\, (\bsp\!\cdot\! \bsq) \, T(\bsp) \, \phi(\bsq) \, \phi(\bsp+\bsq-\bsl) \,.
\label{eq:delta_T2}
\end{align}

Effectively, a fixed lens introduces a correlation between temperature modes that would otherwise be independent. Averaging over the primary CMB fluctuations for fixed lenses, we have at leading order
\begin{equation}
\langle \widetilde{T}(\bsl_1) \widetilde{T}(\bsl_2) \rangle_{\text{CMB}} = (2\pi)^2 \delta^{(2)}(\bsl_1+\bsl_2)C_{\ell_1}^{TT} 
+ f(\bsl_1,\bsl_2)  \phi(\bsl_1+\bsl_2) \, ,
\end{equation}
where the lensing response function
\begin{equation}
f(\bsl_1,\bsl_2) \equiv (\bsl_1+\bsl_2)\cdot\left(\bsl_1 C_{\ell_1}^{TT} + \bsl_2 C_{\ell_2}^{TT} \right) \,, 
\label{eq:small_f}
\end{equation}
and $C_\ell^{TT}$ is the angular power spectrum of the unlensed CMB.

It is sometimes useful to consider a non-perturbative extension of the lensing response function, defined by
\begin{equation}\label{eq:non_pert_response_func}
\left\langle \frac{\delta}{\delta \phi(\boldsymbol{L})}
\left[\widetilde{T}(\bsl_1)\widetilde{T}(\bsl_2)\right]\right\rangle = \tilde{f}(\bsl_1,\bsl_2) \delta^{(2)}(\bsL-\bsl_1-\bsl_2) \, ,
\end{equation}
where the expectation value is now over both the unlensed CMB and the lenses. It is shown in Ref.~\cite{lewis:shape_lensing_bispec} that the non-perturbative response function has the same form as Eq.~\eqref{eq:small_f} but with the unlensed CMB power spectrum replaced by the spectrum $\tilde{C}_\ell^{T\nabla T}$, which describes the correlation between the lensed temperature and the lensed gradient of the temperature. 
The non-perturbative response function is useful as it captures (some) beyond-leading-order terms due to lensing in the connected moments of the lensed CMB, whose biases can easily be mitigated by an appropriate modification of the normalisation of the quadratic estimator used for lensing reconstruction. Throughout this work, we will adopt the non-perturbative approach and use Eq.~\eqref{eq:non_pert_response_func}, which we rename $f(\bsl_1,\bsl_2)$ for convenience.

From the resulting off-diagonal terms in the 2-pt correlation function, we can introduce an estimator to reconstruct the lensing potential from CMB maps that is quadratic in the anisotropies:
\begin{equation}
    \hat{\phi}(\bsL) = A_L \int_{\bsl_1,\bsl_2}\, (2\pi)^2 \delta^{(2)}(\bsL-\bsl_1-\bsl_2)F(\bsl_1,\bsl_2) \, \widetilde{T}(\bsl_1)\widetilde{T}(\bsl_2) \,.
    \label{eq:lensing_potential_estimator}
\end{equation}
Here, the weight function
\begin{equation} \label{eq:optimal_filt}
    F(\bsl_1,\bsl_2) = \frac{f(\bsl_1,\bsl_2)}{2 C^{TT,\, \mathrm{tot}}_{\ell_1} C^{TT,\, \mathrm{tot}}_{\ell_2}}
\end{equation}
is chosen to minimize the Gaussian variance of the estimator under the constraint that the estimator is unbiased (see below).

The power spectrum $C^{TT,\, \mathrm{tot}}_{\ell}$ denotes the total temperature power in the map, including CMB, noise, and foreground power. Given that the goal of this work is to give a first estimate of the noise-biases, we generally do not consider any particular CMB experimental specifications and we assume that the total power spectrum is given by the lensed power spectrum, $C^{TT,\, \mathrm{tot}}_{\ell} = C^{\tilde{T}\tilde{T}}_{\ell}$.

The normalization of the quadratic estimator $A_L$ is derived assuming that the resulting lensing map is unbiased, i.e., 
$\langle \delta \hat{\phi}(\bsL) / \delta \phi(\bsL') \rangle = \delta^{(2)}(\bsL-\bsL')$ (cf.~Eq.~\ref{eq:non_pert_response_func});
it is given by
\begin{equation}
    A_L = \left[ \int_{\bsl_1} \, F(\bsl_1,\bsL-\bsl_1)f(\bsl_1,\bsL-\bsl_1)\right]^{-1} \,.
    \label{eq:norm_qest}
\end{equation}

Finally, using Eq.~\eqref{eq:def_kappa} in Fourier space, we may write the quadratic estimator for the lensing convergence field as
\begin{equation}
    \hat{\kappa}(\bsL) = \frac{1}{2}L^2\hat{\phi}(\bsL) = A_L^{\kappa}\int_{\bsl_1,\bsl_2}\, (2\pi)^2 \delta^{(2)}(\bsL-\bsl_1-\bsl_2)F(\bsl_1,\bsl_2) \, \widetilde{T}(\bsl_1)\widetilde{T}(\bsl_2)\, ,
    \label{eq:lensing_convergence_estimator}
\end{equation}
where $A_L^{\kappa} = L^2 A_L/2$. 


\section{Noise-biases in the reconstructed lensing bispectrum} \label{sec:noisebiases_comp}

It is well known that statistical reconstruction noise, arising from chance fluctuations in the CMB fields that mimic the effects of lensing, is present in the OH quadratic estimator. This leads to ``noise-biases'' when using the $n$-point correlation functions of the reconstructed $\hat{\kappa}$ to estimate the correlation functions of the underlying lensing convergence (see, e.g.,~\cite{hanson:lens_rec} for the power spectrum case).
In this paper, we calculate these biases for the first time for the bispectrum of the reconstructed convergence. Quantifying the biases carefully is crucial as their magnitude is typically much larger than the signal we aim to detect. For power spectrum reconstruction on data, the biases are now commonly removed with realization-dependent methods that mix data and simulations~\cite{hanson:lens_rec,namikawa:bias_hardened,schmittfull:cross_corr_cmb_galaxy}. These methods reduce sensitivity to any mismatch between the statistical properties of the simulations and reality (e.g., due to complex instrumental noise properties) and also reduce covariance between multipoles of the reconstructed power spectrum. Similar methods have been developed for a general $n$-pt function and need to be tested for the bispectrum. However, as a first step here we shall simply compute the dominant biases in the ideal case of an isotropic (and, generally, noiseless) survey.

A generic $n$-pt function of the lensing convergence field estimator reads as
\begin{align}\label{eq:nthpointfunction}
    \langle\hat{\kappa}(\bsL_1)\hat{\kappa}(\bsL_2)  \ldots \hat{\kappa}(\bsL_n)\rangle_{{\rm CMB}, \phi} & = \int'_{\bsl_1,\bsl_2,\dots,\,\bsl_n} F(\bsl_1,\bar{\bsL}_{11}) \, F(\bsl_2,\bar{\bsL}_{22}) \ldots F(\bsl_n,\bar{\bsL}_{nn}) \nonumber\\
        &\qquad\qquad\qquad\times\langle\widetilde{T}_1\widetilde{T}_{11}\widetilde{T}_2\widetilde{T}_{22} \cdots \widetilde{T}_n\widetilde{T}_{nn} \rangle_{{\rm CMB}, \phi}\, ,
\end{align}
where the hat symbols refers to the reconstructed observable and
\begin{equation}\label{eq:shorthand_norm_int}
    \int'_{\bsl_1,\bsl_2,\dots,\,\bsl_n}\equiv\left(\prod_{i=1}^3  A_{L_i}^{\kappa}\right)  \, \int_{\bsl_1,\bsl_2,\dots,\,\bsl_n}.
\end{equation}
The average is performed over many realizations of either the CMB or the lensing potential, as indicated by the suffix `${\rm CMB}, \phi$'. From now on, we drop this notation the sake of brevity. 

In order to compute Eq.~\eqref{eq:nthpointfunction}, we need to evaluate the $2n$-pt function of the lensed temperature fields
\begin{equation} \label{eq:lensed_2n_point}
\langle\widetilde{T}(\bsl_1) \, \widetilde{T}(\bsL_1 - \bsl_1) \, \widetilde{T}(\bsl_2) \, \widetilde{T}(\bsL_2 - \bsl_2) \, ... \, \widetilde{T}(\bsl_n) \, \widetilde{T}(\bsL_n - \bsl_n) \rangle \, ,
\end{equation}
which can be decomposed in connected $m$-pt functions ($m\leq 2n$).
In the case of the 3-pt function of $\hat{\kappa}$, the reconstructed lensing signal receives contributions from the 6-pt function of lensed CMB modes
\begin{align}\label{eq:decomp_6points}
    \langle \widetilde{T}\,\widetilde{T}\, \widetilde{T}\, \widetilde{T} \,\widetilde{T}\,\widetilde{T}\rangle = \, &\langle\widetilde{T}\,\widetilde{T}\rangle_C \, \langle\widetilde{T}\,\widetilde{T}\rangle_C \, \langle\widetilde{T}\,\widetilde{T}\rangle_C + \langle\widetilde{T}\,\widetilde{T}\,\widetilde{T}\,\widetilde{T}\rangle_C \,  \langle\widetilde{T}\,\widetilde{T}\rangle_C \nonumber\\
        &+\langle\widetilde{T}\,\widetilde{T}\,\widetilde{T}\rangle_C \,  \langle\widetilde{T}\,\widetilde{T}\,\widetilde{T}\rangle_C+ \langle\widetilde{T}\,\widetilde{T}\,\widetilde{T}\,\widetilde{T}\,\widetilde{T}\,\widetilde{T}\rangle_C +(\mathrm{perm.})\, ,
\end{align}
where `perm.' stands for permutations over the multipoles $\bsl_i$. 
In the following, we shall ignore the 3-pt function of the lensed CMB, $\langle \tilde{T}\tilde{T}\tilde{T}\rangle_C$. This is only sourced by the correlation between lensing and the late-time integrated Sachs--Wolfe contribution to the unlensed CMB, and should be negligible for the small-scale CMB modes that dominate lensing reconstruction.

When expanding the connected moments in Eq.~\eqref{eq:decomp_6points} perturbatively in $\phi$ (using Eq.~\eqref{eq:expansion_fourier_theta}), we obtain the signal bispectrum and a number of noise-biases:
\begin{equation} \label{eq:kkk_noise_biase_and_signal}
    \begin{split}
        \langle\hat{\kappa}(\bsL_1)\hat{\kappa}(\bsL_2)\hat{\kappa}(\bsL_3)\rangle = (2 \pi)^2 \delta(\bsL_{123})\bigg[B_\kappa (\bsL_1,\bsL_2,\bsL_3) + N_\mathrm{B}^{(0)}+ N_\mathrm{B}^{(1)} +  N_\mathrm{B}^{(3/2)} + N_\mathrm{B}^{(2)}\bigg] \,.
    \end{split}
\end{equation}
The signal term $B_\kappa$ arises from the $\mathcal{O}(\phi^3)$ part of the connected 6-pt function of the lensed CMB in which the unlensed CMB fields within each quadratic estimator are contracted. This involves terms like
\begin{equation}
\langle \widetilde{T}_1 \widetilde{T}_{11} \widetilde{T}_2 \widetilde{T}_{22} \widetilde{T}_3 \widetilde{T}_{33} \rangle_C \supset 
 \langle\wick{\c1 \delta T_1 \c1 T_{11}\; \c1 \delta T_2 \c1 T_{22} \; \c1 \delta T_3 \c1 T_{33}}\rangle_\phi \, ,
\end{equation}
where the indicated (Gaussian) contractions are over the unlensed CMB and the remaining expectation over $\phi$ returns the bispectrum.
For the noise-biases in Eq.~\eqref{eq:kkk_noise_biase_and_signal}, the superscripts denote the leading order at which they arise in an expansion in $\phi$, with $N^{(j/2)}_\text{B} = \mathcal{O}(\phi^j)$.
Similarly to lensing power spectrum reconstruction, $N_\mathrm{B}^{(0)}$ arises from the totally disconnected term in the decomposition of Eq.~\eqref{eq:decomp_6points} and would be present even in the absence of lensing. Here, we compute it non-perturbatively using the lensed temperature power spectrum. 
The noise-bias $N_\mathrm{B}^{(1)}$ comes from the connected 4-pt function up to order $\mathcal{O}(\phi^2)$, namely linear in the lensing potential power spectrum (thus $j=2$). For instance, one such contributing term is
\begin{equation} 
\label{eq:example_connected}
\langle \widetilde{T}\widetilde{T}\widetilde{T}\widetilde{T}\rangle_C \langle \widetilde{T}\widetilde{T} \rangle_C \supset
\langle\wick{\c1 \delta T \c1 T\;\c1 \delta T \c1 T}\rangle_\phi \langle \widetilde{T}  \widetilde{T} \rangle_C \, ,
\end{equation}
where, again, the contractions over the unlensed CMB are shown explicitly.
The terms of order $\mathcal{O}(\phi^3)$ in the connected 4- and 6-pt functions are captured by $N_{\mathrm{B}}^{(3/2)}$, which is linear in the lensing potential bispectrum ($j=3$). Finally, $N_{\mathrm{B}}^{(2)}$ arises from the part of the connected 6-pt function that is quadratic in the lensing potential power spectrum ($j=4$) and from corrections at this order to the connected 4-pt function. These
terms are therefore of order $\mathcal{O}(\phi^4)$. See Tab.~\ref{tab:noise_bias} for a summary.

In the following we review two complementary approaches to evaluate analytically these noise-biases, the Wick-expansion and a Feynman diagram approach, and show the equivalence of the two methods by evaluating the $N^{(0)}_\mathrm{B}$ and $N^{(1)}_\mathrm{B}$ noise-biases. We leave the computation of $N_{\mathrm{B}}^{(2)}$ for future work. The largest terms that are formally $\mathcal{O}(\phi^4)$ should already be captured by our use of non-perturbative response functions in the calculation of $N^{(1)}$~\cite{lewis:shape_lensing_bispec} and the lensed CMB power spectra for $\langle\widetilde{T}\widetilde{T}\rangle_C$ terms, and our numerical results in Sec.~\ref{sec:noisebiases_comp} show that the residual $N^{(2)}_\text{B}$ noise-bias is much smaller than the other bias terms .

\begin{table}[htbp]
\renewcommand{\arraystretch}{1.6} 
\centering
\begin{tabular}{cccl}
\textbf{}  & \multicolumn{3}{c}{\textbf{Noise-biases}}  \\ \cline{2-4} 
\multicolumn{1}{c|}{} & \multicolumn{1}{c|}{$\langle (\widetilde{T})^2\rangle_C \langle(\widetilde{T})^2\rangle_C \langle(\widetilde{T})^2 \rangle_C$} & \multicolumn{1}{c|}{$\langle(\widetilde{T})^4\rangle_C \langle(\widetilde{T})^2\rangle_C$}&\multicolumn{1}{c|}{ $\langle(\widetilde{T})^6\rangle_C$} \\ \hline
\multicolumn{1}{|c|}{all orders} & \multicolumn{1}{c|}{$N_\mathrm{B}^{(0)}$} & \multicolumn{1}{c|}{\cellcolor[HTML]{C0C0C0}} & \multicolumn{1}{c|}{\cellcolor[HTML]{C0C0C0}} \\ \hline
\multicolumn{1}{|c|}{$\mathcal{O}(\phi^2)$} & \multicolumn{1}{c|}{\cellcolor[HTML]{C0C0C0}} & \multicolumn{1}{c|}{$N_\mathrm{B}^{(1)}$} & \multicolumn{1}{c|}{\cellcolor[HTML]{C0C0C0}} \\ \cline{1-1} \cline{3-4} 
\multicolumn{1}{|c|}{$\mathcal{O}(\phi^3)$} & \multicolumn{1}{c|}{\cellcolor[HTML]{C0C0C0}} & \multicolumn{2}{c|}{$N_\mathrm{B}^{(3/2)}$} \\ \cline{1-1} \cline{3-4} 
\multicolumn{1}{|c|}{$\mathcal{O}(\phi^4)$} & \multicolumn{1}{c|}{\multirow{-3}{*}{\cellcolor[HTML]{C0C0C0}}} & \multicolumn{2}{c|}{$N_\mathrm{B}^{(2)}$} \\ \hline
\end{tabular}
\caption{Noise-bias terms of $\langle\hat{\kappa}\hat{\kappa}\hat{\kappa}\rangle$ and their origin in terms of the decomposition \eqref{eq:decomp_6points}.}
\label{tab:noise_bias}
\end{table}
\subsection{Wick-expansion for correlation functions} 
\label{subsec:Wick}
In general, an $n$-pt function as in Eq.~\eqref{eq:lensed_2n_point} can be computed with Wick's theorem in the case that the unlensed CMB and $\phi$ are Gaussian.
In this case, the only fundamental contraction is the 2-pt function defined in Eq.~\eqref{eq:def_power_spectrum}:
\begin{equation} \label{eq:Wc2T}
\contraction[1ex]{}{X}{(\bsl_1) \,\,}{X}
X(\bsl_1) \,\, X(\bsl_2) \,, \quad X = \{T, \phi\}.
\end{equation}
On the other hand, if some level of non-Gaussianity arises in either the temperature modes\footnote{We shall ignore this possibility in the rest of this paper.} or the lensing potential $\phi$, resulting in non-zero bispectra~\eqref{eq:def_bispectrum}, we have an additional fundamental contraction 
\begin{equation} \label{eq:Wc3t}
\contraction[1ex]{}{\phi}{(\bsl_1)\,\,}{X}
\contraction[1ex]{}{\phi}{(\bsl_1)\,\, T(\bsl_2) \,\,}{X}
X(\bsl_1) \,\, X(\bsl_2) \,\, X(\bsl_3)  \,, \quad X = \{T, \phi\}.
\end{equation}
Then, the lensed temperature $2n$-pt function \eqref{eq:lensed_2n_point} can be evaluated by expanding the lensed temperature modes in terms of the lensing potential and by taking all the terms allowed by the fundamental contractions Eqs.~(\ref{eq:Wc2T}--\ref{eq:Wc3t}).

\subsubsection*{$N^{(0)}_\mathrm{B}$ noise-bias: Wick approach.}
The $N_\mathrm{B}^{(0)}$ noise-bias comes from the totally disconnected term of the decomposition~\eqref{eq:decomp_6points} and reflects the non-Gaussian nature of the reconstruction noise in $\hat{\kappa}$. Here, we consider the full (non-perturbative) term
\begin{equation} 
    \begin{split}
        \langle \widetilde{T}(\bsl_1)\widetilde{T}(\bsL_1-\bsl_1)\rangle_C \, \langle \widetilde{T}(\bsl_2)\widetilde{T}(\bsL_2-\bsl_2)\rangle_C \, \langle \widetilde{T}(\bsl_3)\widetilde{T}(\bsL_3-\bsl_3)\rangle_C +\mathrm{(perms.)}.\, 
    \end{split}
\end{equation}
In total, there are 15 permutations that need to be taken into account, given that we must combine six momenta in groups of two. However, seven of these permutations involve at least one contraction over fields in the same quadratic estimator and are therefore proportional to $\delta(\bsL_i)$. For the $\bsL_i \neq 0$, we are left with eight terms, for example,
\begin{equation}
    \langle \widetilde{T}_1\widetilde{T}_3\rangle_C \, \langle \widetilde{T}_2\widetilde{T}_{11}\rangle_C \, \langle \widetilde{T}_{22}\widetilde{T}_{33}\rangle_C \propto \delta^{(2)} (\bsl_1 + \bsl_3)\delta^{(2)} (\bsl_2 + \bar{\bsL}_{11}) \delta^{(2)} (\bar{\bsL}_{22} + \bar{\bsL}_{33})C^{\widetilde{T}\widetilde{T}}_{\ell_1} \, C^{\widetilde{T}\widetilde{T}}_{\ell_2} \, C^{\widetilde{T}\widetilde{T}}_{|\bar{\bsL}_{22}|} \, .
\end{equation}
Using the fact that the integrand in Eq.~\eqref{eq:nthpointfunction} is symmetrical with respect to an exchange of momenta $\bsl_i \leftrightarrow \bsL_i - \bsl_i$, it can be shown that the remaining terms are equivalent and so the lowest-order noise-bias is
\begin{align} \label{eq:N0_wick}
N^{(0)}_\mathrm{B} = 8\int'_{\bsl_1} \, &F(\bsl_1,\bsL_1-\bsl_1) \, F(\bsL_1-\bsl_1, \bsL_3 + \bsl_1) \, F(-\bsl_1,\bsL_3+\bsl_1) \nonumber \\
&\times C^{\widetilde{T}\widetilde{T}}_{\ell_1} \, C^{\widetilde{T}\widetilde{T}}_{|\bsL_1 - \bsl_1|} \, C^{\widetilde{T}\widetilde{T}}_{|\bsL_3 + \bsl_1|} \, .
\end{align}
Note that in the presence of instrumental noise, the lensed power spectra here would be replaced by $C_\ell^{TT,\text{tot}}$.
\subsubsection*{$N^{(1)}_\mathrm{B}$ noise-bias: Wick approach.}
 
The $N_\mathrm{B}^{(1)}$ noise-bias arises from the $4_C \times 2_C$ term in Eq.~\eqref{eq:decomp_6points} and its leading-order contribution is $\mathcal{O}(\phi^2)$.Note that terms of this order also appear in 
$N^{(0)}_\mathrm{B}$, but have already been accounted for there by using the lensed temperature power spectra in Eq.~\eqref{eq:N0_wick}.
Therefore, we are left with terms in $4_C \times 2_C$ such as
\begin{equation} \label{eq:connectted_to_N1}
\langle \widetilde{T}(\bsl_1)\widetilde{T}(\bsL_1 - \bsl_1)\widetilde{T}(\bsl_2)\widetilde{T}(\bsL_2 - \bsl_2)\rangle_C \,  \langle \widetilde{T}(\bsl_3)\widetilde{T}(\bsL_3 - \bsl_3)\rangle_C +\mathrm{(perms.)} \, .
\end{equation}
Let us first briefly show the computation of the connected 4-pt function (details can be found in Ref.~\cite{kesden:lensing_reconstruction}). At leading order, we have
\begin{align}
\langle \widetilde{T}(\bsl_1) \widetilde{T}(\bsl_2) \widetilde{T}(\bsl_3)  \widetilde{T}(\bsl_4) \rangle_C &= \frac{1}{2} \langle \wick{\c1 T(\bsl_1) \c2 T(\bsl_2) \c1 \delta T(\bsl_3) \c2 \delta T(\bsl_4)}\rangle_\phi + (\text{23 perms.}) \nonumber \\     
&= \frac{1}{8} f(\bsl_1,\bsl_3)f(\bsl_2,\bsl_4)\langle \phi(\bsl_1+\bsl_3)\phi(\bsl_2+\bsl_4) \rangle + (\text{23 perms.}) \nonumber \\
&= \frac{1}{8} (2\pi)^2 \delta^{(2)}(\bsl_{1234})f(\bsl_1,\bsl_3)f(\bsl_2,\bsl_4) C^{\phi\phi}_{|\bsl_1+\bsl_3|} C^{\phi\phi}_{|\bsl_1+\bsl_3|} + (\text{23 perms.}) \, ,
\end{align}
where the contractions indicated in the top line are over the unlensed CMB and
we have used\footnote{In practice, we use the non-perturbative response functions to account for some higher-order terms in $\phi$. In particular, for a given pair of lensed temperature fields, say $\widetilde{T}(\bsl_1) \widetilde{T}(\bsl_3)$, whose unlensed fields are contracted, the non-perturbative response functions account for contractions over additional pairs of $\phi$ fields within $\widetilde{T}(\bsl_1) \widetilde{T}(\bsl_3)$ leaving one $\phi$ field uncontracted (i.e., as an external leg).}
\begin{equation}
\wick{\c1 \delta T(\bsl_1) \c1 T(\bsl_2)} + \wick{\c1 T(\bsl_1) \c1 \delta T(\bsl_2)} = f(\bsl_1,\bsl_2)\phi(\bsl_1+\bsl_2) \, .
\end{equation}
Gathering the permutations, we have
\begin{equation}
\langle \widetilde{T}(\bsl_1) \widetilde{T}(\bsl_2) \widetilde{T}(\bsl_3)  \widetilde{T}(\bsl_4) \rangle_C = (2\pi)^2 \delta^{(2)}(\bsl_{1234}) \mathcal{T}(\bsl_1,\bsl_2,\bsl_3,\bsl_4) \, ,
\end{equation}
where the trispectrum is
\begin{multline}
\mathcal{T}(\bsl_1,\bsl_2,\bsl_3,\bsl_4) = C^{\phi\phi}_{|\bsl_1+\bsl_2|} f(\bsl_1,\bsl_2)f(\bsl_3,\bsl_4) + C^{\phi\phi}_{|\bsl_1+\bsl_3|} f(\bsl_1,\bsl_3)f(\bsl_2,\bsl_4) \\
+ C^{\phi\phi}_{|\bsl_1+\bsl_4|}
f(\bsl_1,\bsl_4)f(\bsl_2,\bsl_3) \, .
\end{multline}

Given this result, we may identify all the terms contributing to Eq.~\eqref{eq:connectted_to_N1} when summing over all the possible permutations of the momenta. In principle, we have 15 terms, but three of them are proportional to the $\delta(\bsL_i)$ since the $2_C$ term involves fields within the same quadratic estimator (as in the term shown explicitly in Eq.~\ref{eq:connectted_to_N1}). Then, thanks to the symmetry of the integrand in Eq.~\eqref{eq:nthpointfunction} with respect to the exchange of $\bsl_i$ and $\bsL_i - \bsl_i$, only three of the remaining 12 terms turn out to be independent (with a factor of 4) and are given by 
\begin{align}\label{eq:N1_partial}
     N^{(1)}_\mathrm{B} &= 4\int'_{\bsl_1,\, \bsl_2} \, F(\bsl_1,\bar{\bsL}_{11}) \, F(\bsl_2,\bar{\bsL}_{22}) \, F(-\bsl_2,\bsL_3+\bsl_2)\, C^{\widetilde{T}\widetilde{T}}_{\ell_2} \Big[ C^{\phi\phi}_{L_1} \, f(\bsl_1,\bar{\bsL}_{11}) \, f(\bar{\bsL}_{22},\bsL_3+\bsl_2)\nonumber\\
        &\quad+\,C^{\phi\phi}_{|\bsl_1+\bar{\bsL}_{22}|} \, f(\bsl_1,\bar{\bsL}_{22}) \, f(\bar{\bsL}_{11},\bsL_3+\bsl_2) \,+\, C^{\phi\phi}_{|\bsl_1+\bsl_2+\bsL_3|} \, f(\bsl_1,\bsL_3+\bsl_2) \, f(\bar{\bsL}_{11},\bar{\bsL}_{22})\Big]\nonumber\\
        &\quad  +\, (\bsL_1 \leftrightarrow \bsL_2)  + (\bsL_1 \leftrightarrow \bsL_3) .
\end{align}
This expression can be further simplified by a change of variables $\bsl_1' = \bsL_1 - \bsl_1$ and $\bsl_2' = \bsL_2 - \bsl_2$ (then renaming $\bsl_1' = \bsl_1$, $\bsl_2' = \bsl_2$) and applying the triangle condition $\bsL_1+\bsL_2+\bsL_3 = 0$. Ultimately we obtain 
\begin{equation} \label{eq:N1bperm}
     N^{(1)}_\mathrm{B}= N^{(1)}_\mathrm{B}(\bsL_1,\bsL_2,\bsL_3)+N^{(1)}_\mathrm{B}(\bsL_2,\bsL_1,\bsL_3)+N^{(1)}_\mathrm{B}(\bsL_3,\bsL_2,\bsL_1) \, ,
\end{equation}
with 
\begin{align}\label{eq:N1_final}
    N^{(1)}_\mathrm{B}(\bsL_1,\bsL_2,\bsL_3) = &4 \int'_{\bsl_1,\, \bsl_2} \, F(\bsl_1,\bsL_1-\bsl_1) \, F(\bsl_2,\bsL_2-\bsl_2) F(\bsL_2-\bsl_2,\bsL_1+\bsl_2)\,\nonumber \\ 
        &\qquad\qquad\qquad\times  C^{\widetilde{T}\widetilde{T}}_{|\bsL_2-\bsl_2|}\Big[C^{\phi\phi}_{L_1} \, f(\bsl_1,\bsL_1-\bsl_1) \, f(\bsl_2,-\bsL_1-\bsl_2)\,\nonumber\\ 
        &\qquad\qquad\qquad\; +\, 2  C^{\phi\phi}_{|\bsl_1+\bsl_2|} \, f(\bsL_1-\bsl_1,-\bsL_1-\bsl_2) \, f(\bsl_1,\bsl_2)\Big] \,  \,.
\end{align}
By making the change of variables $\bsl'_1 = \bsL_1 - \bsl_1$ and $\bsl'_2 = -\bsl_2 - \bsL_1$, it can be shown that $N^{(1)}_\mathrm{B}(\bsL_1,\bsL_2,\bsL_3)$ is symmetric under exchange of $\bsL_2$ and $\bsL_3$ when the triangle condition $\bsL_1+\bsL_2+\bsL_3 = 0$ is imposed. This symmetry ensures that the right-hand side of Eq.~\eqref{eq:N1bperm} is symmetric under all permutations of $\bsL_1$, $\bsL_2$ and $\bsL_3$ when the triangle condition holds.
\subsection{Feynman rules} \label{subsec:Feynman}

Similarly to the perturbative computation of scattering amplitudes that is based on the wide application of the Wick's theorem, one can introduce a set of Feynman rules from which Feynman diagrams can be constructed, each associated to a possible contribution to the $n$-pt functions in Eq.~\eqref{eq:nthpointfunction}. We will not give the explicit derivation of these rules, and we refer the reader to the original publications \cite{jenkins:feynman_diagrams1,jenkins:feynman_diagrams2} for more details. In summary, this method is applied as follows: 

\begin{itemize}
\item First, for each capital momentum $\bsL_i$ that appears in the $n$-pt function  \eqref{eq:nthpointfunction} we draw a ``rectangular'' block, each including two vertexes that correspond to the lensed temperatures. To it, we can associate the following Feynman rule\footnote{Note that this Feynman rule is slightly different by what shown in the original reference. This is because the lensing convergence field has a different normalization convention than the lensing potential field.} 
\begin{equation}\label{eq_fund_BB}
\begin{tikzpicture}[x=0.75pt,y=0.75pt,yscale=-1,xscale=1]
\draw   (129.67,6.41) -- (129.67,95.74) -- (100.33,95.74) -- (100.33,6.41) -- cycle ;
\draw   (108.59,21.41) .. controls (108.59,17.91) and (111.42,15.08) .. (114.92,15.08) .. controls (118.42,15.08) and (121.26,17.91) .. (121.26,21.41) .. controls (121.26,24.91) and (118.42,27.74) .. (114.92,27.74) .. controls (111.42,27.74) and (108.59,24.91) .. (108.59,21.41) -- cycle ;
\draw   (108.59,80.08) .. controls (108.59,76.58) and (111.42,73.74) .. (114.92,73.74) .. controls (118.42,73.74) and (121.26,76.58) .. (121.26,80.08) .. controls (121.26,83.58) and (118.42,86.41) .. (114.92,86.41) .. controls (111.42,86.41) and (108.59,83.58) .. (108.59,80.08) -- cycle ;

\draw    (61,21.08) -- (91.33,21.08) ;
\draw [shift={(94.33,21.08)}, rotate = 180] [fill={rgb, 255:red, 0; green, 0; blue, 0 }  ][line width=0.08]  [draw opacity=0] (10.72,-5.15) -- (0,0) -- (10.72,5.15) -- (7.12,0) -- cycle    ;
\draw    (61.67,80.74) -- (92,80.74) ;
\draw [shift={(95,80.74)}, rotate = 180] [fill={rgb, 255:red, 0; green, 0; blue, 0 }  ][line width=0.08]  [draw opacity=0] (10.72,-5.15) -- (0,0) -- (10.72,5.15) -- (7.12,0) -- cycle    ;

\draw (10.33,12.73) node [anchor=north west][inner sep=0.75pt]    {$\boldsymbol{L_{i} -\ell }$};
\draw (40.67,72.73) node [anchor=north west][inner sep=0.75pt]    {$\boldsymbol{\ell }$};
\draw (148.67,38.07) node [anchor=north west][inner sep=0.75pt]    {$=A_{L_{i}}^{\kappa } \, F(\boldsymbol{\ell },\boldsymbol{L_{i} -\ell })$};
\end{tikzpicture}
\end{equation}
where  $F(\bsl,\bsL_i-\bsl)$ is the (optimal) filter in Eq.~\eqref{eq:optimal_filt}. Therefore, when computing the 3-pt correlation function of $\hat{\kappa}$, we have three external fundamental blocks, for a total of 6 circular vertexes. Each vertex corresponds to a lensed temperature mode and the total number of vertexes coincides with the number of lensed temperature modes in the expectation value of Eq.~\eqref{eq:nthpointfunction}.

\item Coming out of each vertex, we can draw one straight line (that corresponds to the unlensed temperature field) and an arbitrary number of wiggly lines representing the order in the lensing potential field expansion. The corresponding Feynman rule for the $n$-wiggly lines vertex is
\begin{equation} \label{eq:unl_pot_vert}
\begin{tikzpicture}[x=0.75pt,y=0.75pt,yscale=-1,xscale=1]
\draw   (97.54,102.74) .. controls (97.54,99.25) and (100.47,96.41) .. (104.09,96.41) .. controls (107.7,96.41) and (110.64,99.25) .. (110.64,102.74) .. controls (110.64,106.24) and (107.7,109.08) .. (104.09,109.08) .. controls (100.47,109.08) and (97.54,106.24) .. (97.54,102.74) -- cycle ;
\draw    (108.33,108.41) -- (185,169.08) ;
\draw [shift={(146.67,138.74)}, rotate = 218.35] [fill={rgb, 255:red, 0; green, 0; blue, 0 }  ][line width=0.08]  [draw opacity=0] (10.72,-5.15) -- (0,0) -- (10.72,5.15) -- (7.12,0) -- cycle    ;
\draw  [dash pattern={on 0.84pt off 2.51pt}]  (109.64,103.74) -- (199.5,47.67) ;
\draw    (108.33,97.41) .. controls (108.36,95.06) and (109.56,93.89) .. (111.91,93.92) .. controls (114.27,93.95) and (115.46,92.78) .. (115.48,90.42) .. controls (115.51,88.06) and (116.7,86.89) .. (119.06,86.92) .. controls (121.41,86.95) and (122.6,85.78) .. (122.63,83.43) .. controls (122.66,81.07) and (123.85,79.9) .. (126.21,79.93) .. controls (128.56,79.96) and (129.75,78.79) .. (129.78,76.44) .. controls (129.81,74.08) and (131,72.91) .. (133.36,72.94) .. controls (135.71,72.97) and (136.9,71.8) .. (136.93,69.45) .. controls (136.96,67.09) and (138.15,65.92) .. (140.51,65.95) .. controls (142.87,65.98) and (144.06,64.81) .. (144.08,62.45) .. controls (144.11,60.09) and (145.3,58.93) .. (147.66,58.96) .. controls (150.02,58.99) and (151.21,57.82) .. (151.23,55.46) .. controls (151.26,53.1) and (152.45,51.94) .. (154.81,51.97) .. controls (157.17,52) and (158.36,50.83) .. (158.38,48.47) .. controls (158.41,46.11) and (159.6,44.95) .. (161.96,44.98) .. controls (164.32,45.01) and (165.51,43.84) .. (165.53,41.48) .. controls (165.56,39.12) and (166.75,37.95) .. (169.11,37.98) -- (172.5,34.67) -- (172.5,34.67) ;
\draw [shift={(140.42,66.04)}, rotate = 135.64] [fill={rgb, 255:red, 0; green, 0; blue, 0 }  ][line width=0.08]  [draw opacity=0] (10.72,-5.15) -- (0,0) -- (10.72,5.15) -- (7.12,0) -- cycle    ;
\draw    (109.64,102.74) .. controls (110.81,100.7) and (112.42,100.27) .. (114.46,101.44) .. controls (116.5,102.61) and (118.11,102.18) .. (119.29,100.14) .. controls (120.46,98.09) and (122.07,97.66) .. (124.12,98.83) .. controls (126.16,100) and (127.77,99.57) .. (128.94,97.53) .. controls (130.12,95.49) and (131.73,95.06) .. (133.77,96.23) .. controls (135.82,97.4) and (137.43,96.97) .. (138.6,94.92) .. controls (139.78,92.88) and (141.39,92.45) .. (143.43,93.62) .. controls (145.47,94.79) and (147.08,94.36) .. (148.25,92.32) .. controls (149.42,90.27) and (151.03,89.84) .. (153.08,91.01) .. controls (155.12,92.18) and (156.73,91.75) .. (157.91,89.71) .. controls (159.08,87.67) and (160.69,87.24) .. (162.73,88.41) .. controls (164.78,89.58) and (166.39,89.15) .. (167.56,87.1) .. controls (168.74,85.06) and (170.35,84.63) .. (172.39,85.8) .. controls (174.44,86.97) and (176.04,86.54) .. (177.21,84.49) .. controls (178.39,82.45) and (180,82.02) .. (182.04,83.19) .. controls (184.08,84.36) and (185.69,83.93) .. (186.87,81.89) .. controls (188.04,79.84) and (189.65,79.41) .. (191.7,80.58) .. controls (193.74,81.75) and (195.35,81.32) .. (196.52,79.28) .. controls (197.7,77.24) and (199.31,76.81) .. (201.35,77.98) -- (202.5,77.67) -- (202.5,77.67) ;
\draw [shift={(156.07,90.21)}, rotate = 164.89] [fill={rgb, 255:red, 0; green, 0; blue, 0 }  ][line width=0.08]  [draw opacity=0] (10.72,-5.15) -- (0,0) -- (10.72,5.15) -- (7.12,0) -- cycle    ;
\draw    (109.64,104.74) .. controls (111.41,103.19) and (113.07,103.29) .. (114.63,105.06) .. controls (116.19,106.83) and (117.85,106.93) .. (119.62,105.38) .. controls (121.39,103.83) and (123.05,103.93) .. (124.61,105.7) .. controls (126.16,107.47) and (127.82,107.57) .. (129.59,106.02) .. controls (131.36,104.47) and (133.02,104.57) .. (134.58,106.34) .. controls (136.14,108.11) and (137.8,108.21) .. (139.57,106.65) .. controls (141.34,105.1) and (143,105.2) .. (144.56,106.97) .. controls (146.12,108.74) and (147.78,108.84) .. (149.55,107.29) .. controls (151.32,105.74) and (152.98,105.84) .. (154.54,107.61) .. controls (156.1,109.38) and (157.76,109.48) .. (159.53,107.93) .. controls (161.3,106.37) and (162.96,106.47) .. (164.52,108.24) .. controls (166.08,110.01) and (167.74,110.11) .. (169.51,108.56) .. controls (171.28,107.01) and (172.94,107.11) .. (174.5,108.88) .. controls (176.06,110.65) and (177.72,110.75) .. (179.49,109.2) .. controls (181.26,107.65) and (182.92,107.75) .. (184.48,109.52) .. controls (186.04,111.29) and (187.7,111.39) .. (189.47,109.84) .. controls (191.24,108.28) and (192.9,108.38) .. (194.46,110.15) .. controls (196.02,111.92) and (197.68,112.02) .. (199.45,110.47) -- (202.5,110.67) -- (202.5,110.67) ;
\draw [shift={(156.07,107.71)}, rotate = 183.65] [fill={rgb, 255:red, 0; green, 0; blue, 0 }  ][line width=0.08]  [draw opacity=0] (10.72,-5.15) -- (0,0) -- (10.72,5.15) -- (7.12,0) -- cycle    ;
\draw    (57.67,102.41) -- (88,102.41) ;
\draw [shift={(91,102.41)}, rotate = 180] [fill={rgb, 255:red, 0; green, 0; blue, 0 }  ][line width=0.08]  [draw opacity=0] (10.72,-5.15) -- (0,0) -- (10.72,5.15) -- (7.12,0) -- cycle    ;

\draw (39.67,92.73) node [anchor=north west][inner sep=0.75pt]    {$\boldsymbol{\ell }$};
\draw (190,160.4) node [anchor=north west][inner sep=0.75pt]    {$\boldsymbol{m}$};
\draw (207.33,102.73) node [anchor=north west][inner sep=0.75pt]    {$\boldsymbol{k_{1}}$};
\draw (209,67.07) node [anchor=north west][inner sep=0.75pt]    {$\boldsymbol{k_{2}}$};
\draw (177.67,18.07) node [anchor=north west][inner sep=0.75pt]    {$\boldsymbol{k_{n}}$};
\draw (256.33,89.07) node [anchor=north west][inner sep=0.75pt]    {$=\ \prod _{i} -(\boldsymbol{m\ \cdotp \ k_{i}})$};

\end{tikzpicture}
\end{equation}
where $\bsl= \boldsymbol{m} + \sum_i \bsk_i$ due to the momentum conserving delta-function at the vertex. 
\item  Next, we define the expression of the propagators, which correspond to contractions between either straight or wiggly lines. Assuming Gaussian distributed lensing potential and unlensed temperature fields, we get the following Feynman rules
\begin{equation} \label{eq:propagators}
\begin{tikzpicture}[x=0.75pt,y=0.75pt,yscale=-1,xscale=1]

\draw    (127.95,365.89) -- (226,366.08) ;
\draw [shift={(176.98,365.99)}, rotate = 180.11] [fill={rgb, 255:red, 0; green, 0; blue, 0 }  ][line width=0.08]  [draw opacity=0] (10.72,-5.15) -- (0,0) -- (10.72,5.15) -- (7.12,0) -- cycle    ;
\draw    (128.5,406) .. controls (130.18,404.35) and (131.85,404.37) .. (133.5,406.05) .. controls (135.15,407.73) and (136.82,407.75) .. (138.5,406.1) .. controls (140.19,404.45) and (141.85,404.47) .. (143.5,406.16) .. controls (145.15,407.84) and (146.82,407.86) .. (148.5,406.21) .. controls (150.18,404.56) and (151.85,404.58) .. (153.5,406.26) .. controls (155.15,407.94) and (156.82,407.96) .. (158.5,406.31) .. controls (160.18,404.66) and (161.85,404.68) .. (163.5,406.36) .. controls (165.15,408.05) and (166.81,408.07) .. (168.5,406.42) .. controls (170.18,404.77) and (171.85,404.79) .. (173.5,406.47) .. controls (175.15,408.15) and (176.82,408.17) .. (178.5,406.52) .. controls (180.18,404.87) and (181.85,404.89) .. (183.5,406.57) .. controls (185.15,408.25) and (186.82,408.27) .. (188.5,406.62) .. controls (190.19,404.97) and (191.85,404.99) .. (193.5,406.68) .. controls (195.15,408.36) and (196.82,408.38) .. (198.5,406.73) .. controls (200.18,405.08) and (201.85,405.1) .. (203.5,406.78) .. controls (205.15,408.46) and (206.82,408.48) .. (208.5,406.83) .. controls (210.19,405.18) and (211.85,405.2) .. (213.5,406.89) .. controls (215.15,408.57) and (216.82,408.59) .. (218.5,406.94) .. controls (220.18,405.29) and (221.84,405.31) .. (223.49,406.99) -- (224.5,407) -- (224.5,407) ;
\draw [shift={(176.5,406.5)}, rotate = 180.6] [fill={rgb, 255:red, 0; green, 0; blue, 0 }  ][line width=0.08]  [draw opacity=0] (10.72,-5.15) -- (0,0) -- (10.72,5.15) -- (7.12,0) -- cycle    ;
\draw   (226,366.08) .. controls (226,362.14) and (229.13,358.95) .. (232.98,358.95) .. controls (236.84,358.95) and (239.96,362.14) .. (239.96,366.08) .. controls (239.96,370.01) and (236.84,373.21) .. (232.98,373.21) .. controls (229.13,373.21) and (226,370.01) .. (226,366.08) -- cycle ;
\draw   (225,406.08) .. controls (225,402.14) and (228.13,398.95) .. (231.98,398.95) .. controls (235.84,398.95) and (238.96,402.14) .. (238.96,406.08) .. controls (238.96,410.01) and (235.84,413.21) .. (231.98,413.21) .. controls (228.13,413.21) and (225,410.01) .. (225,406.08) -- cycle ;
\draw   (114.54,406) .. controls (114.54,402.06) and (117.66,398.87) .. (121.52,398.87) .. controls (125.37,398.87) and (128.5,402.06) .. (128.5,406) .. controls (128.5,409.94) and (125.37,413.13) .. (121.52,413.13) .. controls (117.66,413.13) and (114.54,409.94) .. (114.54,406) -- cycle ;
\draw   (113.99,365.89) .. controls (113.99,361.96) and (117.11,358.76) .. (120.97,358.76) .. controls (124.83,358.76) and (127.95,361.96) .. (127.95,365.89) .. controls (127.95,369.83) and (124.83,373.02) .. (120.97,373.02) .. controls (117.11,373.02) and (113.99,369.83) .. (113.99,365.89) -- cycle ;

\draw (257.67,351.41) node [anchor=north west][inner sep=0.75pt]    {$=C^{TT}_{\ell} \, ,$};
\draw (257.67,393.08) node [anchor=north west][inner sep=0.75pt]    {$=C^{\phi\phi}_{\ell} \, .$};
\draw (170.67,336.67) node [anchor=north west][inner sep=0.75pt]    {$\boldsymbol{\ell}$};
\draw (169.67,376.67) node [anchor=north west][inner sep=0.75pt]    {$\boldsymbol{\ell}$};
\end{tikzpicture}
\end{equation}
\item In the presence of a non-zero bispectrum for the lensing potential, we get the following additional Feynman rule
\begin{equation}
\begin{tikzpicture}[x=0.75pt,y=0.75pt,yscale=-1,xscale=1]

\draw    (213.5,115.33) .. controls (211.23,115.95) and (209.78,115.12) .. (209.16,112.85) .. controls (208.54,110.58) and (207.09,109.75) .. (204.82,110.37) .. controls (202.55,110.99) and (201.1,110.16) .. (200.48,107.89) .. controls (199.86,105.62) and (198.41,104.79) .. (196.14,105.41) .. controls (193.87,106.03) and (192.42,105.2) .. (191.79,102.93) .. controls (191.17,100.66) and (189.72,99.83) .. (187.45,100.45) .. controls (185.18,101.07) and (183.73,100.24) .. (183.11,97.97) .. controls (182.49,95.7) and (181.04,94.87) .. (178.77,95.49) .. controls (176.5,96.11) and (175.05,95.28) .. (174.43,93.01) .. controls (173.81,90.74) and (172.36,89.91) .. (170.09,90.53) .. controls (167.82,91.15) and (166.37,90.32) .. (165.75,88.05) .. controls (165.13,85.78) and (163.68,84.95) .. (161.41,85.56) .. controls (159.14,86.18) and (157.69,85.35) .. (157.06,83.08) .. controls (156.44,80.81) and (154.99,79.98) .. (152.72,80.6) .. controls (150.45,81.22) and (149,80.39) .. (148.38,78.12) .. controls (147.76,75.85) and (146.31,75.02) .. (144.04,75.64) .. controls (141.77,76.26) and (140.32,75.43) .. (139.7,73.16) -- (136.5,71.33) -- (136.5,71.33) ;
\draw [shift={(175,93.33)}, rotate = 29.74] [fill={rgb, 255:red, 0; green, 0; blue, 0 }  ][line width=0.08]  [draw opacity=0] (10.72,-5.15) -- (0,0) -- (10.72,5.15) -- (7.12,0) -- cycle    ;
\draw    (37.5,69.33) .. controls (39.16,67.66) and (40.83,67.65) .. (42.5,69.31) .. controls (44.17,70.96) and (45.84,70.95) .. (47.5,69.28) .. controls (49.16,67.61) and (50.83,67.6) .. (52.5,69.26) .. controls (54.17,70.91) and (55.84,70.9) .. (57.5,69.23) .. controls (59.16,67.56) and (60.83,67.55) .. (62.5,69.2) .. controls (64.17,70.86) and (65.84,70.85) .. (67.5,69.18) .. controls (69.16,67.51) and (70.83,67.5) .. (72.5,69.15) .. controls (74.17,70.81) and (75.84,70.8) .. (77.5,69.13) .. controls (79.16,67.46) and (80.83,67.45) .. (82.5,69.1) .. controls (84.17,70.76) and (85.84,70.75) .. (87.5,69.08) .. controls (89.16,67.41) and (90.83,67.4) .. (92.5,69.05) .. controls (94.17,70.7) and (95.84,70.69) .. (97.5,69.02) .. controls (99.16,67.35) and (100.83,67.34) .. (102.5,69) .. controls (104.17,70.65) and (105.84,70.64) .. (107.5,68.97) .. controls (109.16,67.3) and (110.83,67.29) .. (112.5,68.95) .. controls (114.17,70.6) and (115.84,70.59) .. (117.5,68.92) .. controls (119.16,67.25) and (120.83,67.24) .. (122.5,68.9) .. controls (124.17,70.55) and (125.84,70.54) .. (127.5,68.87) .. controls (129.16,67.2) and (130.83,67.19) .. (132.5,68.84) -- (134.5,68.83) -- (134.5,68.83) ;
\draw [shift={(86,69.08)}, rotate = 179.7] [fill={rgb, 255:red, 0; green, 0; blue, 0 }  ][line width=0.08]  [draw opacity=0] (10.72,-5.15) -- (0,0) -- (10.72,5.15) -- (7.12,0) -- cycle    ;
\draw    (217.5,29.33) .. controls (216.67,31.54) and (215.16,32.24) .. (212.95,31.41) .. controls (210.74,30.59) and (209.23,31.28) .. (208.4,33.49) .. controls (207.58,35.7) and (206.07,36.39) .. (203.86,35.57) .. controls (201.65,34.74) and (200.14,35.43) .. (199.31,37.64) .. controls (198.48,39.85) and (196.97,40.54) .. (194.76,39.72) .. controls (192.55,38.9) and (191.04,39.59) .. (190.21,41.8) .. controls (189.38,44.01) and (187.87,44.7) .. (185.66,43.88) .. controls (183.45,43.05) and (181.94,43.74) .. (181.12,45.95) .. controls (180.29,48.16) and (178.78,48.85) .. (176.57,48.03) .. controls (174.36,47.21) and (172.85,47.9) .. (172.02,50.11) .. controls (171.19,52.32) and (169.68,53.01) .. (167.47,52.19) .. controls (165.26,51.36) and (163.75,52.05) .. (162.92,54.26) .. controls (162.1,56.47) and (160.59,57.16) .. (158.38,56.34) .. controls (156.17,55.52) and (154.66,56.21) .. (153.83,58.42) .. controls (153,60.63) and (151.49,61.32) .. (149.28,60.5) .. controls (147.07,59.67) and (145.56,60.36) .. (144.73,62.57) .. controls (143.9,64.78) and (142.39,65.47) .. (140.18,64.65) -- (136.5,66.33) -- (136.5,66.33) ;
\draw [shift={(177,47.83)}, rotate = 335.45] [fill={rgb, 255:red, 0; green, 0; blue, 0 }  ][line width=0.08]  [draw opacity=0] (10.72,-5.15) -- (0,0) -- (10.72,5.15) -- (7.12,0) -- cycle    ;
\draw   (134.5,68.83) .. controls (134.5,67.45) and (135.4,66.33) .. (136.5,66.33) .. controls (137.6,66.33) and (138.5,67.45) .. (138.5,68.83) .. controls (138.5,70.21) and (137.6,71.33) .. (136.5,71.33) .. controls (135.4,71.33) and (134.5,70.21) .. (134.5,68.83) -- cycle ;
\draw   (217.5,29.33) .. controls (217.5,25.4) and (220.63,22.21) .. (224.48,22.21) .. controls (228.34,22.21) and (231.46,25.4) .. (231.46,29.33) .. controls (231.46,33.27) and (228.34,36.46) .. (224.48,36.46) .. controls (220.63,36.46) and (217.5,33.27) .. (217.5,29.33) -- cycle ;
\draw   (213.5,115.33) .. controls (213.5,111.4) and (216.63,108.21) .. (220.48,108.21) .. controls (224.34,108.21) and (227.46,111.4) .. (227.46,115.33) .. controls (227.46,119.27) and (224.34,122.46) .. (220.48,122.46) .. controls (216.63,122.46) and (213.5,119.27) .. (213.5,115.33) -- cycle ;
\draw   (23.54,69.33) .. controls (23.54,65.4) and (26.66,62.21) .. (30.52,62.21) .. controls (34.37,62.21) and (37.5,65.4) .. (37.5,69.33) .. controls (37.5,73.27) and (34.37,76.46) .. (30.52,76.46) .. controls (26.66,76.46) and (23.54,73.27) .. (23.54,69.33) -- cycle ;

\draw (69.67,40.4) node [anchor=north west][inner sep=0.75pt]    {$\boldsymbol{\ell _{1}}$};
\draw (166.67,14.4) node [anchor=north west][inner sep=0.75pt]    {$\boldsymbol{\ell _{2}}$};
\draw (158.67,100.4) node [anchor=north west][inner sep=0.75pt]    {$\boldsymbol{\ell _{3}}$};
\draw (252.33,67.73) node [anchor=north west][inner sep=0.75pt]    {$=B_\phi(\ell_1,\ell_2,\ell_3)$};
\end{tikzpicture}
\label{eq:3propagators}
\end{equation}
where the momenta $\bsl_i$ are incoming.
\item The sum of the momenta coming inside each vertex must give zero due to momentum conservation. As a result, not all the small momenta $\bsl_j$ are independent, but some of them turn out to be fixed in terms of the capital momenta $\bsL_i$ and other small independent momenta $\bsl_{i}$. For each of these one has to perform an integration $\int \mathrm{d^2} \bsl_i/(2\pi)^2$. Outside the momenta integration, a factor $(2\pi)^2 \delta(\sum_i \bsL_i)$ has to be added to the final result.
\end{itemize}

With these fundamental Feynman rules, one can compute the $N^{(j/2)}$ contribution ($j \geq 0 \in N$)\footnote{Notice that there is a $N^{(1/2)}$ term which is proportional to the cross-correlation $C^{T\phi}_\ell$, but we ignore such contributions given that both $C^{T\phi}_\ell/C^{TT}_\ell \ll 1$ and $C^{T\phi}_\ell/C^{\phi\phi}_\ell \ll 1$.} to the $n$-pt function of the estimated $\hat\kappa$ by writing down all the possible diagrams with $j$ total wiggly lines coming off the lensed vertexes in the building blocks. 
These rules are particularly convenient when dealing with the computation of $j/2>1$ noise-biases, where one must include the $\mathcal{O}(\phi^2)$ or even $\mathcal{O}(\phi^3)$ terms of the expansion~\eqref{eq:expansion_fourier_theta}, for which the Wick expansion becomes challenging. Moreover, the Feynman diagrams allow us to straightforwardly generalize the computations when including the CMB polarization fields. We refer to Refs.~\cite{jenkins:feynman_diagrams1,jenkins:feynman_diagrams2} for more details in this regard. 

Besides the fundamental Feynman rules, in Ref.~\cite{jenkins:feynman_diagrams2} some more composite ones have been derived. In the following, we report one which is useful for the computation of $N_\mathrm{B}^{(3/2)}$ (see App.~\ref{appen:N32}). More specifically, block \eqref{eq:f0} corresponds to all the possible ways in which we can connect two lensed temperature vertexes with only 1 uncontracted wiggly line coming off one of the two vertexes, diagrammatically 
\begin{equation} \label{eq:f0}
\centering
\begin{tikzpicture}[x=0.75pt,y=0.75pt,yscale=-1,xscale=1]

\draw   (330.76,48.79) .. controls (330.76,45.48) and (333.4,42.79) .. (336.64,42.79) .. controls (339.89,42.79) and (342.52,45.48) .. (342.52,48.79) .. controls (342.52,52.11) and (339.89,54.8) .. (336.64,54.8) .. controls (333.4,54.8) and (330.76,52.11) .. (330.76,48.79) -- cycle ;
\draw    (286.78,48.88) .. controls (288.45,50.55) and (288.45,52.21) .. (286.78,53.88) .. controls (285.11,55.55) and (285.11,57.21) .. (286.78,58.88) .. controls (288.45,60.55) and (288.45,62.21) .. (286.78,63.88) .. controls (285.11,65.55) and (285.11,67.21) .. (286.78,68.88) .. controls (288.45,70.55) and (288.45,72.21) .. (286.78,73.88) .. controls (285.11,75.55) and (285.11,77.21) .. (286.78,78.88) .. controls (288.45,80.55) and (288.45,82.21) .. (286.78,83.88) .. controls (285.11,85.55) and (285.11,87.21) .. (286.78,88.88) .. controls (288.45,90.55) and (288.45,92.21) .. (286.78,93.88) .. controls (285.11,95.55) and (285.11,97.21) .. (286.78,98.88) .. controls (288.45,100.55) and (288.45,102.21) .. (286.78,103.88) -- (286.78,105.11) -- (286.78,105.11) ;
\draw    (295.85,72.11) -- (295.85,91.54) ;
\draw [shift={(295.85,94.54)}, rotate = 270] [fill={rgb, 255:red, 0; green, 0; blue, 0 }  ][line width=0.08]  [draw opacity=0] (7.14,-3.43) -- (0,0) -- (7.14,3.43) -- (4.74,0) -- cycle    ;
\draw    (322.31,42.79) -- (301.75,42.79) ;
\draw [shift={(298.75,42.79)}, rotate = 360] [fill={rgb, 255:red, 0; green, 0; blue, 0 }  ][line width=0.08]  [draw opacity=0] (7.14,-3.43) -- (0,0) -- (7.14,3.43) -- (4.74,0) -- cycle    ;
\draw    (250.74,42.17) -- (269.21,42.17) ;
\draw [shift={(272.21,42.17)}, rotate = 180] [fill={rgb, 255:red, 0; green, 0; blue, 0 }  ][line width=0.08]  [draw opacity=0] (7.14,-3.43) -- (0,0) -- (7.14,3.43) -- (4.74,0) -- cycle    ;
\draw   (230.67,48.97) .. controls (230.67,45.65) and (233.3,42.97) .. (236.55,42.97) .. controls (239.8,42.97) and (242.43,45.65) .. (242.43,48.97) .. controls (242.43,52.28) and (239.8,54.97) .. (236.55,54.97) .. controls (233.3,54.97) and (230.67,52.28) .. (230.67,48.97) -- cycle ;
\draw   (64.25,46.08) .. controls (64.25,43.96) and (65.99,42.24) .. (68.13,42.24) .. controls (70.27,42.24) and (72,43.96) .. (72,46.08) .. controls (72,48.2) and (70.27,49.92) .. (68.13,49.92) .. controls (65.99,49.92) and (64.25,48.2) .. (64.25,46.08) -- cycle ;
\draw    (29,49.92) .. controls (30.67,51.59) and (30.67,53.25) .. (29,54.92) .. controls (27.33,56.59) and (27.33,58.25) .. (29,59.92) .. controls (30.67,61.59) and (30.67,63.25) .. (29,64.92) .. controls (27.33,66.59) and (27.33,68.25) .. (29,69.92) .. controls (30.67,71.59) and (30.67,73.25) .. (29,74.92) -- (29,76.11) -- (29,76.11) ;
\draw   (156.25,46.08) .. controls (156.25,43.96) and (157.99,42.24) .. (160.13,42.24) .. controls (162.27,42.24) and (164,43.96) .. (164,46.08) .. controls (164,48.2) and (162.27,49.92) .. (160.13,49.92) .. controls (157.99,49.92) and (156.25,48.2) .. (156.25,46.08) -- cycle ;
\draw   (117,45.85) .. controls (117,43.61) and (118.79,41.79) .. (121,41.79) .. controls (123.21,41.79) and (125,43.61) .. (125,45.85) .. controls (125,48.1) and (123.21,49.92) .. (121,49.92) .. controls (118.79,49.92) and (117,48.1) .. (117,45.85) -- cycle ;
\draw    (33,46.08) -- (64.25,46.08) ;
\draw    (125,45.85) -- (156.25,45.85) ;
\draw    (160,49.92) .. controls (161.67,51.59) and (161.67,53.25) .. (160,54.92) .. controls (158.33,56.59) and (158.33,58.25) .. (160,59.92) .. controls (161.67,61.59) and (161.67,63.25) .. (160,64.92) .. controls (158.33,66.59) and (158.33,68.25) .. (160,69.92) .. controls (161.67,71.59) and (161.67,73.25) .. (160,74.92) -- (160,76.11) -- (160,76.11) ;
\draw   (25,46.08) .. controls (25,43.83) and (26.79,42.01) .. (29,42.01) .. controls (31.21,42.01) and (33,43.83) .. (33,46.08) .. controls (33,48.32) and (31.21,50.14) .. (29,50.14) .. controls (26.79,50.14) and (25,48.32) .. (25,46.08) -- cycle ;
\draw    (242.43,48.97) -- (331,48.97) ;

\draw (250.5,20.32) node [anchor=north west][inner sep=0.75pt]  [font=\small]  {$\boldsymbol{\ell _{1}}$};
\draw (299.82,73.37) node [anchor=north west][inner sep=0.75pt]  [font=\small]  {$\boldsymbol{\ell _{1} +\ell _{2}}$};
\draw (307.76,22.85) node [anchor=north west][inner sep=0.75pt]  [font=\small]  {$\boldsymbol{\ell _{2}}$};
\draw (357.83,38.32) node [anchor=north west][inner sep=0.75pt]    {$=f(\boldsymbol{\ell _{1} ,\ell _{2}})$};
\draw (88,41.92) node [anchor=north west][inner sep=0.75pt]   [align=left] {+};
\draw (188.33,43.32) node [anchor=north west][inner sep=0.75pt]    {$=$};

\end{tikzpicture}
\end{equation}

\begin{figure}
    \centering
    \begin{subfigure}[t]{0.4\textwidth}
        \centering
        \includegraphics[width=\linewidth]{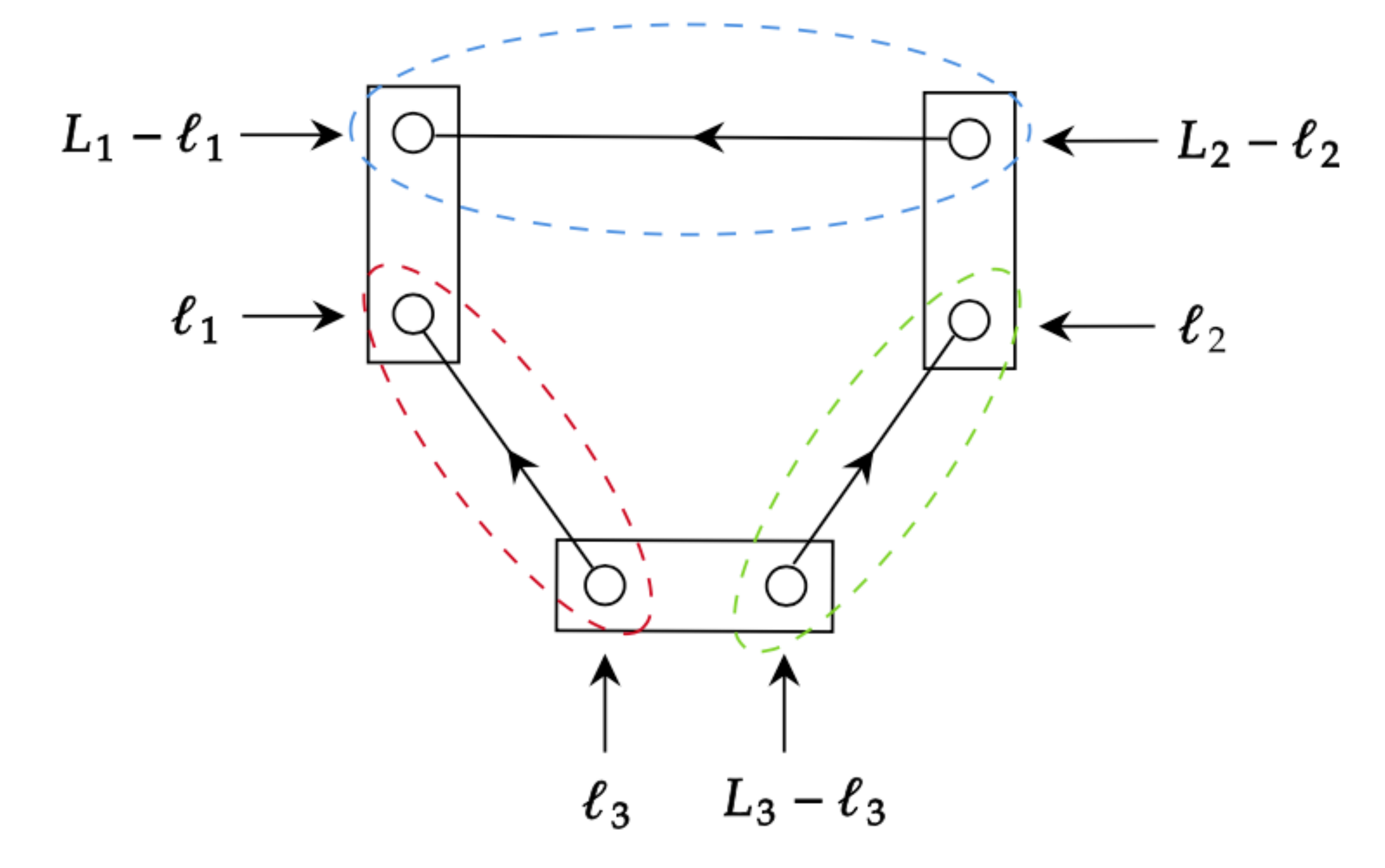} 
        \caption{$2_C\times 2_C\times 2_C$} \label{fig:2x2x2}
    \end{subfigure}
    \hspace{1cm}
    \begin{subfigure}[t]{0.4\textwidth}
        \centering
        \includegraphics[width=\linewidth]{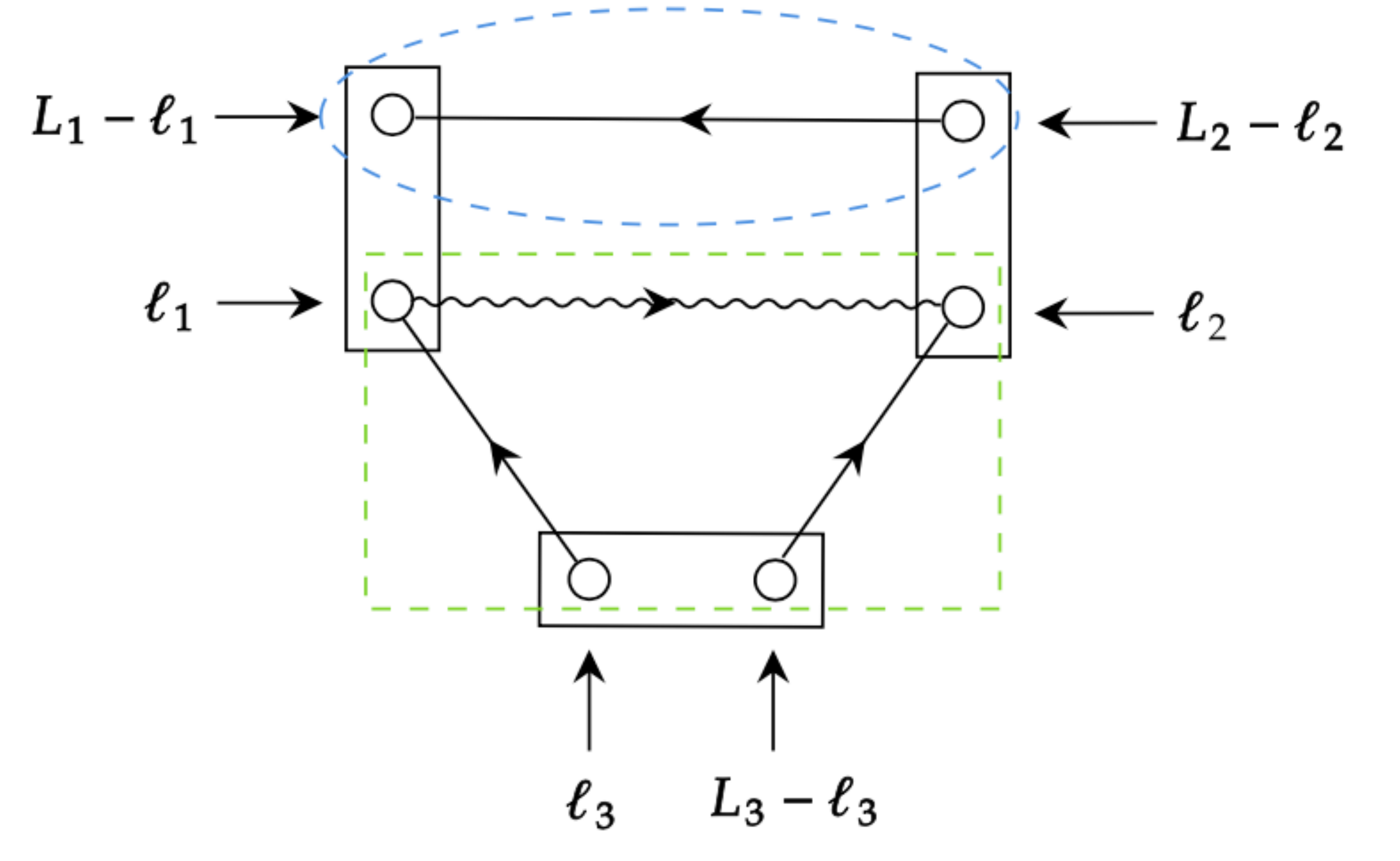} 
        \caption{$4_C \times 2_C$} \label{fig:4x2}
    \end{subfigure}
    
    \vspace{1cm}
    
    \begin{subfigure}[t]{0.4\textwidth}
        \centering
        \includegraphics[width=\linewidth]{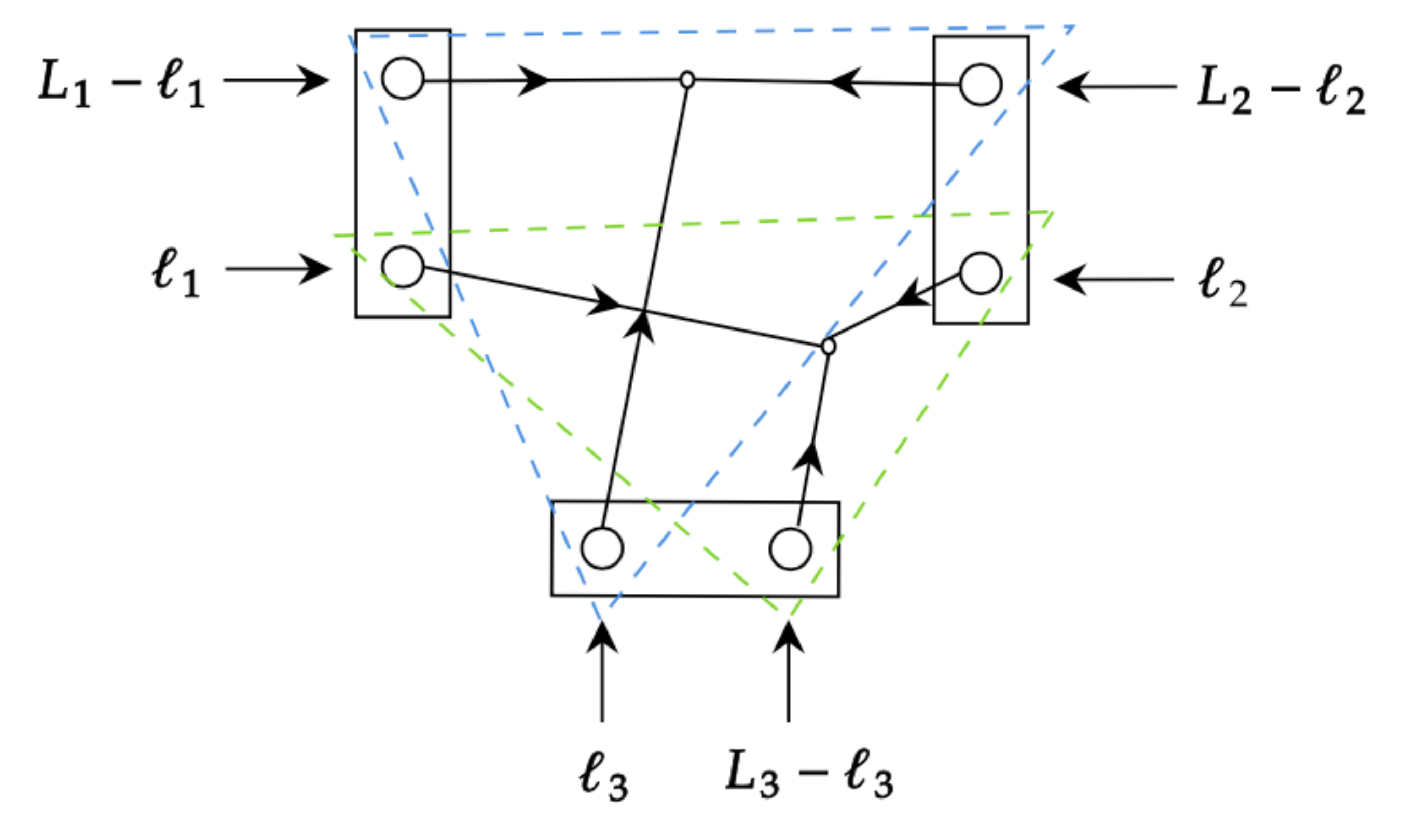} 
        \caption{$3_C \times 3_C$} \label{fig:3x3}
    \end{subfigure}
     \hspace{1cm}
    \begin{subfigure}[t]{0.4\textwidth}
        \centering
        \includegraphics[width=\linewidth]{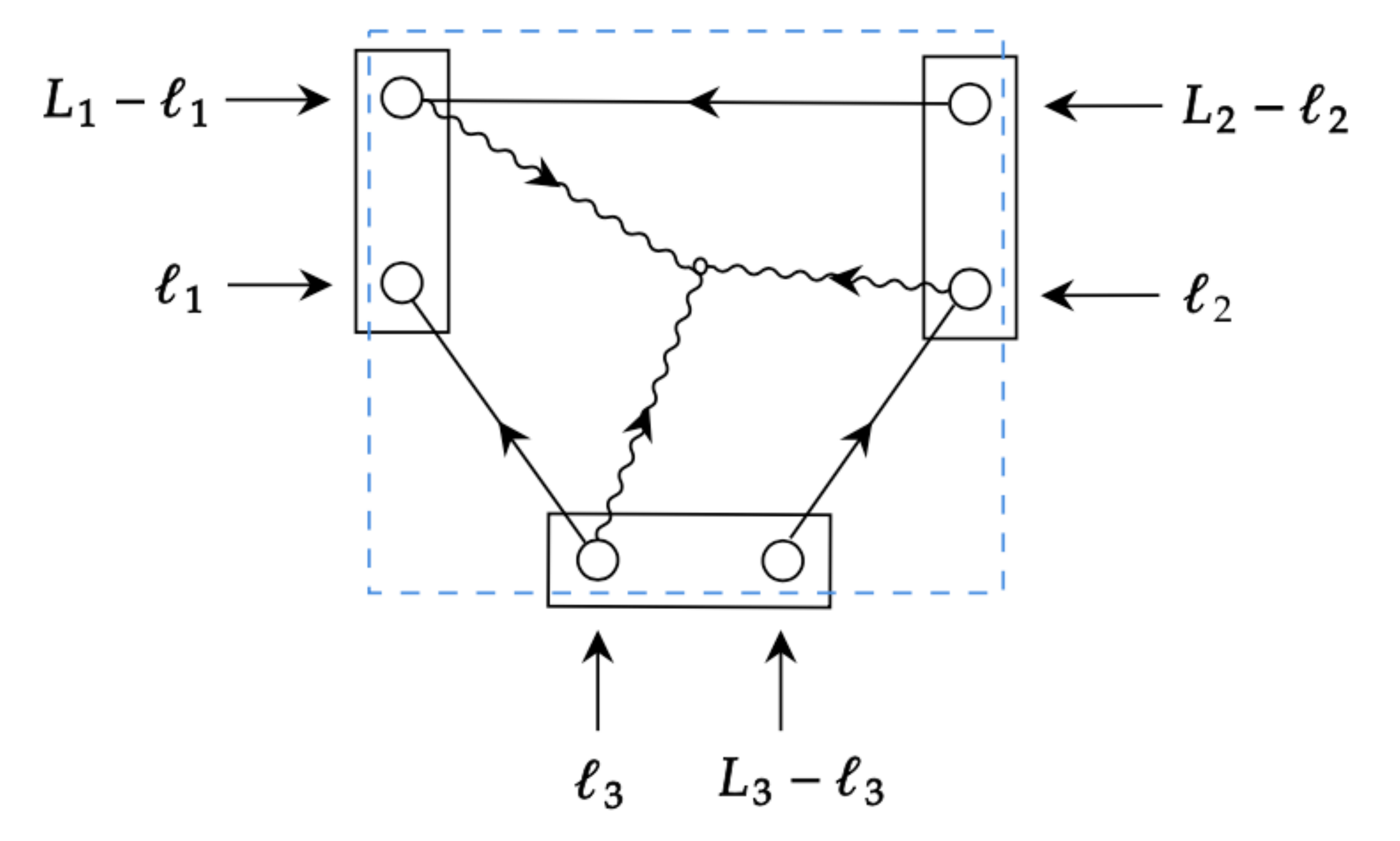} 
        \caption{$6_C$} \label{fig:6}
    \end{subfigure}
    \caption{Example of Feynman diagrams decomposed in the connected components. Each block enclosed by a dotted line of a different color represents a different connected component. } \label{fig:connection}
\end{figure}

Finally, Feynman diagrams can be decomposed in connected contributions, which are defined as subsets of the diagram where the connection between a set of vertexes can be removed only by erasing one or more lines (either wiggly or straight lines). As an example, in Fig.~\ref{fig:connection} we show some examples of Feynman diagrams contributing to each of the terms appearing in the decomposition \eqref{eq:decomp_6points}.

\subsubsection*{$N^{(0)}_\mathrm{B}$ noise-bias: Feynman approach.}

The $N^{(0)}_\mathrm{B}$ noise-bias comes from the bispectrum of the reconstructed $\hat\kappa$ field, thus the diagrams representing its contribution comprise three building blocks~\eqref{eq_fund_BB}. Moreover, $N^{(0)}_\mathrm{B}$ arises from the connected 2-pt function of lensed temperature anisotropies, therefore it features three propagators~\eqref{eq:propagators} corresponding to the lensed temperature power spectra\footnote{Notice that the propagator introduced in Sec.~\ref{subsec:Feynman} corresponds to the \textit{unlensed} temperature power spectrum, however for our purposes of taking the non-perturbated term, here we consider it as lensed.}.
Of the 15 diagrams, the ones that contain an ``interaction'' of the type
\begin{equation}
\begin{tikzpicture}[x=0.75pt,y=0.75pt,yscale=-1,xscale=1]

\draw   (249.67,42.08) -- (249.67,131.41) -- (220.33,131.41) -- (220.33,42.08) -- cycle ;
\draw   (228.59,57.08) .. controls (228.59,53.58) and (231.42,50.74) .. (234.92,50.74) .. controls (238.42,50.74) and (241.26,53.58) .. (241.26,57.08) .. controls (241.26,60.58) and (238.42,63.41) .. (234.92,63.41) .. controls (231.42,63.41) and (228.59,60.58) .. (228.59,57.08) -- cycle ;
\draw   (228.59,115.74) .. controls (228.59,112.25) and (231.42,109.41) .. (234.92,109.41) .. controls (238.42,109.41) and (241.26,112.25) .. (241.26,115.74) .. controls (241.26,119.24) and (238.42,122.08) .. (234.92,122.08) .. controls (231.42,122.08) and (228.59,119.24) .. (228.59,115.74) -- cycle ;

\draw    (234.92,63.41) -- (234.92,109.41) ;
\draw [shift={(234.92,86.41)}, rotate = 270] [fill={rgb, 255:red, 0; green, 0; blue, 0 }  ][line width=0.08]  [draw opacity=0] (10.72,-5.15) -- (0,0) -- (10.72,5.15) -- (7.12,0) -- cycle    ;
\draw    (179,57.74) -- (209.33,57.74) ;
\draw [shift={(212.33,57.74)}, rotate = 180] [fill={rgb, 255:red, 0; green, 0; blue, 0 }  ][line width=0.08]  [draw opacity=0] (10.72,-5.15) -- (0,0) -- (10.72,5.15) -- (7.12,0) -- cycle    ;
\draw    (179.67,116.41) -- (210,116.41) ;
\draw [shift={(213,116.41)}, rotate = 180] [fill={rgb, 255:red, 0; green, 0; blue, 0 }  ][line width=0.08]  [draw opacity=0] (10.72,-5.15) -- (0,0) -- (10.72,5.15) -- (7.12,0) -- cycle    ;

\draw (120.33,48.4) node [anchor=north west][inner sep=0.75pt]    {$\boldsymbol{L_{i} -\ell _{i}}$};
\draw (155.67,107.4) node [anchor=north west][inner sep=0.75pt]    {$\boldsymbol{\ell _{i}}$};
\end{tikzpicture}
\end{equation}
will not contribute, given that the sum of the momenta must be zero at each vertex. The remaining 8 diagrams are equivalent because each block is symmetrical for an exchange of momenta $\bsl_i \leftrightarrow \bsL_i - \bsl_i$.
\begin{figure}[htbp]
        \centering
        \includegraphics[width=0.7\linewidth]{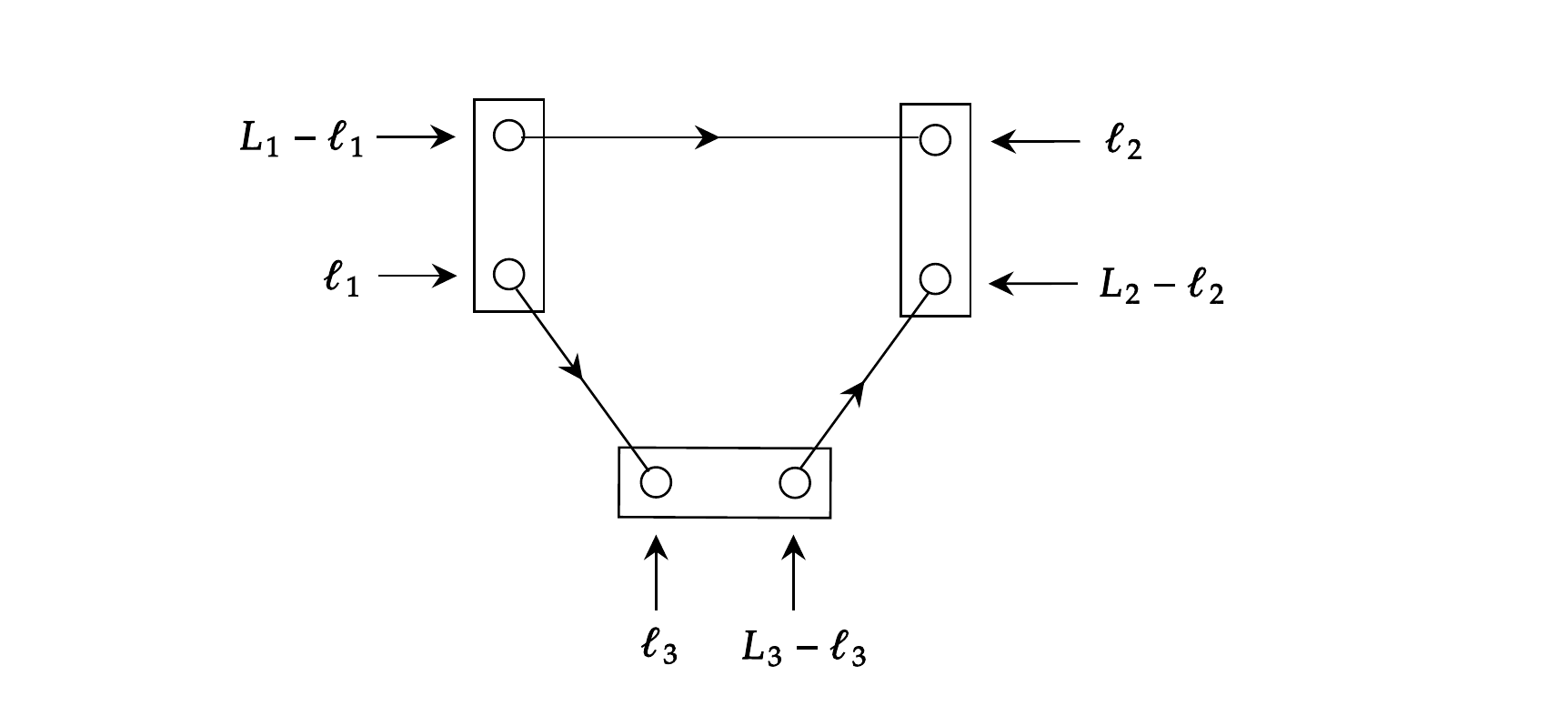} 
    \caption{The only independent Feynman diagram contributing to the $N_\mathrm{B}^{(0)}$ noise-bias of $\langle \hat{\kappa} \hat{\kappa} \hat{\kappa} \rangle$.} \label{fig:N0kkk}
\end{figure} 

Therefore, we are left with one diagram (for example, the one depicted in Fig.~\ref{fig:N0kkk}), which, according to the Feynman rules, corresponds to 
\begin{align} \label{eq:N0_feyn}
N^{(0)}_\mathrm{B} = 8\int'_{\bsl_1} \, &F(\bsl_1,\bsL_1-\bsl_1) \, F(\bsL_1-\bsl_1, \bsL_3 + \bsl_1) \, F(-\bsl_1,\bsL_3+\bsl_1)  \nonumber \\
&\times C^{\widetilde{T}\widetilde{T}}_{\ell_1} \, C^{\widetilde{T}\widetilde{T}}_{|\bsL_1 - \bsl_1|} \, C^{\widetilde{T}\widetilde{T}}_{|\bsL_3 + \bsl_1|} \, .
\end{align}
where we used the relations $\bsl_2 = \bsL_1 - \bsl_1$ and $\bsl_3 = -\bsl_1$ coming from momentum conservation. This is exactly what we found in Sec.~\ref{subsec:Wick}.

\subsubsection*{$N^{(1)}_\mathrm{B}$ noise-bias: Feynman approach.}

In order to compute $N^{(1)}_\mathrm{B}$ using the Feynman approach, we need to add two wiggly lines coming off two vertexes corresponding to the $\phi$ fields involved. 

\begin{figure}[htbp]
    \centering
    \begin{subfigure}[t]{0.4\textwidth}
        \centering
        \includegraphics[width=\linewidth]{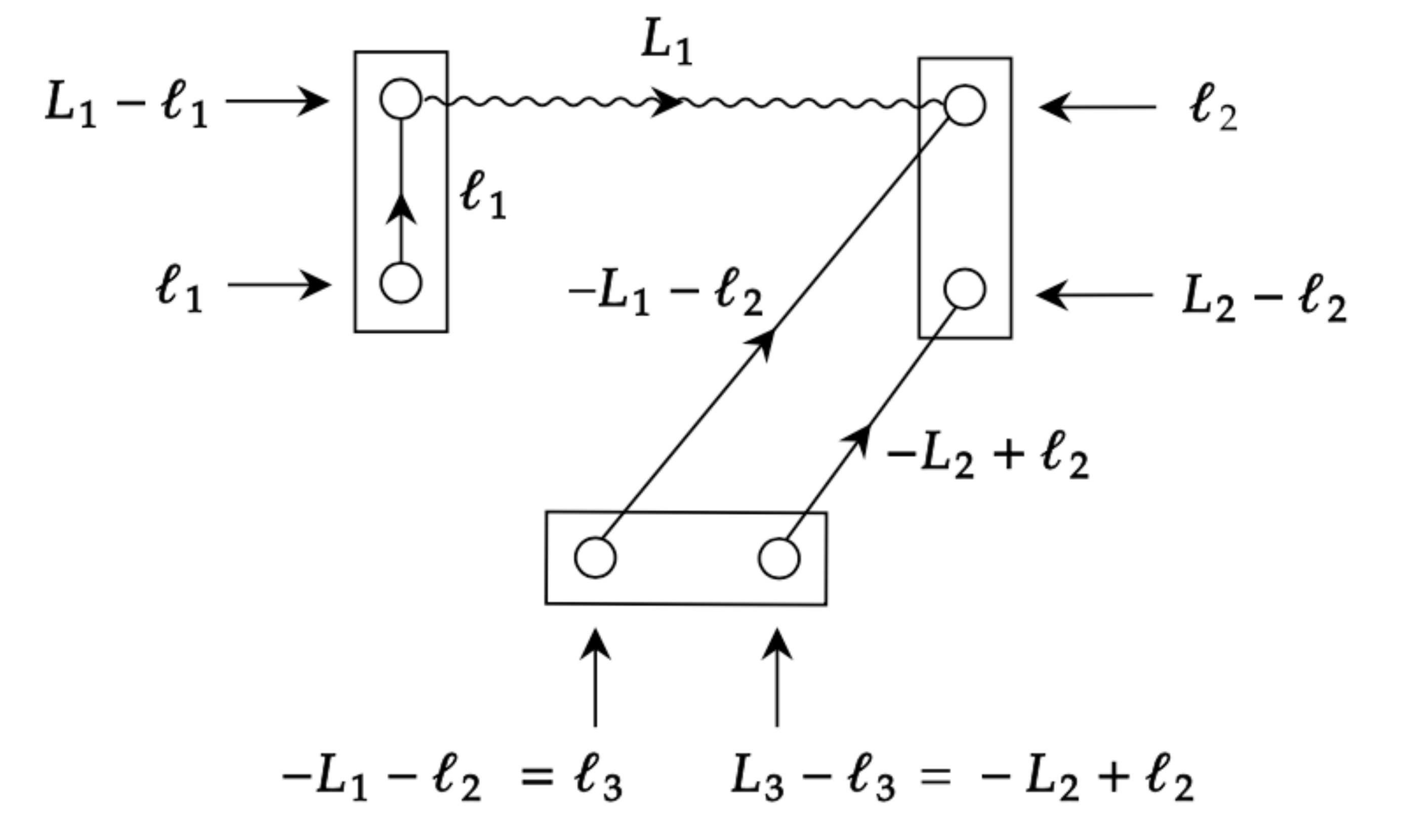} 
        \caption{$N^{(1)}_a$} \label{fig:N1_a}
    \end{subfigure}
    \hspace{1cm}
    \begin{subfigure}[t]{0.4\textwidth}
        \centering
        \includegraphics[width=\linewidth]{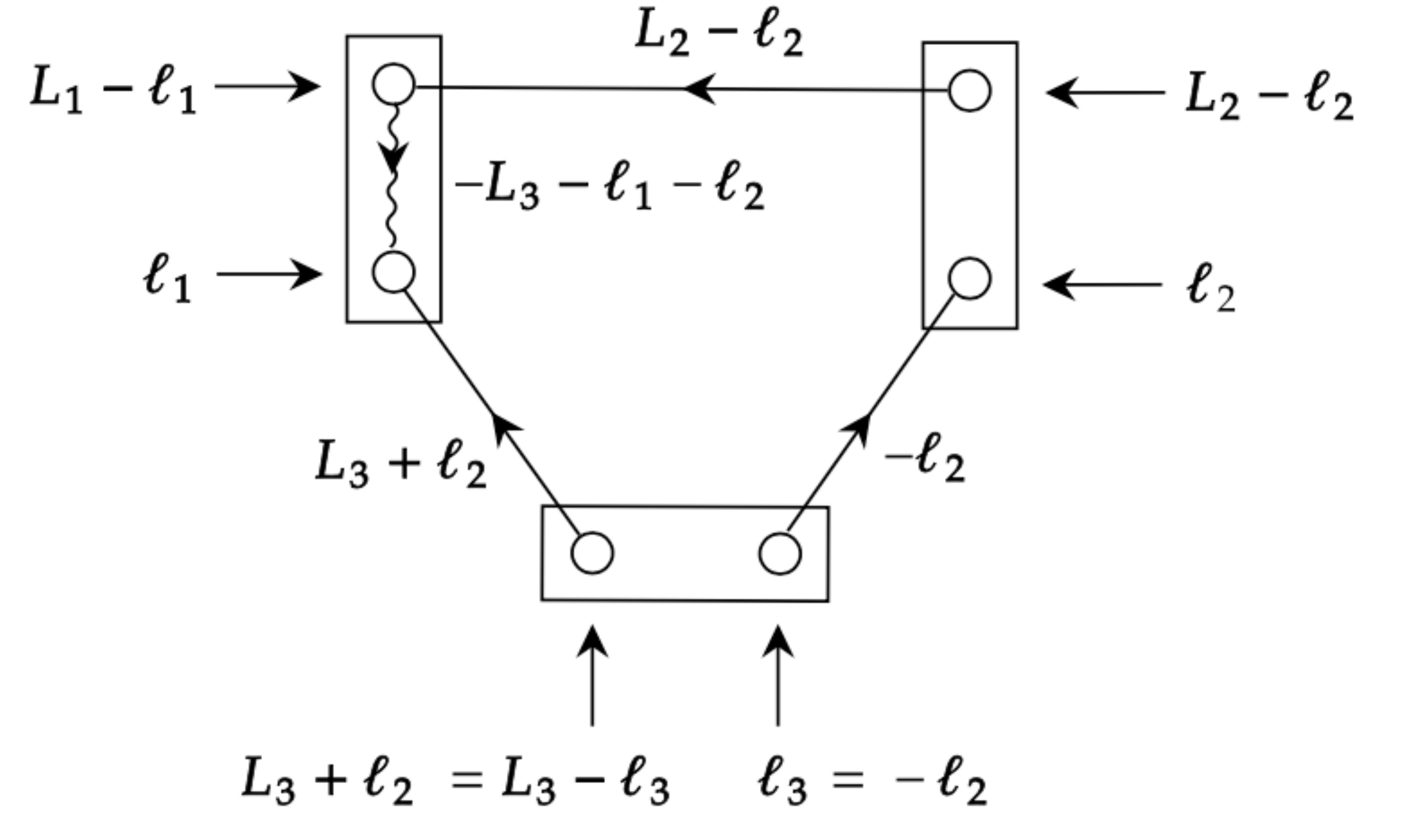} 
        \caption{$N^{(1)}_b$} \label{fig:N1_b}
    \end{subfigure}
    
    \vspace{1cm}
    
    \begin{subfigure}[t]{0.4\textwidth}
        \centering
        \includegraphics[width=\linewidth]{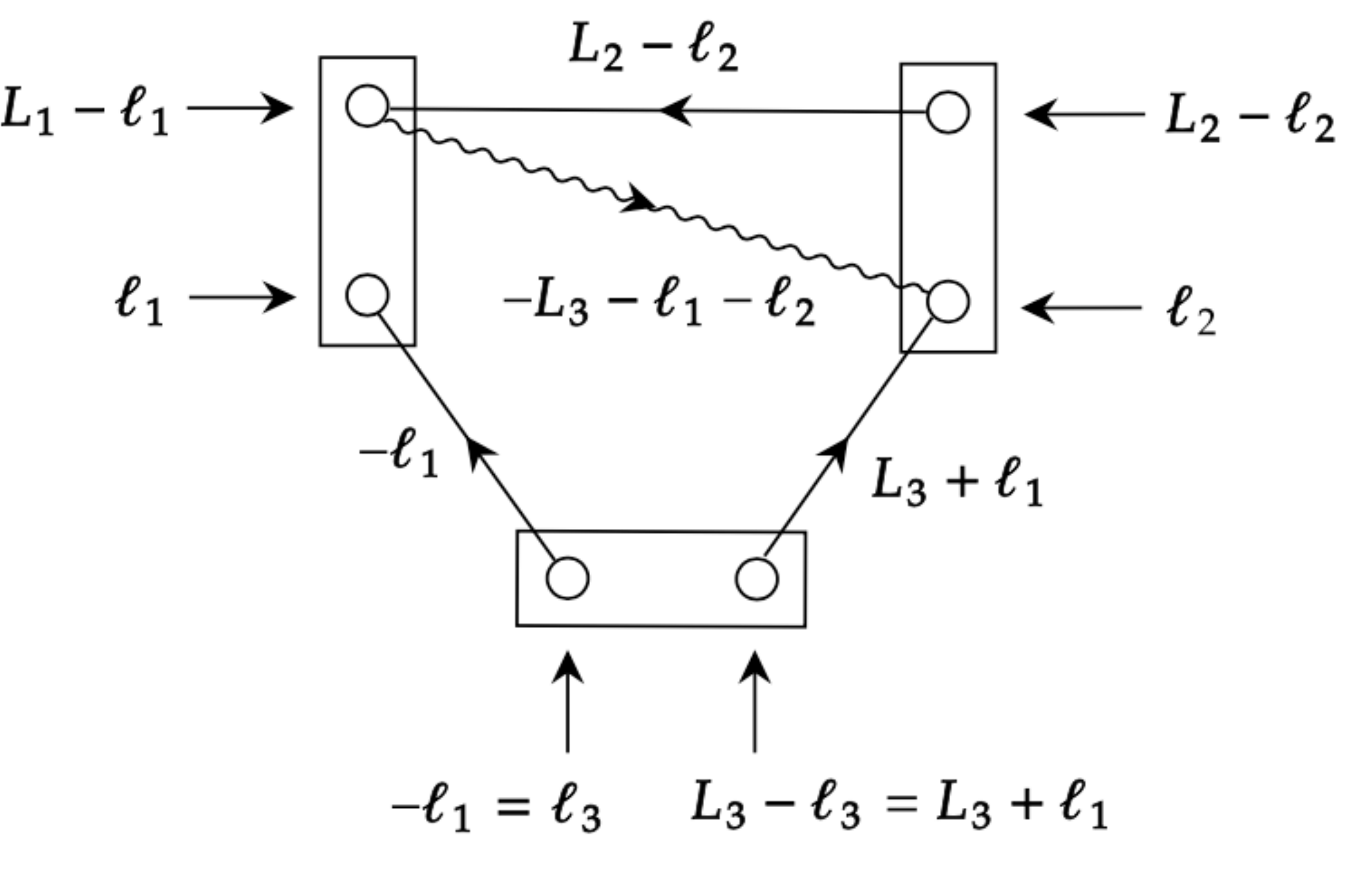} 
        \caption{$N^{(1)}_c$} \label{fig:N1_c}
    \end{subfigure}
     \hspace{1cm}
    \begin{subfigure}[t]{0.5\textwidth}
        \centering
        \includegraphics[width=\linewidth]{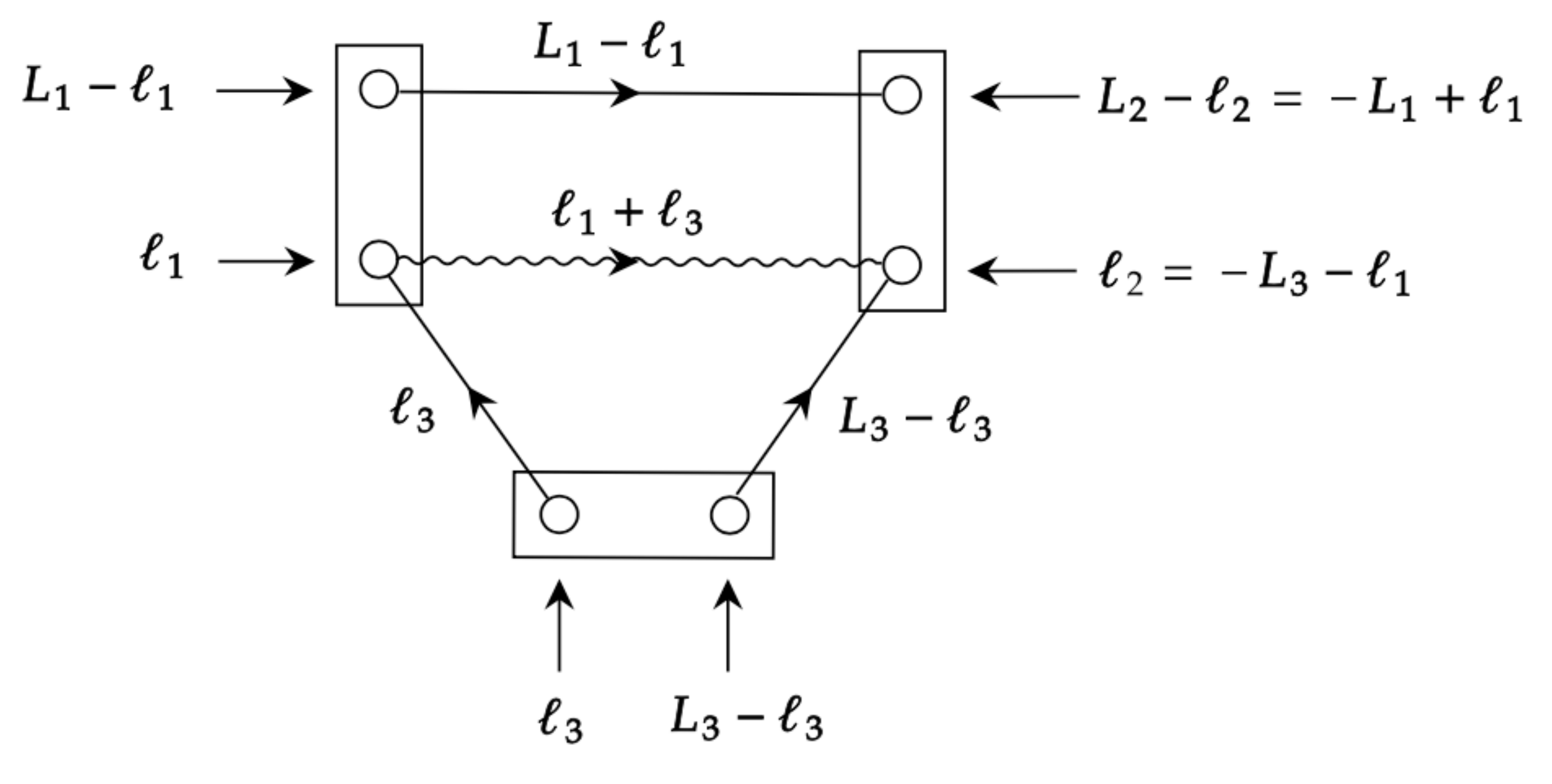} 
        \caption{$N^{(1)}_d$} \label{fig:N1_d}
    \end{subfigure}
    \caption{Independent Feynman diagrams contributing to the $N^{(1)}_\mathrm{B}$ noise-bias. The internal momenta are fixed by the momenta-conserving delta function at each vertex, as explicitly shown in the figure. Note that only two of the three $\ell_i$'s turn out to be independent.} \label{fig:N1kkk}
\end{figure}

Once we have defined the basic elements that constitute the diagrams, it is possible to visualize the symmetries of the system. Indeed, as shown in Fig.~\ref{fig:N1kkk}, we have only 4 independent Feynman diagrams, as all other ones can be obtained by exchanging the vertexes inside the building blocks or by exchanging $\bsL_i$. The power of the Feynman approach lies in the fact that we do not have to go through lengthy calculations to identify the contributing terms.
%
%

Using the Feynman rules outlined in Sec.~\ref{subsec:Feynman}, we can translate each of the diagrams in Fig.~\ref{fig:N1kkk} into the corresponding expressions (which account for symmetry factors)
\begin{align} \label{eq:N1a}
N^{(1)}_a = 8 \, & \int'_{\bsl_1,\, \bsl_2}  \, F(\bsl_1, \bsL_1 - \bsl_1) F(\bsl_2,\bsL_2 - \bsl_2)F(\bsL_2 - \bsl_2, \bsL_1+\bsl_2) \nonumber \\
& \qquad \times  \, (\bsl_1 \cdot \bsL_1) \Big[\bsL_1 \cdot (\bsL_1+\bsl_2)\Big] C^{\phi\phi}_{L_1} \, C^{TT}_{|\bsL_1 + \bsl_2|} C^{TT}_{|\bsL_2 - \bsl_2|} C^{TT}_{\ell_1} + \mbox{perms($\bsL_i$)}\,,
\end{align}
\begin{align} \label{eq:N1b}
N^{(1)}_b = 8 \, & \int'_{\bsl_1,\,\bsl_2}   \, F(\bsl_1, \bsL_1 - \bsl_1) F(\bsl_2,\bsL_2 - \bsl_2) F(\bsL_3 +\bsl_2,- \bsl_2) \nonumber \\
& \qquad\times  \,\Big[(-\bsL_2 + \bsl_2) \cdot (\bsL_3 + \bsl_1 + \bsl_2) \Big]\Big[(\bsL_3 +\bsl_1 + \bsl_2) \cdot (\bsL_3+\bsl_2)\Big] \nonumber \\
& \qquad\times  \,  C^{\phi\phi}_{|\bsL_3 + \bsl_1 + \bsl_2|} \, C^{TT}_{|\bsL_3 + \bsl_2|} C^{TT}_{|\bsL_2 - \bsl_2|} C^{TT}_{\ell_2}   + (\bsL_1 \leftrightarrow \bsL_2)  + (\bsL_1 \leftrightarrow \bsL_3)  \, ,     
\end{align}
\begin{align} \label{eq:N1c}
N^{(1)}_c =  - 8 \, &\int'_{\bsl_1,\,\bsl_2}  \, F(\bsl_1,\bsL_1 - \bsl_1) F(\bsl_2,\bsL_2 - \bsl_2) F(- \bsl_1, \bsL_3 + \bsl_1) \nonumber \\
& \qquad\times \, \Big[(\bsL_2 - \bsl_2) \cdot (\bsL_3 + \bsl_1 + \bsl_2) \Big] \Big[(\bsL_3 + \bsl_1 + \bsl_2) \cdot (\bsL_3 + \bsl_1)\Big] \nonumber \\
& \qquad\times  \, C^{\phi\phi}_{|\bsl_1 + \bsl_2+\bsL_3|} \, C^{TT}_{|\bsL_2 - \bsl_2|} C^{TT}_{|\bsL_3 + \bsl_1|} C^{TT}_{\ell_1} + \mbox{perms($\bsL_i$)} \, ,     
\end{align}
\begin{align} \label{eq:N1d}
N^{(1)}_d =  - 8 \, &\int'_{\bsl_1,\,\bsl_3} \,  F(\bsl_1,\bsL_1 - \bsl_1)  F(\bsl_3,\bsL_3 - \bsl_3) F(\bsL_1 - \bsl_1, \bsL_3 + \bsl_1) \nonumber \\
& \qquad\times  \, \Big[(\bsL_3 - \bsl_3) \cdot (\bsl_1 + \bsl_3) \Big] \Big[(\bsl_1 + \bsl_3) \cdot \bsl_3 \Big]  \nonumber \\
&\qquad\times \,  C^{\phi\phi}_{|\bsl_1 + \bsl_3|} \, C^{TT}_{|\bsL_3 - \bsl_3|} C^{TT}_{|\bsL_1 - \bsl_1|} C^{TT}_{\ell_3} + (\bsL_1 \leftrightarrow \bsL_3)  + (\bsL_2 \leftrightarrow \bsL_3) \, .   
\end{align}
The internal momenta, explicitly shown in Fig.~\ref{fig:N1kkk}, are fixed by momentum conservation at the vertexes. We make a few change of variables to rearrange the terms\footnote{In particular, $N^{(1)}_a$ stays the same. In $N^{(1)}_b$ we made the change of variable $\bsL_3 + \bsl_2 = \bsl_2'$. In $N^{(1)}_c$ we made the double change of variable $\bsL_3 + \bsl_1 = \bsl_2'$, $ \, \bsl_2 = \bsl_1'$. In $N^{(1)}_d$ we made the double change of variable $\bsl_3 = \bsl_1'$, $ \, \bsl_1 = \bsl_2'$.}, we can rewrite some of the contributions as
\begin{align} \label{eq:N1b2}
N^{(1)}_b = 8 \, & \int'_{\bsl_1,\, \bsl_2}   \, F(\bsl_1, \bsL_1 - \bsl_1) F(\bsl_2,\bsL_3 - \bsl_2) F(\bsL_3 -\bsl_2,\bsL_1 + \bsl_2) \nonumber \\
& \times  \, \Big[(\bsL_1 + \bsl_2) \cdot (\bsl_1 + \bsl_2) \Big]  \Big[(\bsl_1 + \bsl_2) \cdot \bsl_2\Big] C^{\phi\phi}_{|\bsl_1 + \bsl_2|} \, C^{TT}_{|\bsL_1 + \bsl_2|} C^{TT}_{|\bsL_3 - \bsl_2|} C^{TT}_{\ell_2}  \nonumber \\
&\,  + (\bsL_1 \leftrightarrow \bsL_2)  + (\bsL_1 \leftrightarrow \bsL_3)  \, ,    
\end{align}
\begin{align} \label{eq:N1c2}
N^{(1)}_c =  - 8 \, &\int'_{\bsl_1,\, \bsl_2}  \, F(\bsl_1,\bsL_2 - \bsl_1) F(\bsl_2,\bsL_3 - \bsl_2) F(\bsL_3 - \bsl_2, \bsL_2 + \bsl_2) \nonumber \\
& \times \, \Big[(\bsL_2 - \bsl_1) \cdot (\bsl_1 + \bsl_2) \Big] \Big[(\bsl_1 + \bsl_2) \cdot \bsl_2 \Big] \, C^{\phi\phi}_{|\bsl_1 + \bsl_2|} \, C^{TT}_{|\bsL_2 - \bsl_1|} C^{TT}_{|\bsL_3 - \bsl_2|} C^{TT}_{\ell_2} \nonumber \\
&+ \mbox{perms($\bsL_i$)} \, ,     
\end{align}
\begin{align} \label{eq:N1d2}
N^{(1)}_d =  -  8 \, &\int'_{\bsl_1,\, \bsl_2}  \,  F(\bsl_1,\bsL_3 - \bsl_1)  F(\bsl_2,\bsL_1 - \bsl_2) F(\bsL_1 - \bsl_2, \bsL_3 + \bsl_2) \nonumber \\
&  \times \, \Big[(\bsL_3 - \bsl_1) \cdot (\bsl_1 + \bsl_2) \Big] \Big[(\bsl_1 + \bsl_2) \cdot \bsl_2 \Big] C^{\phi\phi}_{|\bsl_1 + \bsl_2|} \, C^{TT}_{|\bsL_3 - \bsl_1|} C^{TT}_{|\bsL_1 - \bsl_2|} C^{TT}_{\ell_1} \nonumber \\
& + (\bsL_1 \leftrightarrow \bsL_3)  + (\bsL_2 \leftrightarrow \bsL_3) \, .   
\end{align}
It is now straightforward to verify that summing over all these contributions leads to Eq.~\eqref{eq:N1_final}. 
This can be done by exploiting the definition in Eq.~\eqref{eq:small_f} and expanding Eq.~\eqref{eq:N1_final} in terms of scalar products between momenta and temperature power spectra. Here we show this explicitly for one term as an example. Consider the following contribution to Eq.~\eqref{eq:N1_final}
\begin{equation} \label{eq:N1_example}
    \begin{split}
         4\times\int_{\bsl_1,\, \bsl_2}  \, & F(\bsl_1,\bsL_1-\bsl_1) \, F(\bsl_2,\bsL_2-\bsl_2) \, F(\bsL_2-\bsl_2,\bsL_1+\bsl_2) \\
         &\times \, C^{\phi\phi}_{L_1} \, f(\bsl_1,\bsL_1-\bsl_1) \, f(\bsl_2,-\bsL_1-\bsl_2) \, C^{TT}_{|\bsL_2-\bsl_2|} \, .
    \end{split}
\end{equation} 
By expanding the two response functions, and taking the product, we get
\begin{align} \label{eq:fprod}
f(\bsl_1,\bsL_1-\bsl_1) \, f(\bsl_2,-\bsL_1-\bsl_2) = &\Big[\bsl_1 \cdot \bsL_1\Big]  \Big[\bsL_1 \cdot (\bsL_1+\bsl_2)\Big]  \, C^{TT}_{|\bsL_1 + \bsl_2|} C^{TT}_{\ell_1} \nonumber \\
 & + \Big[\bsL_1 \cdot (\bsL_1-\bsl_1)\Big]  \Big[\bsL_1 \cdot (\bsL_1+\bsl_2)\Big]  \, C^{TT}_{|\bsL_1 + \bsl_2|} C^{TT}_{|\bsL_1 - \bsl_1|} \nonumber \\
& - \Big[\bsl_1 \cdot \bsL_1\Big] \Big[\bsl_2 \cdot \bsL_1\Big]  \, C^{TT}_{\ell_2} C^{TT}_{\ell_1} \nonumber  \\
& -  \Big[\bsL_1 \cdot (\bsL_1-\bsl_1)\Big] \Big[\bsl_2 \cdot \bsL_1\Big]   \, C^{TT}_{\ell_2} C^{TT}_{|\bsL_1 - \bsl_1|} \, .  
\end{align}
If we replace Eq.~\eqref{eq:fprod} into Eq.~\eqref{eq:N1_example}, we get the following 4 terms
\begin{align}
    &\text{(A):}\quad  4\times\int_{\bsl_1,\, \bsl_2}  \,  F(\bsl_1,\bsL_1-\bsl_1) \, F(\bsl_2,\bsL_2-\bsl_2) \, F(\bsL_2-\bsl_2,\bsL_1+\bsl_2)  \nonumber\\
    &\qquad\;\quad\times \,\Big[\bsl_1 \cdot \bsL_1\Big]  \Big[\bsL_1 \cdot (\bsL_1+\bsl_2)\Big]  \, C^{TT}_{|\bsL_1 + \bsl_2|} \, C^{TT}_{\ell_1} \, C^{TT}_{|\bsL_2-\bsl_2|} \,  C^{\phi\phi}_{L_1}  \, ,\label{eq:N1_example1}\\
    &\text{(B):}\quad  4\times\int_{\bsl_1,\, \bsl_2}  \, F(\bsl_1,\bsL_1-\bsl_1) \, F(\bsl_2,\bsL_2-\bsl_2) \, F(\bsL_2-\bsl_2,\bsL_1+\bsl_2)  \nonumber\\
         &\qquad\;\quad \times \,\Big[\bsL_1 \cdot (\bsL_1-\bsl_1)\Big]  \Big[\bsL_1 \cdot (\bsL_1+\bsl_2)\Big]  \, C^{TT}_{|\bsL_1 + \bsl_2|} \, C^{TT}_{|\bsL_1 - \bsl_1|} \, C^{TT}_{|\bsL_2-\bsl_2|} \,  C^{\phi\phi}_{L_1}  \, ,\label{eq:N1_example2}\\
    &\text{(C):}\quad- 4\times\int_{\bsl_1,\, \bsl_2}  \, F(\bsl_1,\bsL_1-\bsl_1) \, F(\bsl_2,\bsL_2-\bsl_2) \, F(\bsL_2-\bsl_2,\bsL_1+\bsl_2)  \nonumber\\
         &\qquad\;\;\qquad\times \,\Big[\bsl_1 \cdot \bsL_1\Big] \Big[\bsl_2 \cdot \bsL_1\Big]  \, C^{TT}_{\ell_2} \, C^{TT}_{\ell_1} \, C^{TT}_{|\bsL_2-\bsl_2|} \,  C^{\phi\phi}_{L_1}  \, ,\label{eq:N1_example3}\\
    &\text{(D):}\quad- 4\times\int_{\bsl_1,\, \bsl_2}  \, F(\bsl_1,\bsL_1-\bsl_1) \, F(\bsl_2,\bsL_2-\bsl_2) \, F(\bsL_2-\bsl_2,\bsL_1+\bsl_2)  \nonumber\\
         &\qquad\;\;\qquad \times \, \Big[\bsL_1 \cdot (\bsL_1-\bsl_1)\Big] \Big[\bsl_2 \cdot \bsL_1\Big]   \, C^{TT}_{\ell_2} \, C^{TT}_{|\bsL_1 - \bsl_1|} \, C^{TT}_{|\bsL_2-\bsl_2|} \,  C^{\phi\phi}_{L_1}  \, .\label{eq:N1_example4}
\end{align}
By a change of variables $\bsl_1' = \bsL_1-\bsl_1$ in the integrals Eq.~\eqref{eq:N1_example2} and Eq.~\eqref{eq:N1_example4}, they become equivalent to Eq.~\eqref{eq:N1_example1} and Eq.~\eqref{eq:N1_example3}, respectively. Therefore, in the end we have the two independent contributions
\begin{equation} \label{eq:N1_example5}
    \begin{split}
    8\times\int_{\bsl_1,\, \bsl_2}  \, &F(\bsl_1,\bsL_1-\bsl_1) \, F(\bsl_2,\bsL_2-\bsl_2) \, F(\bsL_2-\bsl_2,\bsL_1+\bsl_2)  \\
         &\times \,\Big[\bsl_1 \cdot \bsL_1\Big]  \Big[\bsL_1 \cdot (\bsL_1+\bsl_2)\Big]  \, C^{TT}_{|\bsL_1 + \bsl_2|} \, C^{TT}_{\ell_1} \, C^{TT}_{|\bsL_2-\bsl_2|} \,  C^{\phi\phi}_{L_1}  \, ,
    \end{split}
\end{equation}
and
\begin{equation} 
    \begin{split} \label{eq:N1_example6}
   - 8\times\int_{\bsl_1,\, \bsl_2}  \, &F(\bsl_1,\bsL_1-\bsl_1) \, F(\bsl_2,\bsL_2-\bsl_2) \, F(\bsL_2-\bsl_2,\bsL_1+\bsl_2)  \\
         &\times \,\Big[\bsl_1 \cdot \bsL_1\Big] \Big[\bsl_2 \cdot \bsL_1\Big]  \, C^{TT}_{\ell_2} \, C^{TT}_{\ell_1} \, C^{TT}_{|\bsL_2-\bsl_2|} \,  C^{\phi\phi}_{L_1}  \, .
    \end{split}
\end{equation}
Now, it is evident that Eq.~\eqref{eq:N1_example5} gives exactly the first term of Eq.~\eqref{eq:N1a}, without accounting for permutations. When making an additional changes of variables $\bsl_2' = - \bsl_2 -\bsL_1$, Eq.~\eqref{eq:N1_example6} becomes
\begin{equation} \label{eq:N1_example7}
    \begin{split}
    8\times\int_{\bsl_1,\, \bsl_2}  \, &F(\bsl_1,\bsL_1-\bsl_1) \, F(\bsl_2,\bsL_3-\bsl_2) \, F(\bsL_3-\bsl_2,\bsL_1+\bsl_2)  \\
         &\times \,\Big[\bsl_1 \cdot \bsL_1\Big]  \Big[\bsL_1 \cdot (\bsL_1+\bsl_2)\Big]  \, C^{TT}_{|\bsL_1 + \bsl_2|} \, C^{TT}_{\ell_1} \, C^{TT}_{|\bsL_3-\bsl_2|} \,  C^{\phi\phi}_{L_1}  \, ,
    \end{split}
\end{equation}
which corresponds to the permutation $\bsL_2 \leftrightarrow \bsL_3$ in Eq.~\eqref{eq:N1a}. 

Once again, in order to obtain the non-perturbative noise-bias, we replace the unlensed power spectra in the response function Eq.~\eqref{eq:small_f} and the ones explicitly shown in the integrand respectively with the gradient lensed $C_\ell^{\tilde{T}\nabla \tilde{T}}$ and the lensed $C_\ell^{\widetilde{T}\widetilde{T}}$.

As seen in Sec.~\ref{subsec:Wick}, in order to obtain $N^{(1)}_\mathrm{B}$ with the Wick method, we need to compute all the possible contractions of the fields, which involve a large number of permutations. On the other hand, here we show how the relevant symmetries of the calculations become evident with the Feynman approach, allowing us to write down all the relevant terms directly from the diagrams. This makes the Feynman diagram method more efficient than the Wick approach, particularly when the order of the $n$-pt function and of the noise-bias $N^{(j/2)}$ increases. 


\section{Numerical results} \label{sec:comp:N}

In this section we present the results of evaluating the ``Gaussian'' noise-biases $N_\mathrm{B}^{(0)}$ and $N_\mathrm{B}^{(1)}$ numerically. The $N_B^{(3/2)}$ noise-bias appears if there is a non-zero matter bispectrum inducing a lensing bispectrum; we calculate $N_B^{(3/2)}$ with the Feynman formalism in App.~\ref{appen:N32} but leave its numerical evaluation for future work.

We make a number of simplifying assumptions as mentioned earlier: we consider only CMB temperature modes, and we neglect the effect of primordial non-Gaussianities and of the cross-correlations between temperature modes and the lensing potential
\begin{equation}
    B_T(\bm{\ell}_1,\bm{\ell}_2,\bm{\ell}_3) = 0\, , \qquad C^{T\phi}_{\ell} = 0\,.
\end{equation}

We present the numerical evaluation of the two largest noise-biases, $N_\mathrm{B}^{(0)}$ and $N_\mathrm{B}^{(1)}$ in two different slicings (or configurations) of the bispectrum: the equilateral slicing, where $L_1 = L_2 = L_3 \equiv L$, and the folded one, given by $L_1 \equiv L$ and $L_2 = L_3 \equiv L/2$. We do not consider the squeezed limit for two main reasons:
the lensing bispectrum has low signal-to-noise in this configuration; and it is significantly affected by multipole binning (see Ref.~\cite{namikawa:cmb_analyt_sims} for details).

The numerical computation of these noise-biases, which include up to 4D integrals over many multipoles, can be challenging to perform in a reasonable amount of time. This problem was already highlighted in Ref.~\cite{bohm:cmb_bias_bispectrum} for the $N^{(3/2)}$ bias in lensing power spectrum reconstruction. 
Here, we tackle this issue with a state-of-the-art numerical method, the VEGAS algorithm~\cite{lepage:vegas}, which gives Monte Carlo estimates of arbitrary multidimensional integrals by using importance sampling. The algorithm locates those parts of the integrand that contribute most to the final integral through a known or approximate probability distribution function and then concentrates sampling in those areas. This allows the computation of high-dimensional integrals with good precision while keeping the computational cost low.

\begin{figure}[htbp]
        \centering
        \includegraphics[width=\linewidth]{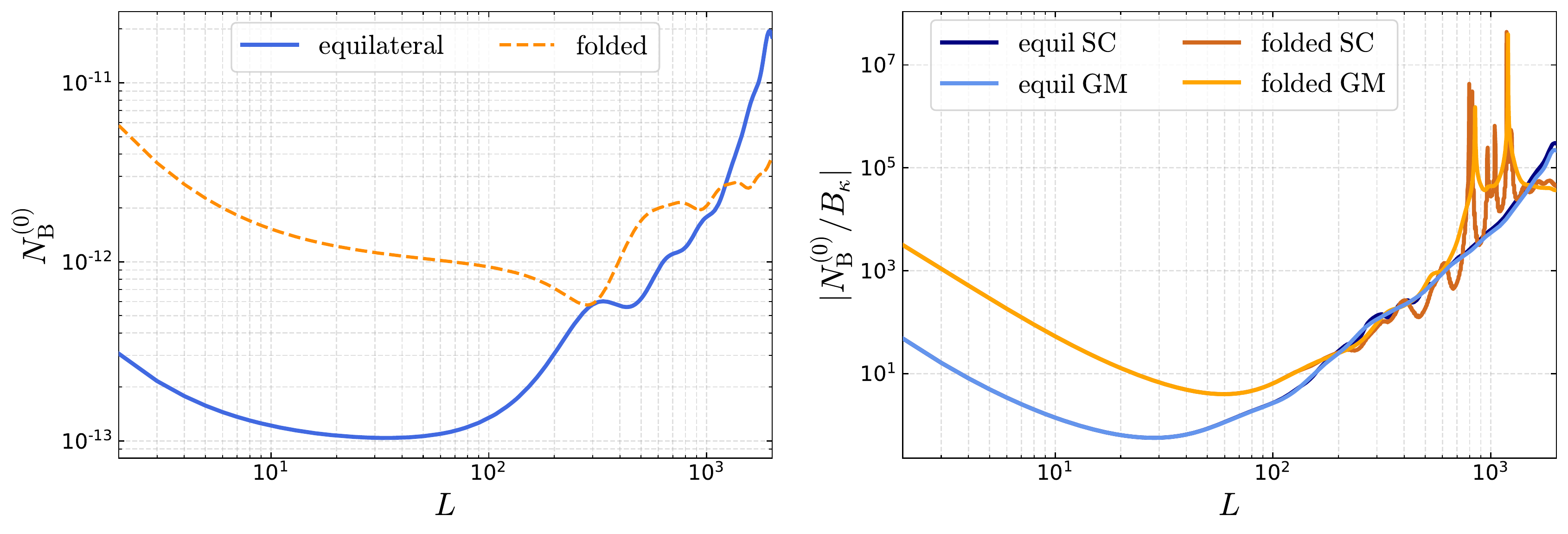} 
    \caption{\textit{Left panel:} Numerical evaluation of $N^{(0)}_\mathrm{B}$ in the equilateral and folded slicings. The dashed line refers to negative values. \textit{Right panel:} Ratio between $N^{(0)}_\mathrm{B}$ and the theoretical $\kappa$-bispectrum. Notice that the noise-bias in the folded slicing is negative, while the signal bispectrum is positive over a large range of $L$ values.} \label{fig:N0_vs_bispec}
\end{figure}

Similarly to the power spectrum reconstruction, $N_\mathrm{B}^{(0)}$ comes from the $2_C\times 2_C\times 2_C$ of the connected 6-pt function~\eqref{eq:decomp_6points}, and is given in Eq.~\eqref{eq:N0_wick}.
In Fig.~\ref{fig:N0_vs_bispec}, we show the magnitude of $N^{(0)}_\mathrm{B}$ in the equilateral and folded slicings, and compare it to the theoretically predicted lensing bispectrum induced by non-linear evolution of the matter field and post-Born corrections~\cite{pratten:post_born_lensing}. For the former, we employ the fitting formulae for the matter bispectrum given by Scoccimaro and Couchman in Ref.~\cite{scoccimarro:fitting}, hereafter denoted by ``SC'', and by Gil-Mar\'in et al. in Ref.~\cite{gilmarin:fitting}, denoted below by ``GM'' \footnote{Recently, Ref.~\cite{BiHalofit:2019} provides a more accurate fitting formula for the matter bispectrum calibrated by high-resolution cosmological N-body simulations. Although such an accurate fitting formula is desirable in practice, the difference between the new and SC/GM fitting formula is not so large compared to the noise bias. Since we are interested in the level of noise bias, this paper focuses on the SC and GM formulas.}.
We find that the reconstructed convergence bispectrum is dominated by $N_\mathrm{B}^{(0)}$ by several orders of magnitude, up to $10^5$ on small scales for the equilateral slicing.

The next-order noise-bias is $N^{(1)}_\mathrm{B}$. The expression in Eq.~\eqref{eq:N1_final} comprises two terms, one ``separable'', where the bias term can be split into a product of 2D integrals giving
\begin{align}\label{eq:N1_sep}
    N^{(1)}_\mathrm{B,\,sep} = &4 C^{\phi\phi}_{L_1} A_{L_1}^\kappa \int_{\bsl_1} \, F(\bsl_1,\bsL_1-\bsl_1) f(\bsl_1,\bsL_1-\bsl_1)\, \nonumber\\
         &\qquad\times A_{L_2}^\kappa A_{L_3}^\kappa \int_{\bsl_2} F(\bsl_2,\bsL_2-\bsl_2) F(\bsL_2-\bsl_2,\bsL_1+\bsl_2) f(\bsl_2,-\bsL_1-\bsl_2)\, C^{\widetilde{T}\widetilde{T}}_{|\bsL_2-\bsl_2|} \nonumber \\
         &\qquad + (\bsL_1 \leftrightarrow \bsL_2) +  (\bsL_1 \leftrightarrow \bsL_3) \, ,
\end{align}
and one ``coupled'', where the full 4D integral has to be evaluated, giving
\begin{align} \label{eq:N1_coupled}
     N^{(1)}_\mathrm{B,\,coupled} = &8 \int'_{\bsl_1,\, \bsl_2}  \, F(\bsl_1,\bsL_1-\bsl_1) \, F(\bsl_2,\bsL_2-\bsl_2) \,F(\bsL_2-\bsl_2,\bsL_1+\bsl_2) \nonumber\\
        &\quad\times f(\bsL_1-\bsl_1,-\bsL_1-\bsl_2)  f(\bsl_1,\bsl_2)\,C^{\phi\phi}_{|\bsl_1+\bsl_2|} C^{\widetilde{T}\widetilde{T}}_{|\bsL_2-\bsl_2|} + (\bsL_1 \leftrightarrow \bsL_2) +  (\bsL_1 \leftrightarrow \bsL_3)  \, .
\end{align}
Note that the first integral in Eq.~\eqref{eq:N1_sep} is simply $1/A_{L_1}^\kappa$, by the normalisation condition~\eqref{eq:norm_qest}, and so cancels with the normalisation $A_{L_1}^\kappa$.

\begin{figure}[htbp]
        \centering
        \includegraphics[width=\linewidth]{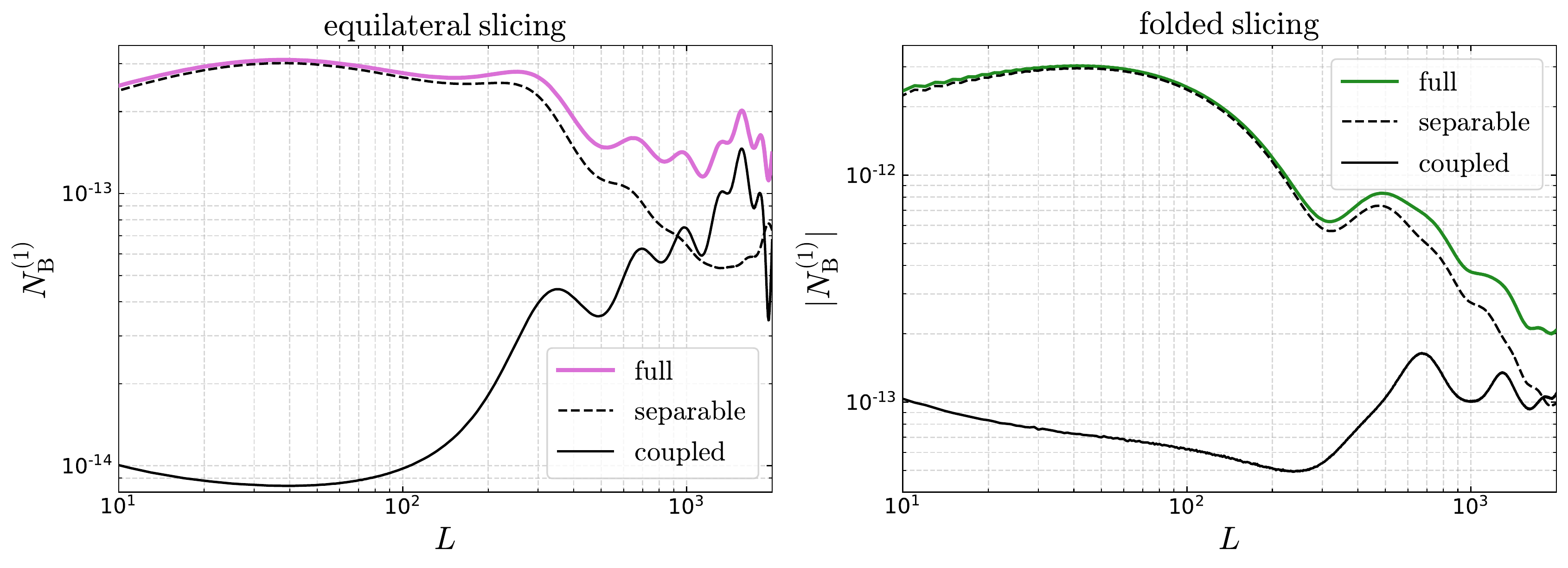} 
    \caption{Numerical evaluation of $N^{(1)}_\mathrm{B,\,sep}$ (dashed black) and $N^{(1)}_\mathrm{B,\,coupled}$ (solid black) for the equilateral (\textit{left}) and folded (\textit{right}) slicings. The total $N^{(1)}_\mathrm{B}$ is also shown in each slicing (magenta for equilateral and green for folded). Note that the noise-bias in the folded slicing is negative.} \label{fig:N1kkk_solutions}
\end{figure}
\begin{figure}[htbp]
        \centering
        \includegraphics[width=\linewidth]{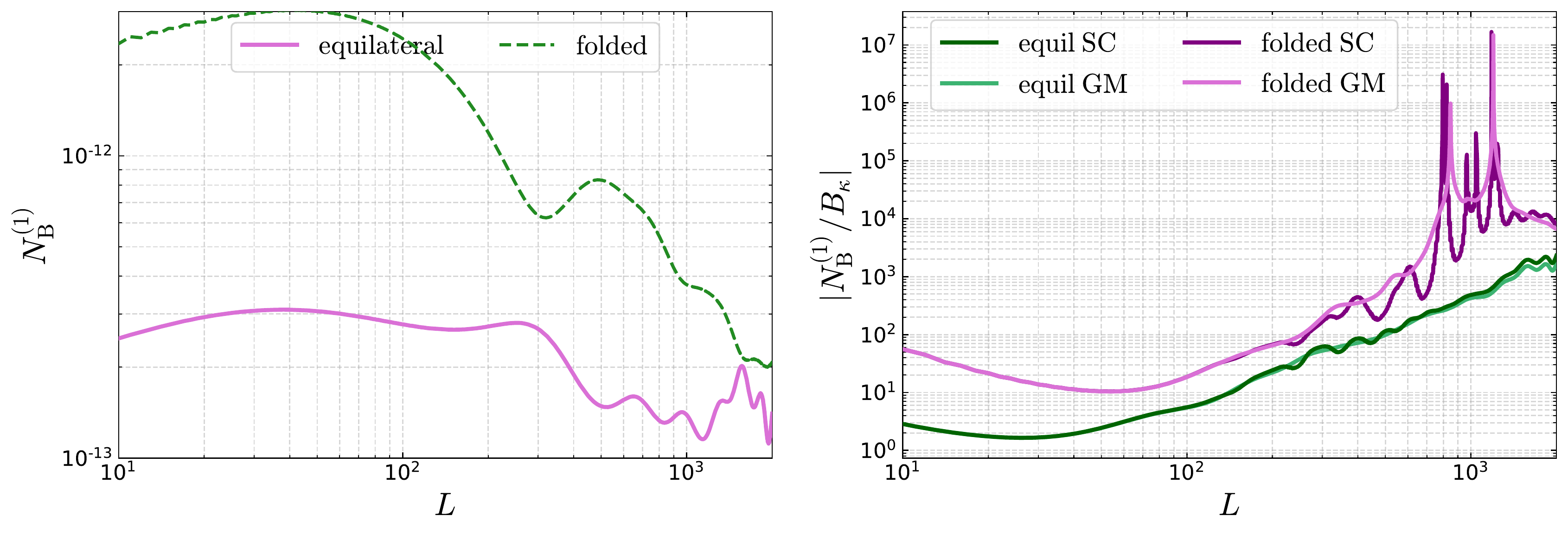}\caption{As Fig.~\ref{fig:N0_vs_bispec}, but for $N^{(1)}_\mathrm{B}$.} \label{fig:N1_vs_bispec}
\end{figure}

As shown in Fig.~\ref{fig:N1kkk_solutions}, for both slicings, the separable solution contributes the most at low multipoles. Intuitively, at low values of $L$ the lensing reconstruction weights couple tightly the small-scale CMB modes that dominate the reconstruction and the separable term~\eqref{eq:N1_sep} allows a larger 4D integration volume than the coupled term~\eqref{eq:N1_coupled}. On the other hand, at high $L$ the lensing weights couple the CMB modes less tightly and the two terms become comparable.
Furthermore, in Fig.~\ref{fig:N1_vs_bispec} we show that $N^{(1)}_\mathrm{B}$ is several orders of magnitude larger than the $\kappa$-bispectrum and, like $N^{(0)}_{\text{B}}$, needs to be carefully accounted for when reconstructing the bispectrum.

It is interesting to note that both $N^{(0)}_\mathrm{B}$ and $N^{(1)}_\mathrm{B}$ are negative in the folded slicing, while the signal itself has positive values over most of the $L$ range (the negative ones are due to Post-Born corrections). While in general, we must subtract any leading-order noise-biases as they are much larger than the signal, in the folded configuration this feature constitutes a consistency check of the reconstruction process. 


\section{Validation with simulations}\label{sec:validation}
In this section, we use simulated temperature maps to validate our calculations of the $N_\mathrm{B}^{(0)}$ and $N_\mathrm{B}^{(1)}$ noise-biases.
The simulation pipeline for each bias consists of:
\begin{enumerate}
    \item generating full-sky temperature maps;
    \item reconstructing $\hat\kappa$ from the above simulated temperature map with the quadratic estimator; and
    \item computing the binned bispectrum of the resulting $\hat\kappa$. 
\end{enumerate}
We use the publicly available code \texttt{cmblensplus} \cite{cmblensplus} for the quadratic estimator and binned bispectrum estimator, whose details are described in Ref.~\cite{namikawa:cmb_analyt_sims}. 

In general, we bin the simulation results as it reduces the computational costs, but we do not bin the analytical results as this would require costly evaluation at many neighbouring triangle configurations.
However, when binning squeezed and folded bispectra, it is known that significant differences between the binned and unbinned analytical results can occur. As shown in~\cite{namikawa:cmb_analyt_sims}, the binned bispectrum of a specific shape is less contaminated by different triangle configurations if the number of multipole bins increases. We discuss the binning dependency for the folded shape in Sec.~\ref{subsec:valid_n0} below.

Finally, in App.~\ref{app:paired_sims} we outline an alternative method for estimating these bias terms that uses paired simulations to isolate the specific Wick contractions that contribute to each term. This has the advantage of allowing each term in the expansion in connected moments, Eq.~\eqref{eq:decomp_6points} to be estimated individually.

\subsection{Validation of $N^{(0)}_\mathrm{B}$}
\label{subsec:valid_n0}

As discussed in Sec.~\ref{sec:comp:N}, $N^{(0)}_\mathrm{B}$ arises from the $2_C\times 2_C\times 2_C$ term in Eq.~\eqref{eq:decomp_6points}.
Therefore, in order to estimate such bias from simulations, it suffices to generate Gaussian temperature maps using a fiducial lensed temperature power spectrum computed with \texttt{CAMB}~\cite{lewis:camb}. We use the \texttt{healpy} package~\cite{zonca:healpy} for generating a full-sky temperature map with the lensed temperature power spectrum. 
Consequently, the bispectrum of the reconstructed convergence field, $\hat{\kappa}$, from each realization is proportional to $N^{(0)}_\mathrm{B}$ only.
\begin{figure}[htbp]
        \centering
        \includegraphics[width=0.8\linewidth]{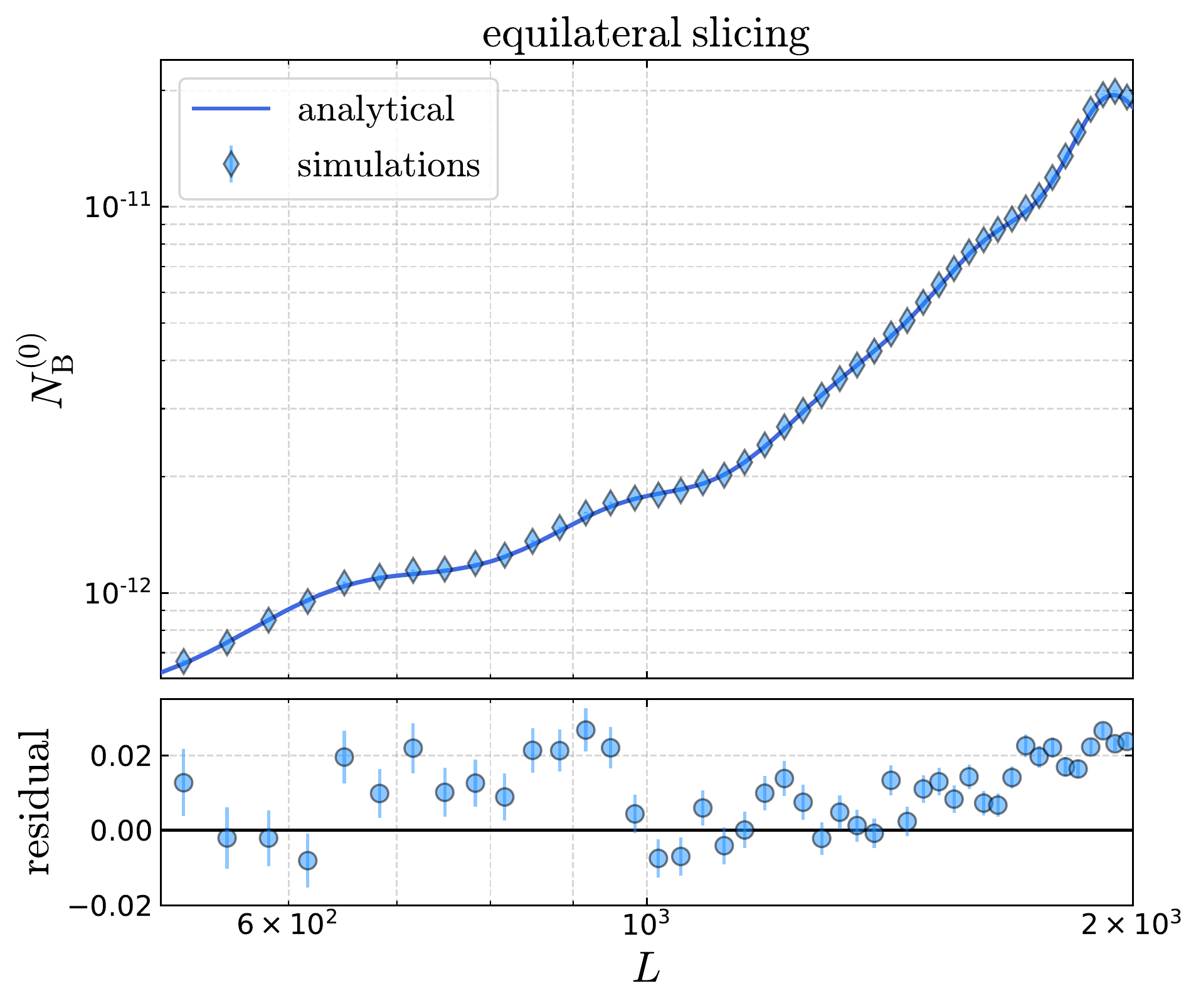} 
    \caption{Numerical evaluation of $N^{(0)}_{\text{B}}$ (Eq.~\ref{eq:N0_wick}) for the equilateral slicing (solid line), compared to binned results averaged over $10^4$ Gaussian simulations.
 The lower panel shows the fractional residuals. The error-bars are very small due to the large number of simulations.}
    \label{fig:n0_vs_sims_equil}
\end{figure}
\begin{figure}[htbp]
        \centering
        \includegraphics[width=0.8\linewidth]{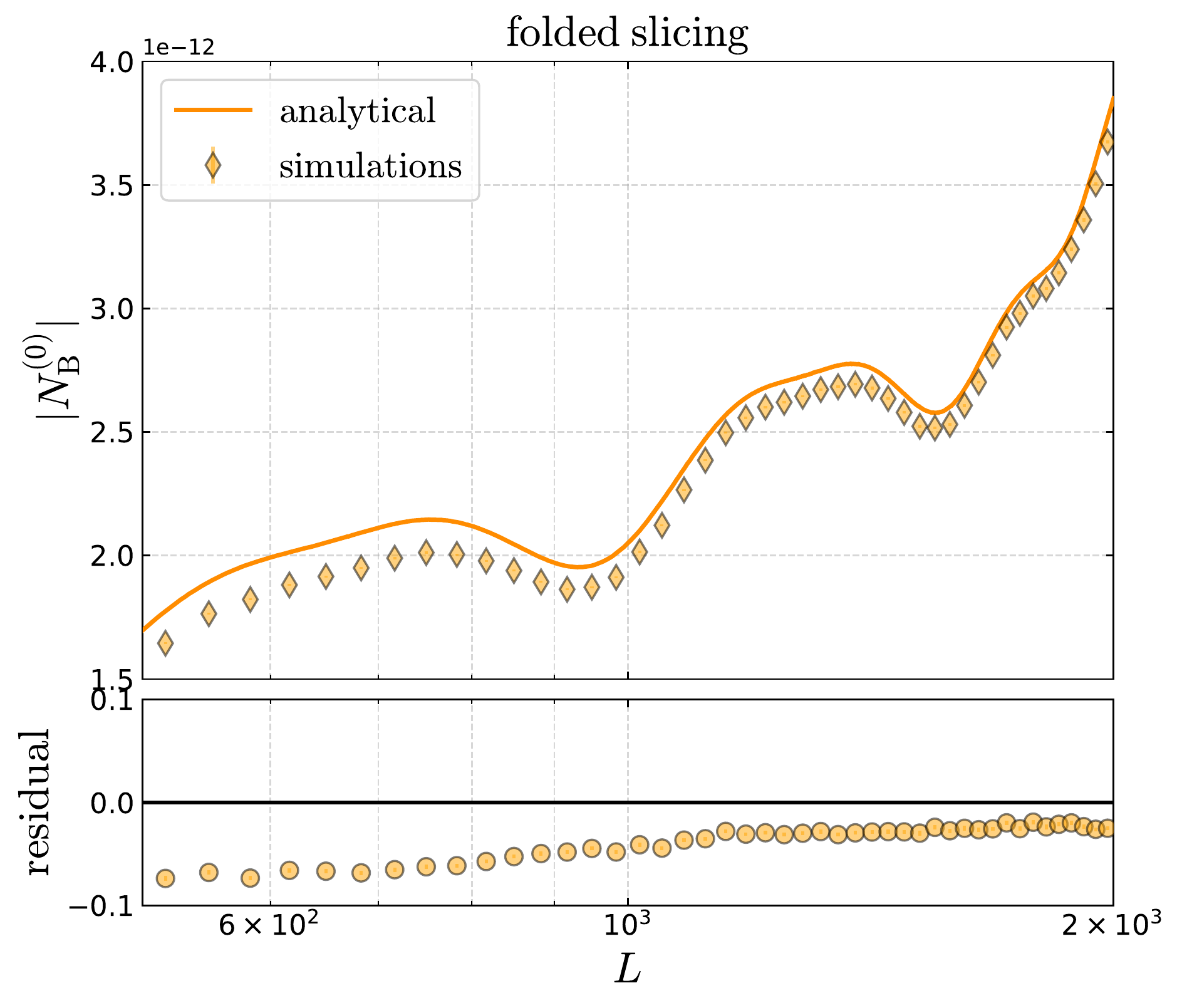} 
    \caption{As Fig.~\ref{fig:n0_vs_sims_equil}, but for the folded slicing.}
    \label{fig:n0_vs_sims_folded}
\end{figure}
\begin{figure}[htbp]
        \centering
        \includegraphics[width=0.8\linewidth]{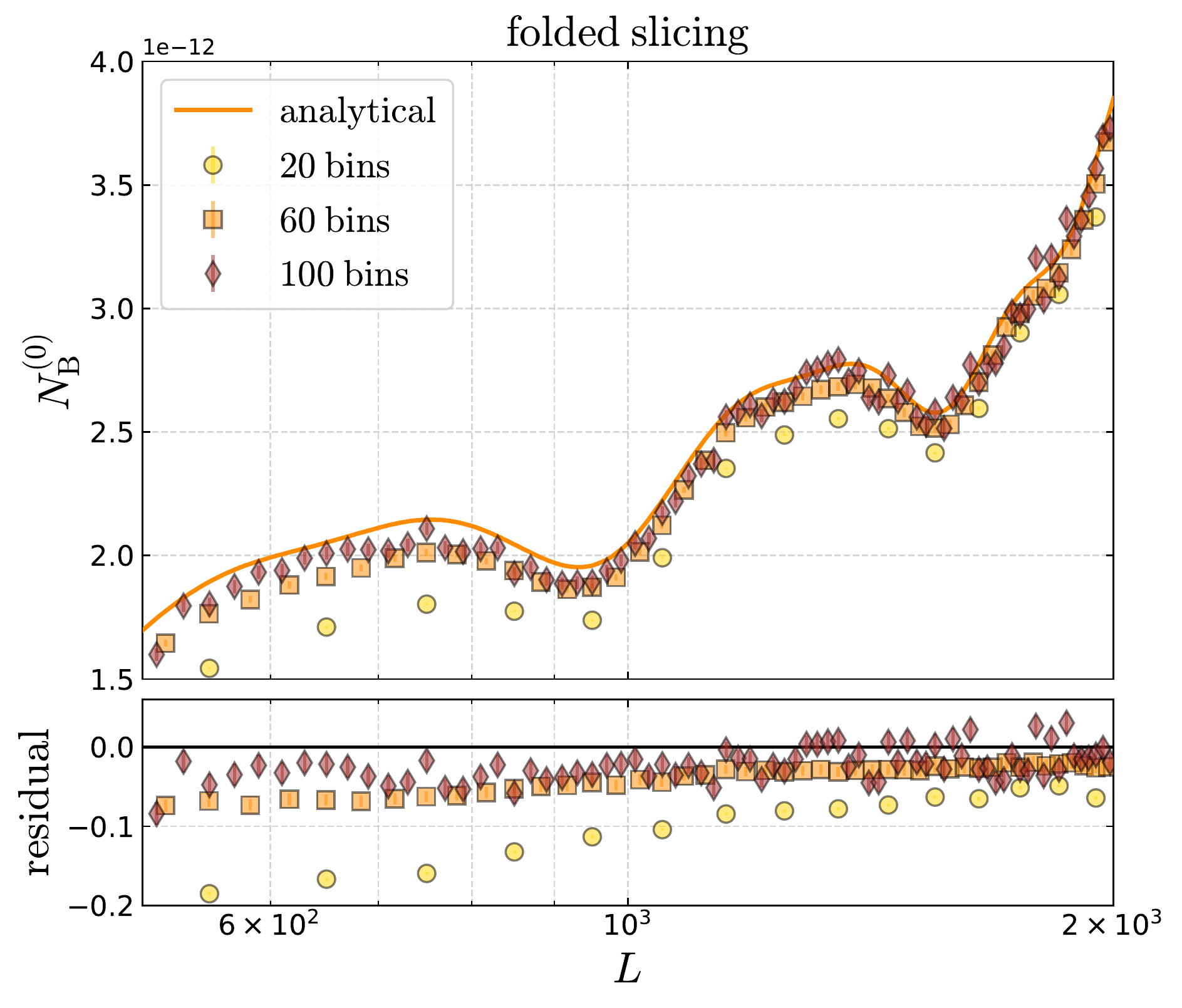} 
    \caption{Simulation results for $N^{(0)}_{\text{B}}$ in the folded slicing computed with the binned bispectrum estimator with increasing numbers of bins. In each case, 800 simulations are used. The lower panel shows the residual plot. It is clear that finer binning improves the match with the (unbinned) analytical results.}
    \label{fig:n0_folded_diffbin}
\end{figure}

In Figs.~\ref{fig:n0_vs_sims_equil} and \ref{fig:n0_vs_sims_folded} we show a comparison between simulations and the numerical evaluation of the analytical result in Eq.~\eqref{eq:N0_wick} for two slicings -- equilateral and folded. We average over $10^4$ realizations to reduce the simulation error in each data point, and we show the fractional residual defined as
\begin{equation}\label{eq:def_residual}
    \text{residual} = \frac{(\text{simulation}-\text{analytical})}{\text{analytical}}.
\end{equation}
We find good agreement between the analytical flat-sky result and the full-sky, binned simulation results in the equilateral shape. We believe that the small residuals, around $2\,\%$, are a consequence of the flat-sky approximation employed in the analytical results. 
The folded slicing is additionally affected by binning, as discussed above. In Fig.~\ref{fig:n0_folded_diffbin} we explicitly show how increasing the number of bins in the bispectrum estimate improves the match between the simulations and our analytical calculations.
Finally, a similar result is obtained using the paired-simulation approach (see App.~\ref{app:N0}).

\subsection{Validation of $N^{(1)}_\mathrm{B}$}
\label{subsec:valid_n1}
In order to test $N^{(1)}_\mathrm{B}$, which explicitly depends on the lensing potential power spectrum, we generate unlensed CMB maps and subsequently lens them with a Gaussian realization of $\phi$. For this step, we employ the publicly available code \texttt{lenspyx}\footnote{\hyperlink{https://github.com/carronj/lenspyx}{https://github.com/carronj/lenspyx}}. 
In principle, the resulting bispectrum contains all the ``Gaussian'' noise-biases\footnote{The signal $B_{\kappa}$ is zero given that we lens with a Gaussian $\phi$, as is $N^{(3/2)}_{\text{B}}$.} 
\begin{equation}
    B_{\hat{\kappa}} = N^{(0)}_\mathrm{B} + N^{(1)}_\mathrm{B} + N^{(2)}_\mathrm{B} + \cdots \, .
\end{equation}
However, the results that we present below (see also App.~\ref{app:paired_sims}) suggest that $N^{(2)}_{\text{B}}$ (and higher-order noise-biases) are small compared to $N^{(1)}_{\text{B}}$.
In this subsection, therefore, we show the residual, $B_{\hat{\kappa}}- N^{(0)}_\mathrm{B}$, and denote it as $N^{(1)}_\mathrm{B}$. 

\begin{figure}[htbp]
        \centering
        \includegraphics[width=0.8\linewidth]{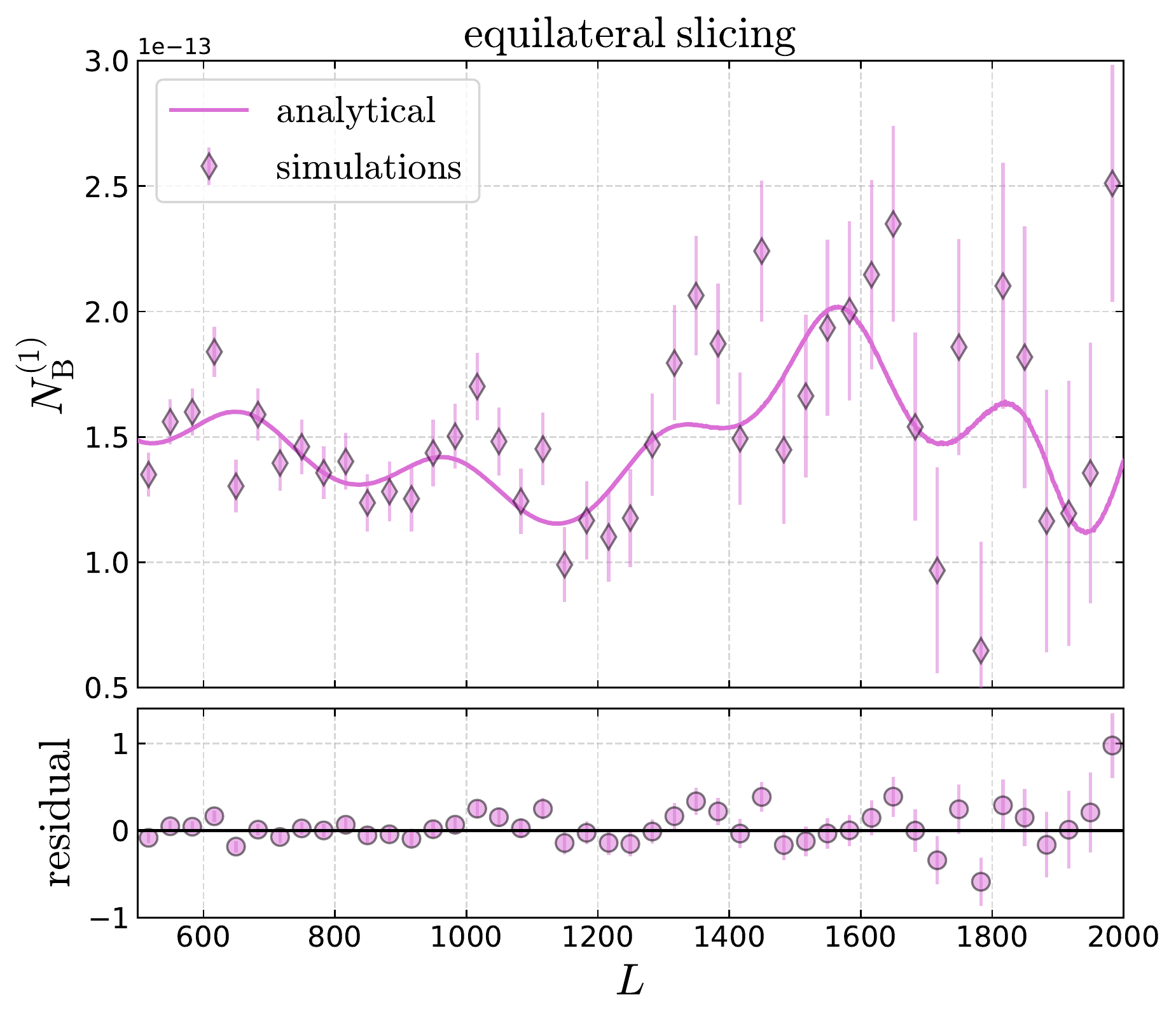} 
    \caption{Numerical evaluation of $N^{(1)}_{\text{B}}$ from Eq.~\eqref{eq:N1_final} for the equilateral slicing compared to binned simulation results averaged over $10^4$ realizations. In the lower panel, we show the fractional residuals.}
    \label{fig:n1_vs_sims_equil}
\end{figure} 
\begin{figure}[htbp]
        \centering
        \includegraphics[width=0.8\linewidth]{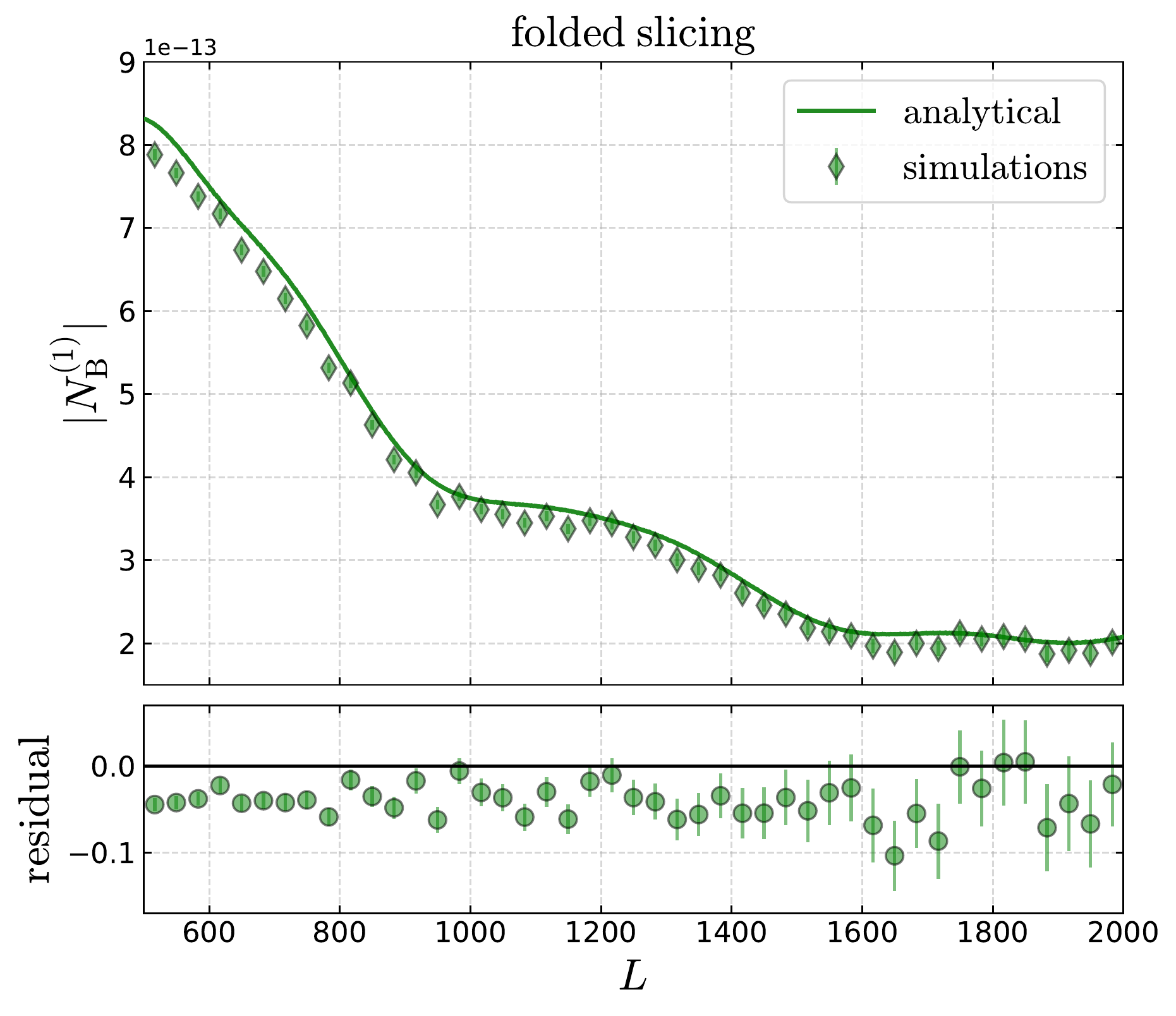} 
    \caption{As Fig.~\ref{fig:n1_vs_sims_equil}, but for the folded slicing.}
    \label{fig:n1_vs_sims_fold}
\end{figure} 
Generally, there are many ways to subtract $N^{(0)}_\mathrm{B}$. For an ideal survey, such as the one presented here, it would suffice to compute $N^{(0)}_\mathrm{B}$ with a lensed power spectrum that agrees well with reality (such as a smoothed version of the empirical power spectrum of our sky).
However, given the discrepancy between flat-sky analytical results and the full-sky simulations as well as the binning effect discussed in Sec.~\ref{subsec:valid_n0}, we subtract a simulation-based\footnote{Specifically, from each realization of the lensed CMB, we generate Gaussian CMB maps drawn from the empirical angular power spectrum measured from that simulation, and we then estimate $N^{(0)}_\mathrm{B}$ from these Gaussian maps as done in Sec.~\ref{subsec:valid_n0}. Note that a small bias is introduced in this estimate of $N^{(0)}_\mathrm{B}$ since its expectation value involves $\langle \hat{C}^{TT}_{\ell_1} \hat{C}^{TT}_{\ell_2} \hat{C}^{TT}_{\ell_3} \rangle$, the average of three products of the empirical power spectrum. This differs from the true value, $\widetilde{C}^{TT}_{\ell_1} \widetilde{C}^{TT}_{\ell_2} \widetilde{C}^{TT}_{\ell_3}$, because of cosmic variance:
\begin{equation*}
\langle \hat{C}^{TT}_{\ell_1} \hat{C}^{TT}_{\ell_2} \hat{C}^{TT}_{\ell_3} \rangle = 
\widetilde{C}^{TT}_{\ell_1} \widetilde{C}^{TT}_{\ell_2} \widetilde{C}^{TT}_{\ell_3} \left[1 + \left(\frac{2}{2\ell_2+1} \delta_{\ell_2 \ell_3} + (\text{2 perms.})\right) + \frac{8}{(2\ell_1+1)^2} \delta_{\ell_1 \ell_2 \ell_3}\right]\, ,
\end{equation*}
where the last term enforces $\ell_1 = \ell_2 = \ell_3$.
However, the difference is suppressed by factors that are $\mathcal{O}(\ell_i)$, which are small for the CMB multipoles that dominate the lensing reconstruction, and by the restricted integration volume from requiring two or three of the $\ell_i$ to be equal. The benefit of subtracting $N^{(0)}_{\text{B}}$
with this approach is that it is unaffected by any difference between the ensemble-average of the angular power spectra of the lensed CMB maps and the fiducial lensed power spectrum, which might arise from small inaccuracies in simulating the lensing operation.
} estimate of $N^{(0)}_\mathrm{B}$. 

In Figs.~\ref{fig:n1_vs_sims_equil} and \ref{fig:n1_vs_sims_fold}, we show the comparison between the evaluation of Eq.~\eqref{eq:N1_final} and the binned simulation results (averaged over $10^4$ realizations) in the equilateral and folded configurations, finding that the analytical curves generally show good agreement with the simulations for both configurations. The equilateral slicing has oscillatory features at high $L$ that cannot be easily captured by the binned bispectrum, and we need a large number of realizations, or an increase in the number of bins, to capture these features accurately. Moreover, the error-bars on the binned simulation result are about an order of magnitude larger than those of the folded configuration. This is likely due to the fact that the equilateral $N^{(0)}_\mathrm{B}$ is larger than the folded one by roughly an order of magnitude, hence when subtracting a simulation-based $N^{(0)}_\mathrm{B}$ any variance in it from the finite number of simulations used will leave additional variance in $ N^{(1)}_\mathrm{B}$. In order to reduce the error in the equilateral slicing, we would need around 100 times more simulations.

In the folded configuration, we find a small mismatch, around $5\,\%$, which likely mostly originates from the binning effect as we discussed in Sec.~\ref{subsec:valid_n0} and/or a small contribution of $N^{(2)}_\mathrm{B}$. 
The good agreement shown in Figs.~\ref{fig:n1_vs_sims_equil} and \ref{fig:n1_vs_sims_fold} suggests that $N^{(2)}_\mathrm{B}$ is much smaller than $N^{(0)}_\mathrm{B}$ and $N^{(1)}_\mathrm{B}$, although further work is needed to assess whether it is negligible compared to the bispectrum signal. 
In App.~\ref{app:paired_sims}, we reach similar conclusions using the alternative, paired-simulation approach. We show that the $4_C\times 2_C$ term is well described by our analytical expression for $N^{(1)}_{\text{B}}$ and that the $6_C$ term (which partly sources $N^{(2)}_{\text{B}}$) is significantly smaller.

\section{Conclusions and future perspectives}\label{sec:conclusions}
In the past decade, weak gravitational lensing of the CMB has been established as a robust probe of late-time cosmological evolution. Data from upcoming CMB experiments are expected to be sensitive to the nonlinear clustering of matter at low redshift, which effectively induces non-Gaussianity in the lensing convergence field. This signal, if detected, provides additional information about structure growth, and has the potential to break degeneracies between cosmological parameters. 

In this paper, we calculated for the first time the noise-biases that arise when estimating the lensing convergence bispectrum from CMB temperature anisotropies via the Hu \&\ Okamoto quadratic estimator. We provided a theoretical model of the leading noise-biases using two methods: the standard Wick contraction of correlation functions; and the Feynman diagram formalism applied to CMB lensing. 
Regarding the latter approach, we adapted the results of Refs.~\cite{jenkins:feynman_diagrams1,jenkins:feynman_diagrams2} to the bispectrum case. The diagrammatic approach allows one to organise very effectively the calculation of the noise-biases up to $j$-orders of the lensing potential $\phi$ and is particularly helpful for higher-order biases.

We further numerically evaluated the leading noise-biases, $N^{(0)}_\mathrm{B}$ and $N^{(1)}_\mathrm{B}$, in the equilateral and folded configurations of the bispectrum. We showed that they represent the largest contribution to the reconstructed bispectrum and dominate over the signal by several orders of magnitude at small scales ($L\gg 100$) for a temperature-based reconstruction. We finally tested our theoretical results against simulations of temperature maps, finding excellent agreement. 

The general scope of this work was to give a first quantification of the reconstruction noise-biases of the CMB lensing bispectrum in view of a potential detection with ground-based experiments. Therefore, we leave a number of generalizations and tests for future work:
\begin{enumerate}
    \item We have limited our analysis to temperature anisotropies only, however polarization will be increasingly important for forthcoming surveys. We should therefore extend our analysis to quantify the noise-biases for reconstructions combining temperature and polarization anisotropies. A promising feature of polarization-based bispectrum reconstruction is that for the $EB$ quadratic estimator, which will dominate in the CMB-S4 era~\cite{cmbs4:science_book}, the $N^{(0)}_{\text{B}}$ noise-bias vanishes by parity.
    \item  While our simulation results suggest that the higher-order $N^{(2)}_{\text{B}}$ bias is much smaller than $N^{(0)}_{\text{B}}$ and $N^{(1)}_{\text{B}}$, further work is needed to determine whether $N^{(2)}_{\text{B}}$ is negligible compared to the signal bispectrum. More converged simulation-based estimates and an analytic understanding of $N^{(2)}_{\text{B}}$ will both be helpful in this regard.
    \item While we have provided a theoretical model of all the terms that contribute to $N^{(3/2)}_\mathrm{B}$, a numerical evaluation of this noise-bias is needed to assess its contribution to the reconstructed bispectrum. Given that it contains the lensing convergence bispectrum, the analysis will depend on the matter bispectrum modelling. Future work should also compare the numerical results against estimates of the same non-Gaussian bias from ray-traced $N$-body simulations.
    \item Our analysis ignores the effects of the curl-modes in the deflection angle introduced by post-Born corrections, therefore it would be interesting to assess how they affect our results.
    \item In the case of the reconstruction of the lensing power spectrum from data, the biases are commonly removed with realization-dependent methods that mix data and simulations. These make the subtraction more robust to mismodelling in the simulations and reduce covariance between the reconstructed power at different multipoles. The realization-dependent estimators for a general $n$-pt correlation function has been developed in Ref.~\cite{namikawa:polyspectra_rd}, but future work should test the bispectrum estimator for specific experiments, such as SO. Given the size of the $N^{(0)}_{\text{B}}$ and $N^{(1)}_{\text{B}}$ biases, achieving highly converged bias corrections will be very important to avoid significant additional scatter in the reconstructed bispectrum.
    \item It will also be important to quantify the impact on bispectrum reconstruction of residuals of extragalactic foregrounds in maps of the temperature anisotropies. As for the reconstructed power spectrum, polarization-based estimators will be significantly less affected by these foreground signals.
\end{enumerate}
Given the potential of the lensing bispectrum as a probe of cosmological evolution, it will be critical to explore all of the work mentioned above. 

\vskip 15 pt
\paragraph{Acknowledgments}

We are grateful to Vanessa Böhm, Julien Carron, William Coulton, Mathew Madhavacheril, Joel Meyers, and Alexander van Engelen for fruitful discussions. A.K., G.O. and P.D.M acknowledge support from the Netherlands organization for scientific research (NWO) VIDI grant (dossier 639.042.730). A.K. also acknowledges the Simons Foundation for support in the latest stage of this work. A.C. acknowledges support from the STFC (grant numbers ST/N000927/1 and ST/S000623/1).
T.N. acknowledges support from JSPS KAKENHI Grant No. JP20H05859 and No. JP22K03682. The Kavli IPMU is supported by World Premier International Research Center Initiative (WPI Initiative), MEXT, Japan. \\
\appendix

\section{Detailed computation of $N_B^{(3/2)}$} \label{appen:N32}
The noise-bias $N_B^{(3/2)}$ receives contribution from connected functions both linear and non-linear in $\phi$. 
In the following we evaluate such terms separately, using the Feynman diagrams.

\subsubsection*{Contributions from $4_C \times 2_C$ (linear order):} 
\begin{figure}
    \centering
    \begin{subfigure}{0.35\textwidth}
        \centering
        \includegraphics[width=\linewidth]{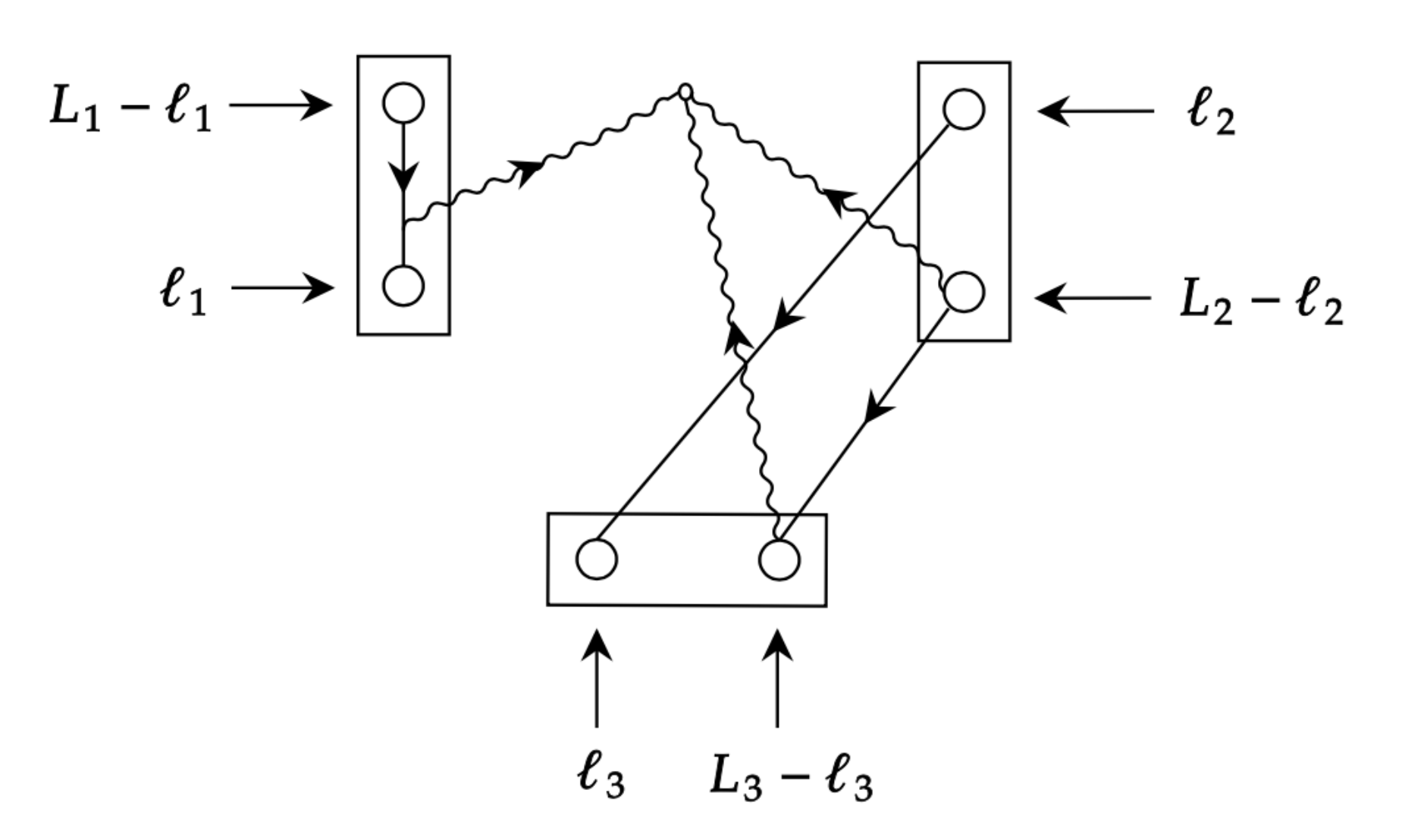} 
        \caption{$N^{(3/2)}_{{\rm lin}, a}$} \label{fig:N321_a}
    \end{subfigure}
    \hspace{1cm}
    \begin{subfigure}{0.35\textwidth}
        \centering
        \includegraphics[width=\linewidth]{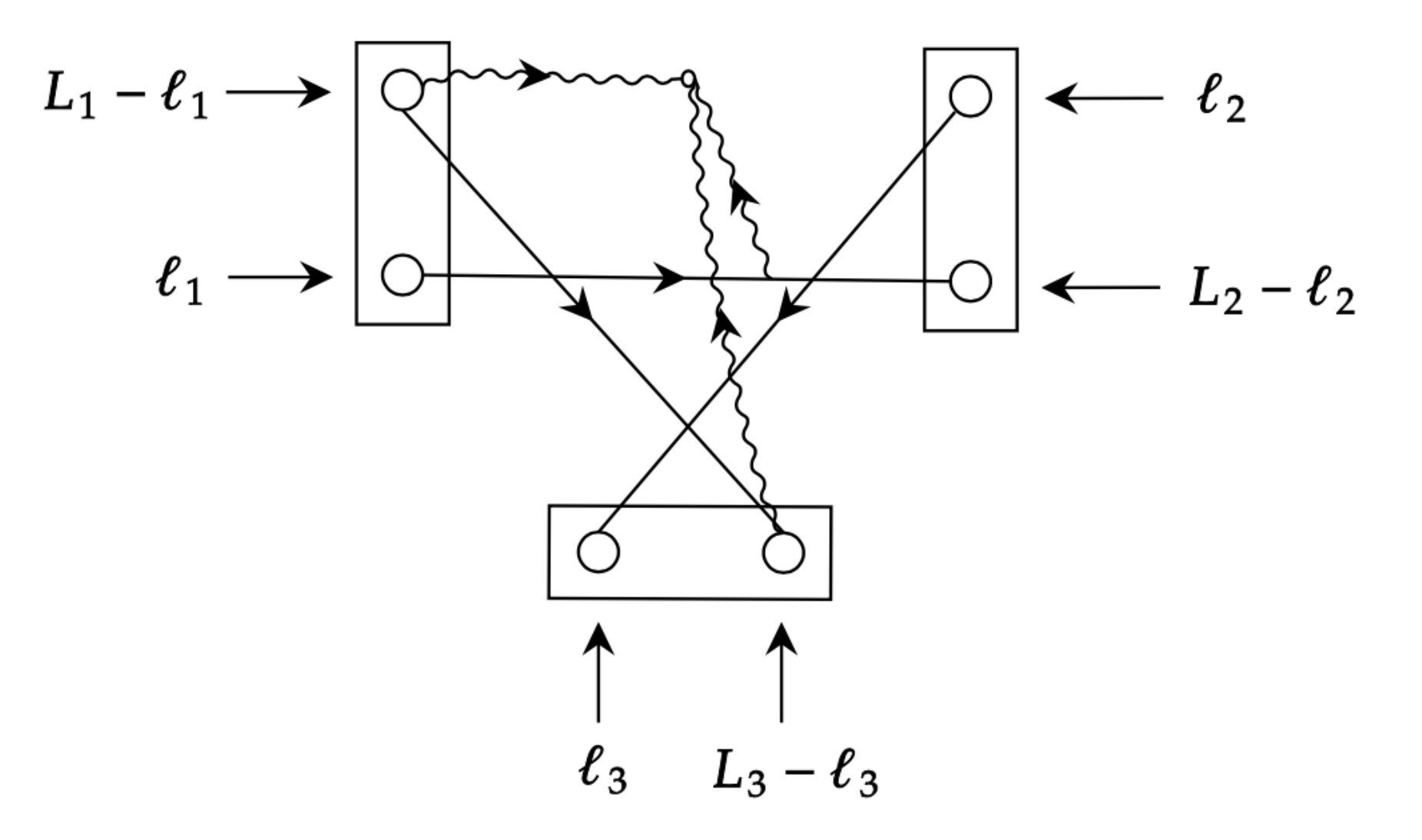} 
        \caption{$N^{(3/2)}_{{\rm lin}, b}$} \label{fig:N321_b}
    \end{subfigure}
    
    \vspace{1cm}
    
    \begin{subfigure}{0.35\textwidth}
        \centering
        \includegraphics[width=\linewidth]{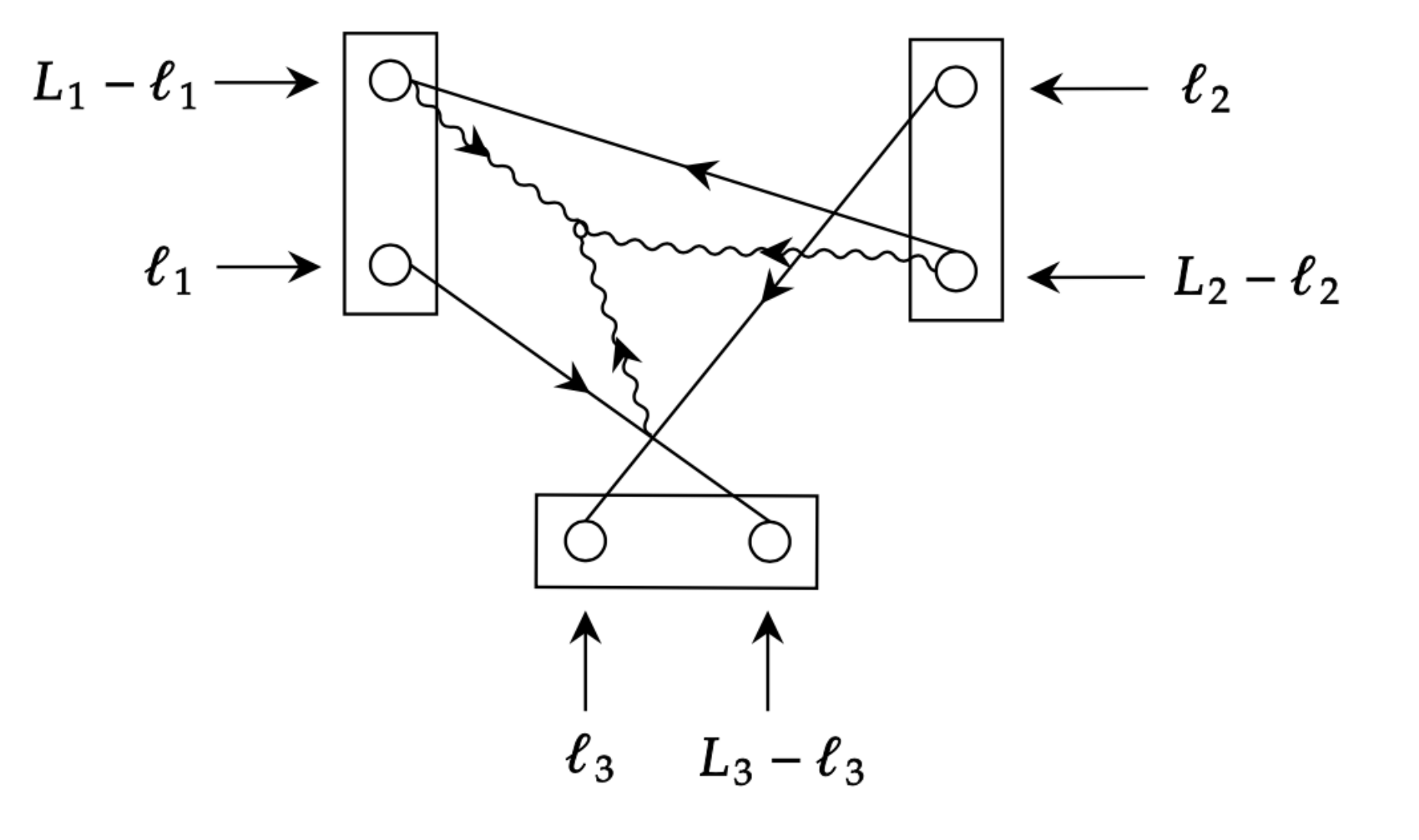} 
        \caption{$N^{(3/2)}_{{\rm lin}, c}$} \label{fig:N321_c}
    \end{subfigure}
     \hspace{1cm}
    \begin{subfigure}{0.35\textwidth}
        \centering
        \includegraphics[width=\linewidth]{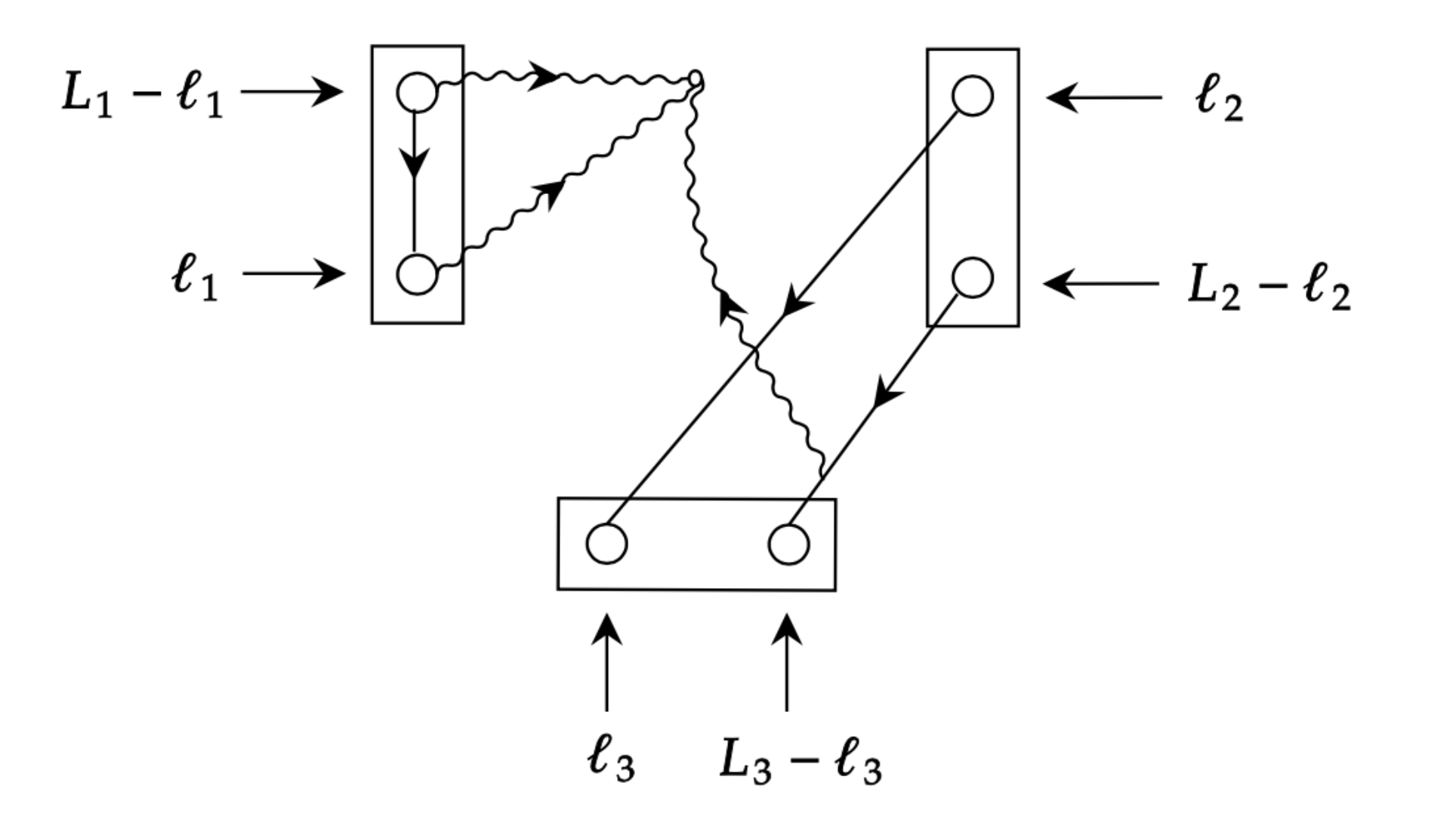} 
        \caption{$N^{(3/2)}_{{\rm lin}, d}$} \label{fig:N321_d}
    \end{subfigure}

    \caption{Independent Feynman diagrams contributing to the $N^{(3/2)}_\mathrm{B}$ noise-bias of $\langle \hat{\kappa} \hat{\kappa} \hat{\kappa}  \rangle$ with no lensed temperature vertexes expanded beyond the linear order \eqref{eq:delta_T}, $4_C \times 2_C$ terms.} \label{fig:N32kkk_linear_4x2}
\end{figure}

At linear order, there are 4 independent Feynman diagrams that correspond to 
\begin{equation}
    \langle\delta T \delta T \delta T  T\rangle_C \,  \langle \widetilde{T}\widetilde{T}\rangle_C\,,
\end{equation}
and are depicted in Fig.~\ref{fig:N32kkk_linear_4x2}. These diagrams correspond to
\begin{equation}
    \begin{split}
        N_{{\rm lin}, a}^{(3/2)} = 4 \int'_{\bsl_1,\bsl_2} & \,  F(\bsl_1,\bsL_1-\bsl_1) \,  F(\bsl_2,\bsL_2-\bsl_2) \, F(-\bsl_2,\bsL_3+\bsl_2) \, C^{\widetilde{T}\widetilde{T}}_{\ell_2}\\
        &\times \mathcal{F}(\bsl_1,\bsL_1-\bsl_1,\bsL_2-\bsl_2,\bsL_3+\bsl_2) + (\bsL_1 \leftrightarrow \bsL_2) + (\bsL_1 \leftrightarrow \bsL_3) \, , 
    \end{split}
\end{equation}
\begin{equation}
    \begin{split}
        N_{{\rm lin}, b}^{(3/2)} = 8 \int'_{\bsl_1,\bsl_2} & \, F(\bsl_1,\bsL_1-\bsl_1) \,  F(\bsl_2,\bsL_2-\bsl_2) \, F(-\bsl_2,\bsL_3+\bsl_2) \, C^{\widetilde{T}\widetilde{T}}_{\ell_2}\\
        &\times \mathcal{F}(\bsl_1,\bsL_2-\bsl_2,\bsL_1-\bsl_1,\bsL_3+\bsl_2) + (\bsL_1 \leftrightarrow \bsL_2) + (\bsL_1 \leftrightarrow \bsL_3) \, , 
    \end{split}
\end{equation}
\begin{equation}
    \begin{split}
        N_{{\rm lin}, c}^{(3/2)} = 8  \int'_{\bsl_1,\bsl_2}& \, F(\bsl_1,\bsL_1-\bsl_1) \,  F(\bsl_2,\bsL_2-\bsl_2) \, F(-\bsl_2,\bsL_3+\bsl_2) \, C^{\widetilde{T}\widetilde{T}}_{\ell_2}\\
        &\times \mathcal{F}(\bsl_1,\bsL_3+\bsl_2,\bsL_1-\bsl_1,\bsL_2-\bsl_2) + (\bsL_1 \leftrightarrow \bsL_2) + (\bsL_1 \leftrightarrow \bsL_3) \, ,
    \end{split}
\end{equation}
\begin{equation}
    \begin{split}
        N_{{\rm lin}, d}^{(3/2)} = 4  \int'_{\bsl_1,\bsl_2} & \, F(\bsl_1,\bsL_1-\bsl_1) \,  F(\bsl_2,\bsL_2-\bsl_2) \, F(-\bsl_2,\bsL_3+\bsl_2) \, C^{\widetilde{T}\widetilde{T}}_{\ell_2}\\
        &\times \mathcal{F}(\bsL_2-\bsl_2,\bsL_3+\bsl_2,\bsl_1,\bsL_1-\bsl_1) + (\bsL_1 \leftrightarrow \bsL_2) + (\bsL_1 \leftrightarrow \bsL_3) \,,
    \end{split}
\end{equation}
where we defined $\mathcal{F}(\bsl_1,\bsl_2,\bsl_3,\bsl_4)= f(\bsl_1,\bsl_2)I(\bsl_3,\bsl_4)$, with
\begin{equation}
    \begin{split}
       I(\bsl_i,\bsl_j)  = \int_{\bsp} \left[\bsp\cdot (\bsl_i-\bsp)\right] \,\left[\bsp\cdot (\bsl_j+\bsp)\right] \, B_\phi (|\bsl_i-\bsp|,|\bsl_j+\bsp|,|\bsl_i+\bsl_j|) \, C^{\widetilde{T}\widetilde{T}}_{\bsp} \, .
    \end{split}
\end{equation}

Therefore, the final contribution from the $4_C \times 2_C$ terms reads as
\begin{equation}
     N^{(3/2)}_{B,\rm lin}= N^{(3/2)}_{B,\rm lin}(\bsL_1,\bsL_2,\bsL_3)+N^{(3/2)}_{B,\rm lin}(\bsL_2,\bsL_1,\bsL_3)+N^{(3/2)}_{B,\rm lin}(\bsL_3,\bsL_1,\bsL_2) \, ,
\end{equation}
where
\begin{align}
        N_{B,\rm lin}^{(3/2)}&(\bsL_1,\bsL_2,\bsL_3) = 4  \int'_{\bsl_1,\bsl_2}\, F(\bsl_1,\bsL_1-\bsl_1) \, F(\bsl_2,\bsL_2-\bsl_2) \, F(-\bsl_2,\bsL_3+\bsl_2) \, C^{\widetilde{T}\widetilde{T}}_{\ell_2}\nonumber\\
        &\qquad \qquad\times \Big[\mathcal{F}(\bsl_1,\bsL_1-\bsl_1,\bsL_2-\bsl_2,\bsL_3+\bsl_2)+2 \mathcal{F}(\bsl_1,\bsL_2-\bsl_2,\bsL_1-\bsl_1,\bsL_3+\bsl_2)\nonumber\\
        & \qquad \qquad + 2\mathcal{F}(\bsl_1,\bsL_3+\bsl_2,\bsL_1-\bsl_1,\bsL_2-\bsl_2)+\mathcal{F}(\bsL_2-\bsl_2,\bsL_3+\bsl_2,\bsl_1,\bsL_1-\bsl_1)\Big] \, .
\end{align}

\subsubsection*{Contributions from $6_C$ (linear order):} 
\begin{figure}
    \centering
    \begin{subfigure}{0.35\textwidth}
        \centering
        \includegraphics[width=\linewidth]{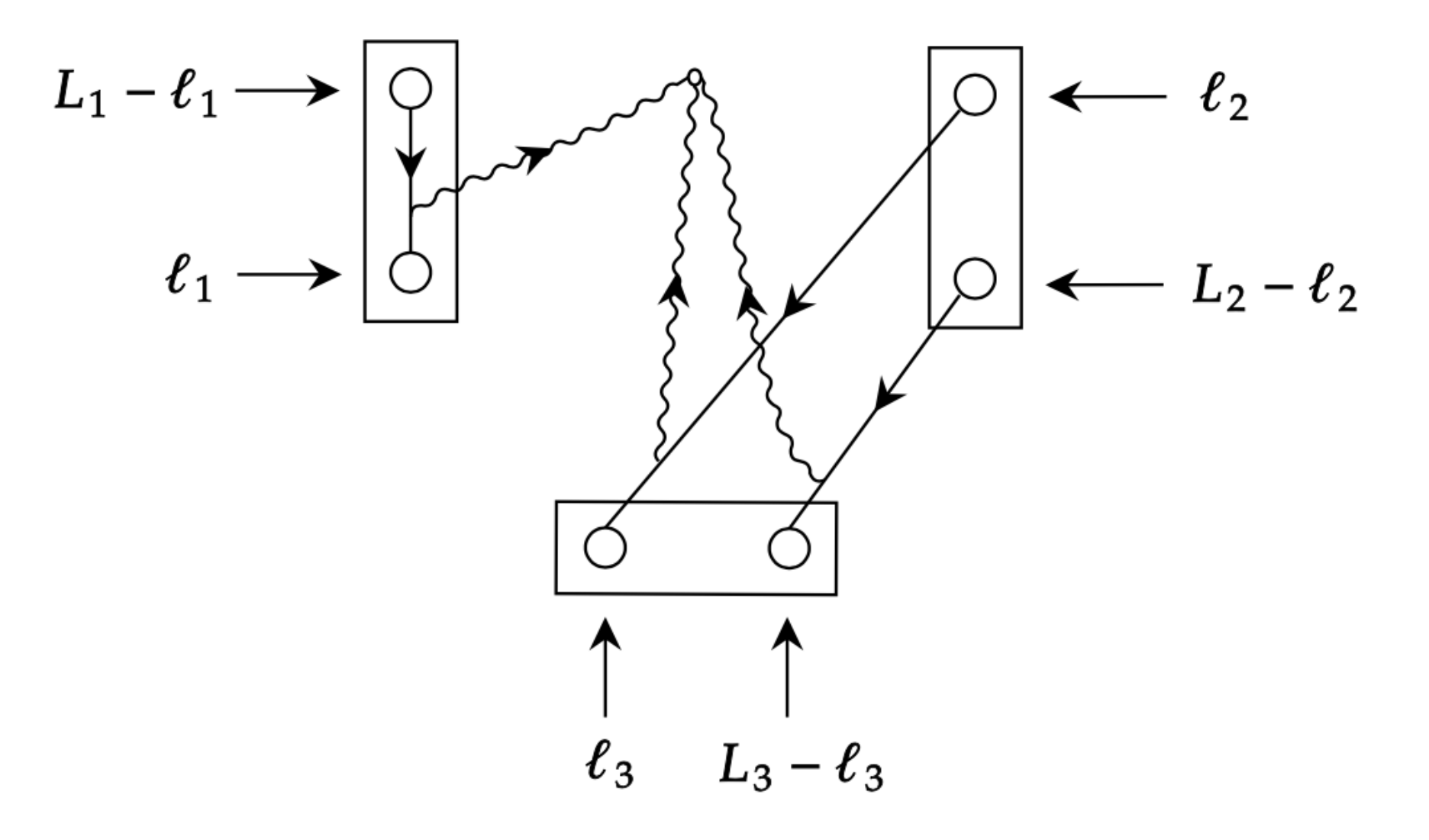} 
        \caption{$N^{(3/2)}_{{\rm lin}, e}$} \label{fig:N321_e}
    \end{subfigure}
    \hspace{1cm}
    \begin{subfigure}{0.35\textwidth}
        \centering
        \includegraphics[width=\linewidth]{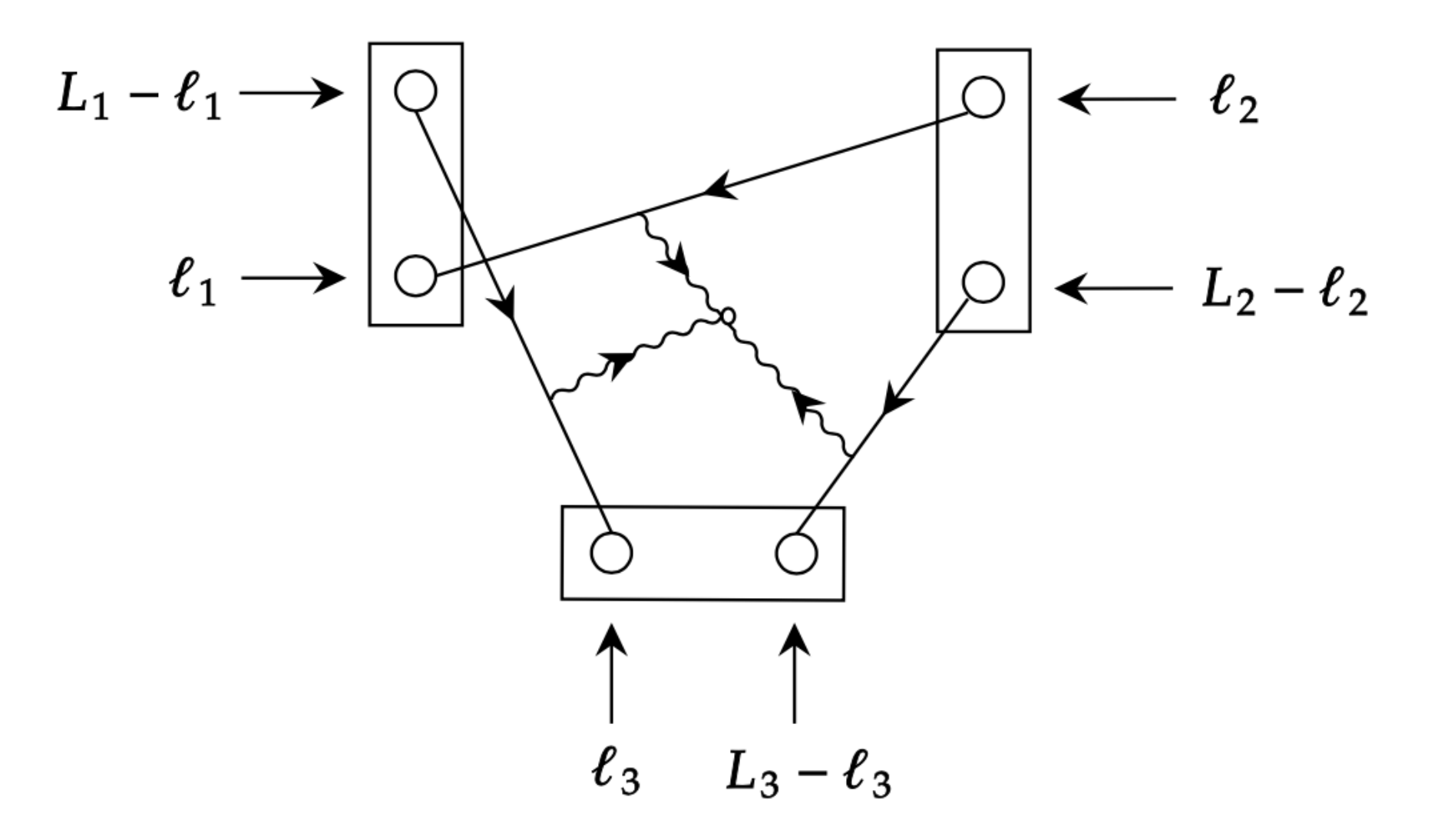} 
        \caption{$N^{(3/2)}_{{\rm lin}, f}$} \label{fig:N321_f}
    \end{subfigure}
    \caption{Independent Feynman diagrams contributing to the $N^{(3/2)}_\mathrm{B}$ noise-bias of $\langle \hat{\kappa} \hat{\kappa} \hat{\kappa}  \rangle$ with no lensed temperature vertexes expanded beyond the linear order \eqref{eq:delta_T}, $6_C$ terms.} \label{fig:N32kkk_linear_6}
\end{figure}

\begin{figure}[htbp]
        \centering
        \includegraphics[width=0.5\linewidth]{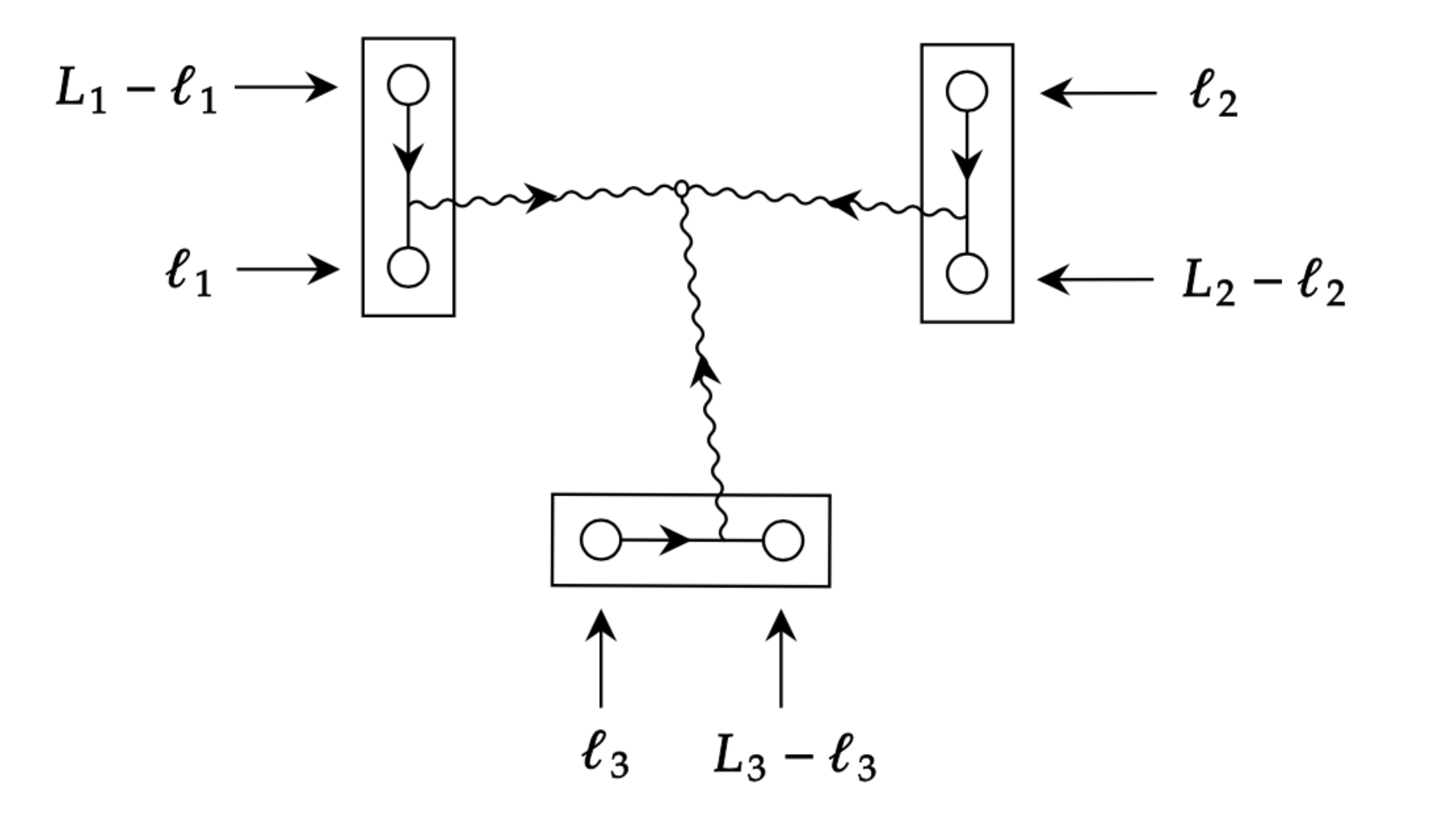} 
        \label{fig:N321_sign}
    
    \caption{Feynman diagram corresponding to the bispectrum signal $B_\kappa$.} \label{fig:N32kkk_signal}
\end{figure}

At linear order, the contribution to $N_\mathrm{B}^{3/2}$ comes from 
\begin{equation}
\langle\delta T \delta T \delta T  T  T T\rangle_C
\end{equation}
In terms of Feynman diagrams, depicted in Fig.~\ref{fig:N32kkk_linear_6}, we have
\begin{align}
        N_{{\rm lin}, e}^{(3/2)}=& 2\int'_{\bsl_1,\bsl_2,\bsl_3} F(\bsl_1,\bsL_1-\bsl_1) \, F(\bsl_2,\bsL_2-\bsl_2)F(\bsl_3,\bsL_3-\bsl_3) \nonumber\\
        &\times \, f(\bsl_1,\bsL_1-\bsl_1) \, f(\bsl_2,\bsl_3) \, f(\bsL_2-\bsl_2,\bsL_3-\bsl_3) \nonumber\\
        &\times B_\phi (L_1,|\bsl_2+\bsl_3|,|\bsL_2-\bsl_2+\bsL_3-\bsl_3|) + (\bsL_1 \leftrightarrow \bsL_2) + (\bsL_1 \leftrightarrow \bsL_3)\, ,
\end{align}
\begin{align}
            N_{{\rm lin}, f}^{(3/2)} =& 8 \int'_{\bsl_1,\bsl_2,\bsl_3} F(\bsl_1,\bsL_1-\bsl_1) \, F(\bsl_2,\bsL_2-\bsl_2)F(\bsl_3,\bsL_3-\bsl_3)\, f(\bsl_1,\bsl_2)\, f(\bsL_1-\bsl_1,\bsl_3)  \nonumber\\
            &\times \, f(\bsL_2-\bsl_2,\bsL_3-\bsl_3) B_\phi (|\bsl_1+\bsl_2|,|\bsL_1-\bsl_1+\bsl_3| , |\bsL_2-\bsl_2+\bsL_3-\bsl_3|).
\end{align}
The expansion of $6_C$ function at linear order also includes the bispectrum signal $B_\kappa$ that we aim to reconstruct, and it is as shown Fig. \ref{fig:N32kkk_signal}. 

\subsubsection*{Contributions from the second order expansion of the lensed temperature:} 

\begin{figure}
    \centering
    \begin{subfigure}{0.35\textwidth}
        \centering
        \includegraphics[width=\linewidth]{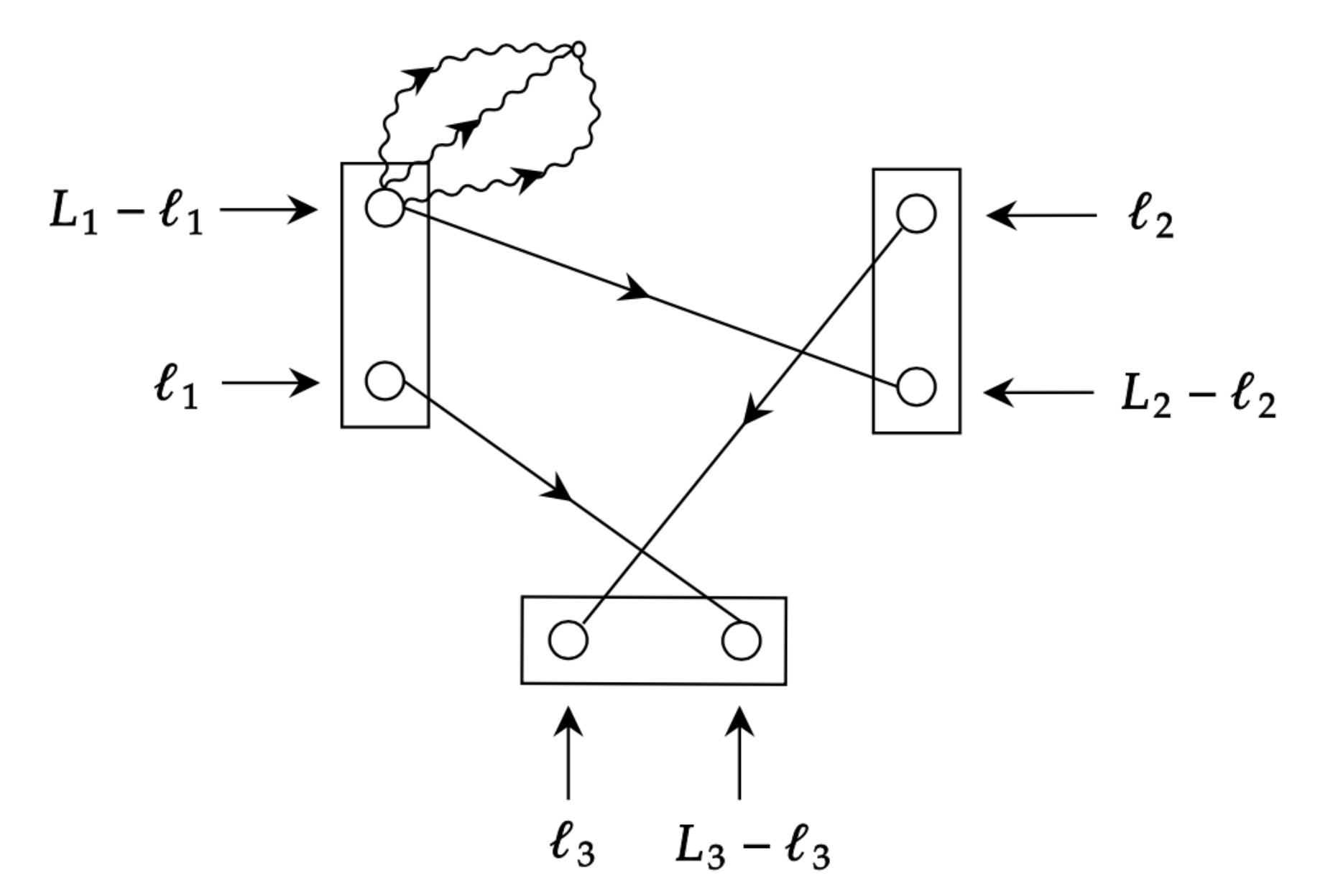} 
        \caption{$N^{(3/2)}_a$} \label{fig:N32_a}
    \end{subfigure}
    \hspace{1cm}
    \begin{subfigure}{0.35\textwidth}
        \centering
        \includegraphics[width=\linewidth]{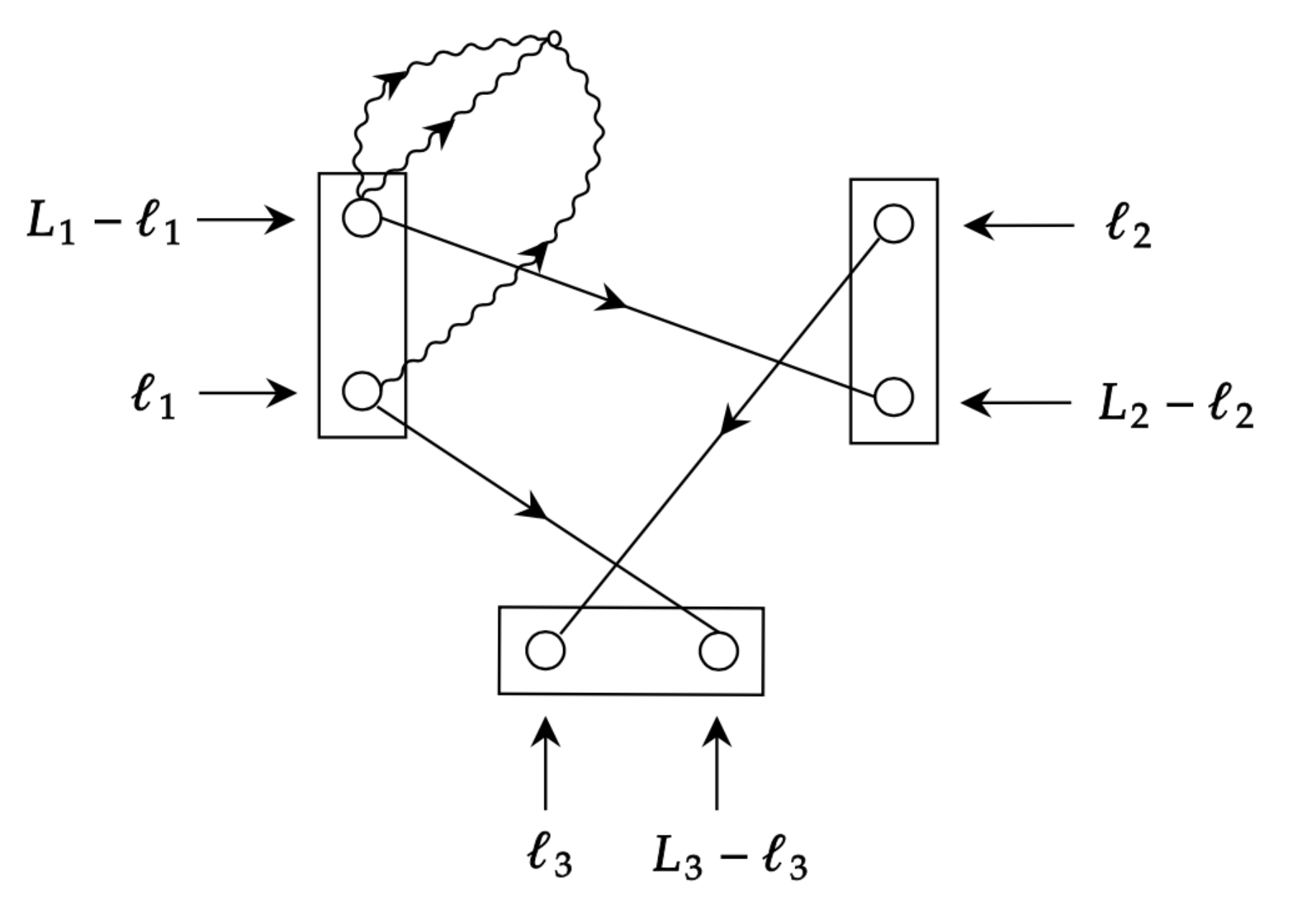} 
        \caption{$N^{(3/2)}_b$} \label{fig:N32_b}
    \end{subfigure}
    
    \vspace{1cm}
    
    \begin{subfigure}{0.35\textwidth}
        \centering
        \includegraphics[width=\linewidth]{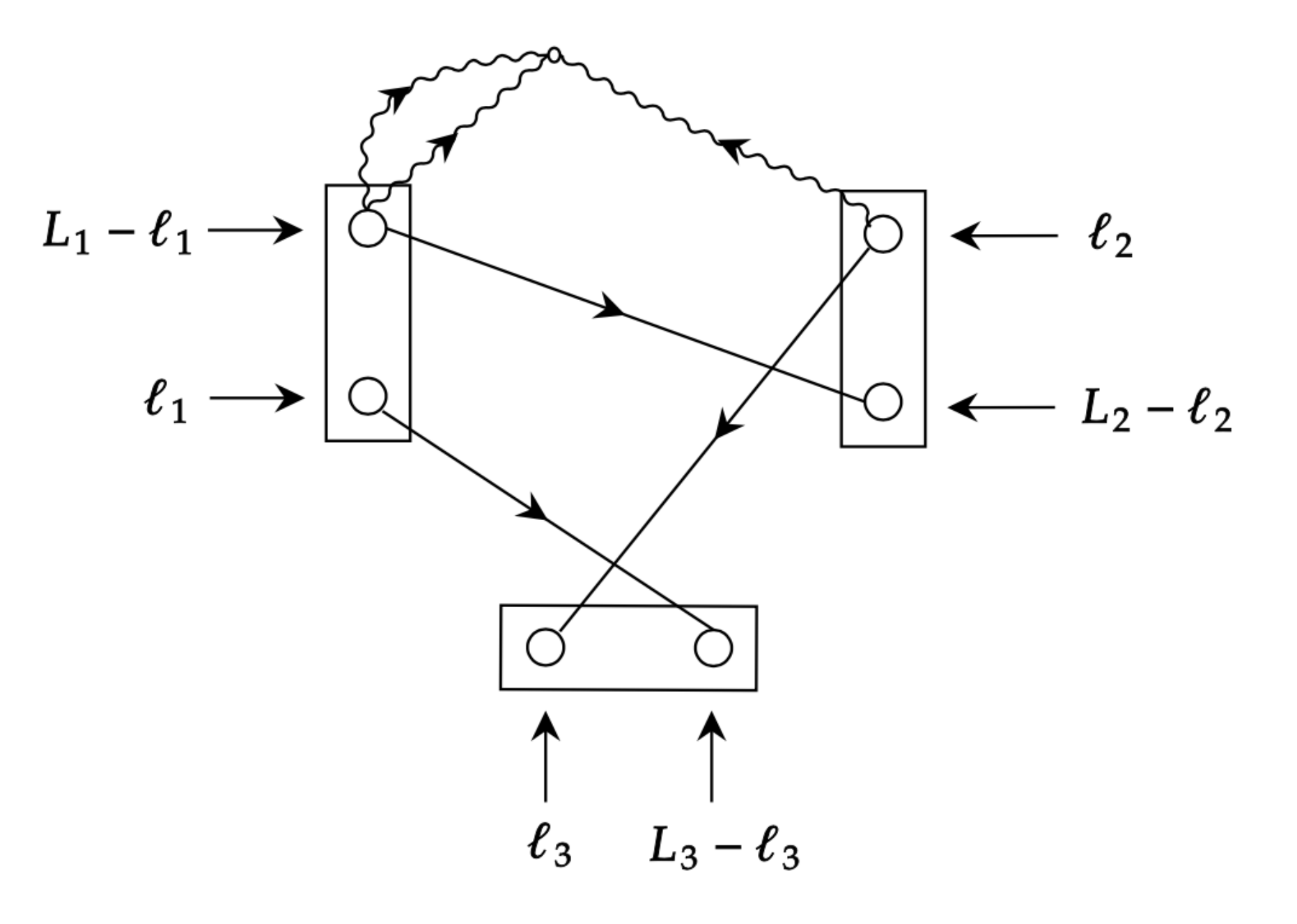} 
        \caption{$N^{(3/2)}_c$} \label{fig:N32_c}
    \end{subfigure}
     \hspace{1cm}
    \begin{subfigure}{0.35\textwidth}
        \centering
        \includegraphics[width=\linewidth]{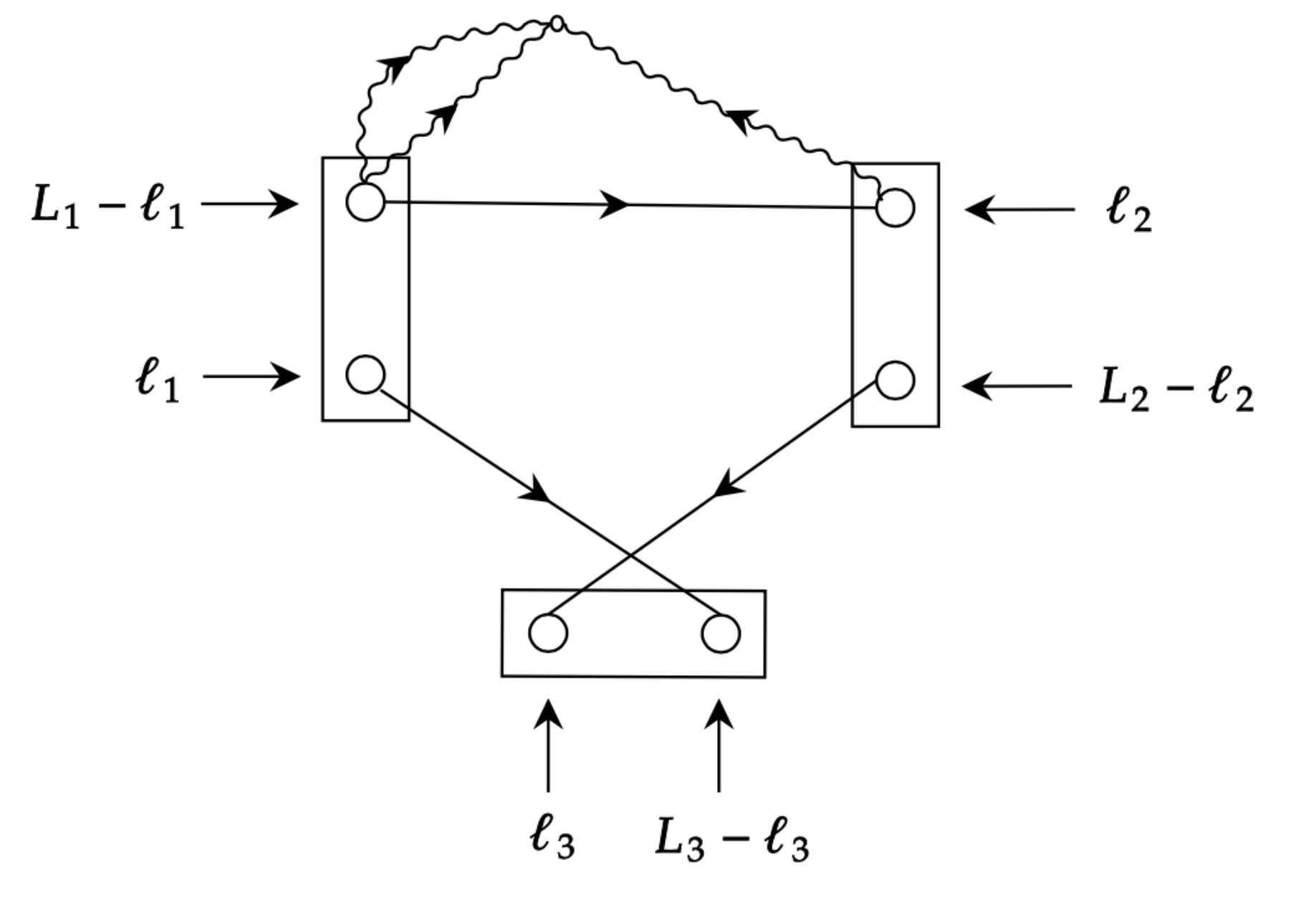} 
        \caption{$N^{(3/2)}_d$} \label{fig:N32_d}
    \end{subfigure}
    
        \vspace{1cm}
    
    \begin{subfigure}{0.35\textwidth}
        \centering
        \includegraphics[width=\linewidth]{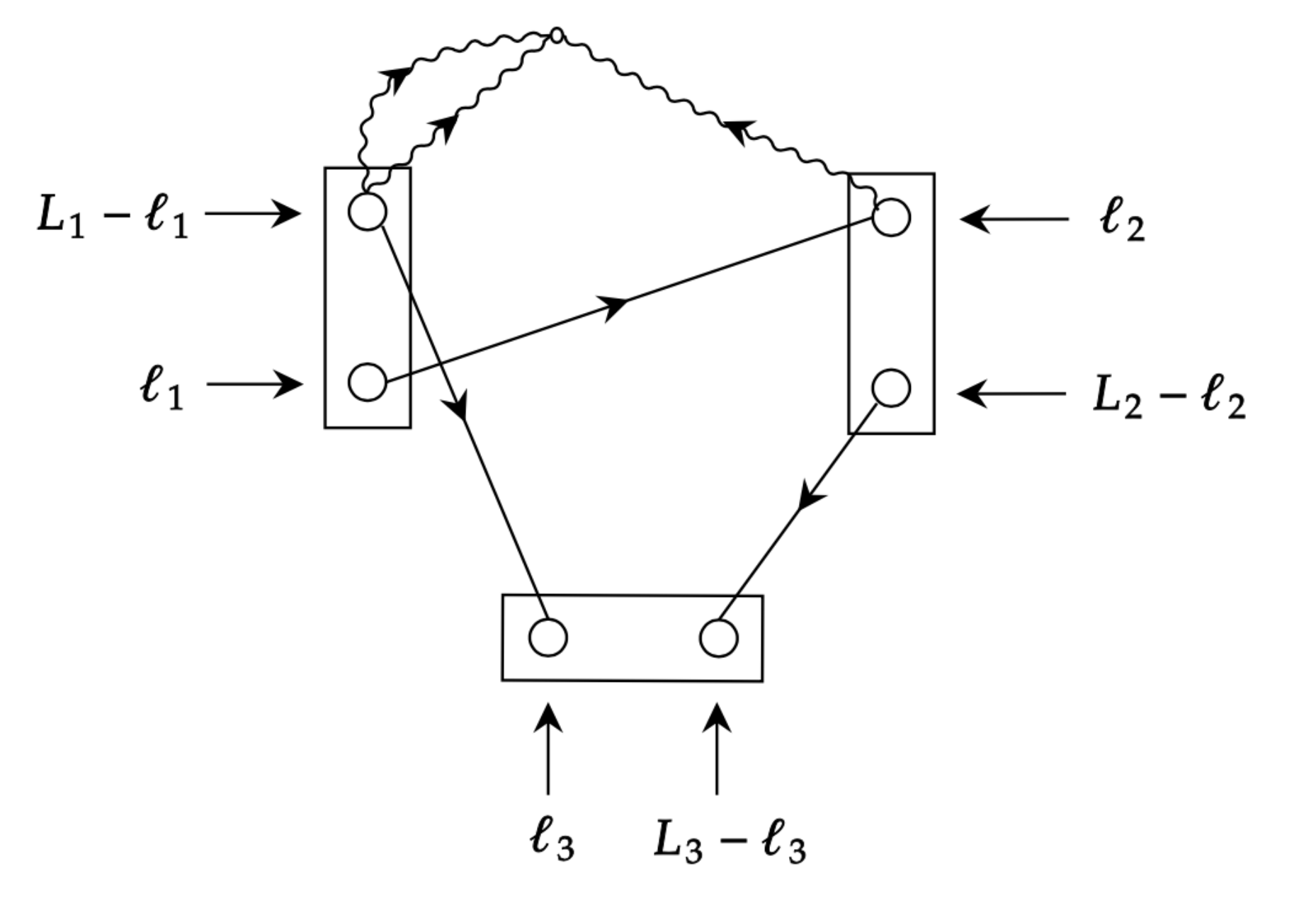} 
        \caption{$N^{(3/2)}_e$} \label{fig:N32_e}
    \end{subfigure}
     \hspace{1cm}
    \begin{subfigure}{0.35\textwidth}
        \centering
        \includegraphics[width=\linewidth]{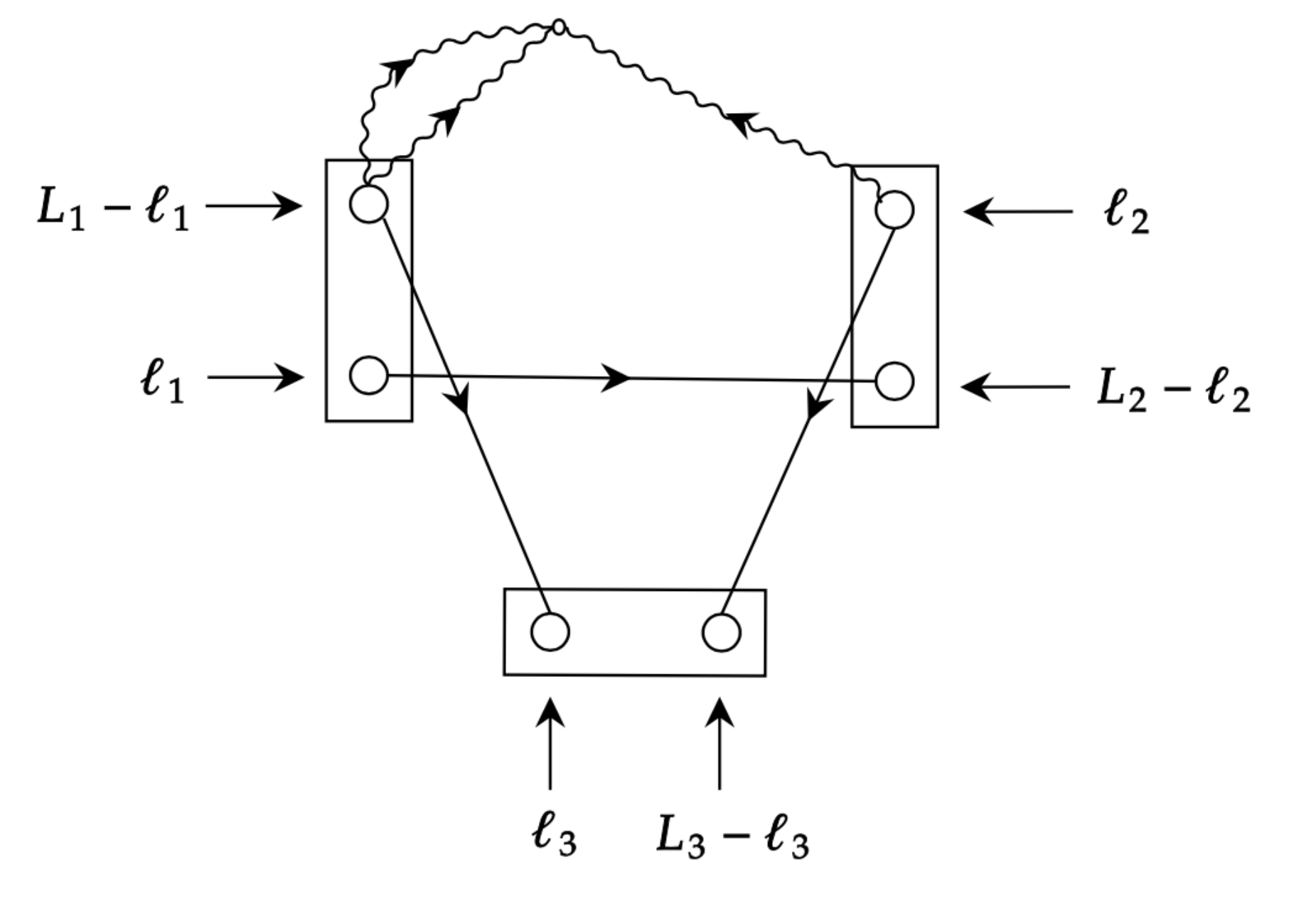} 
        \caption{$N^{(3/2)}_f$} \label{fig:N32_f}
    \end{subfigure}
    
        \vspace{1cm}
    
    \begin{subfigure}{0.35\textwidth}
        \centering
        \includegraphics[width=\linewidth]{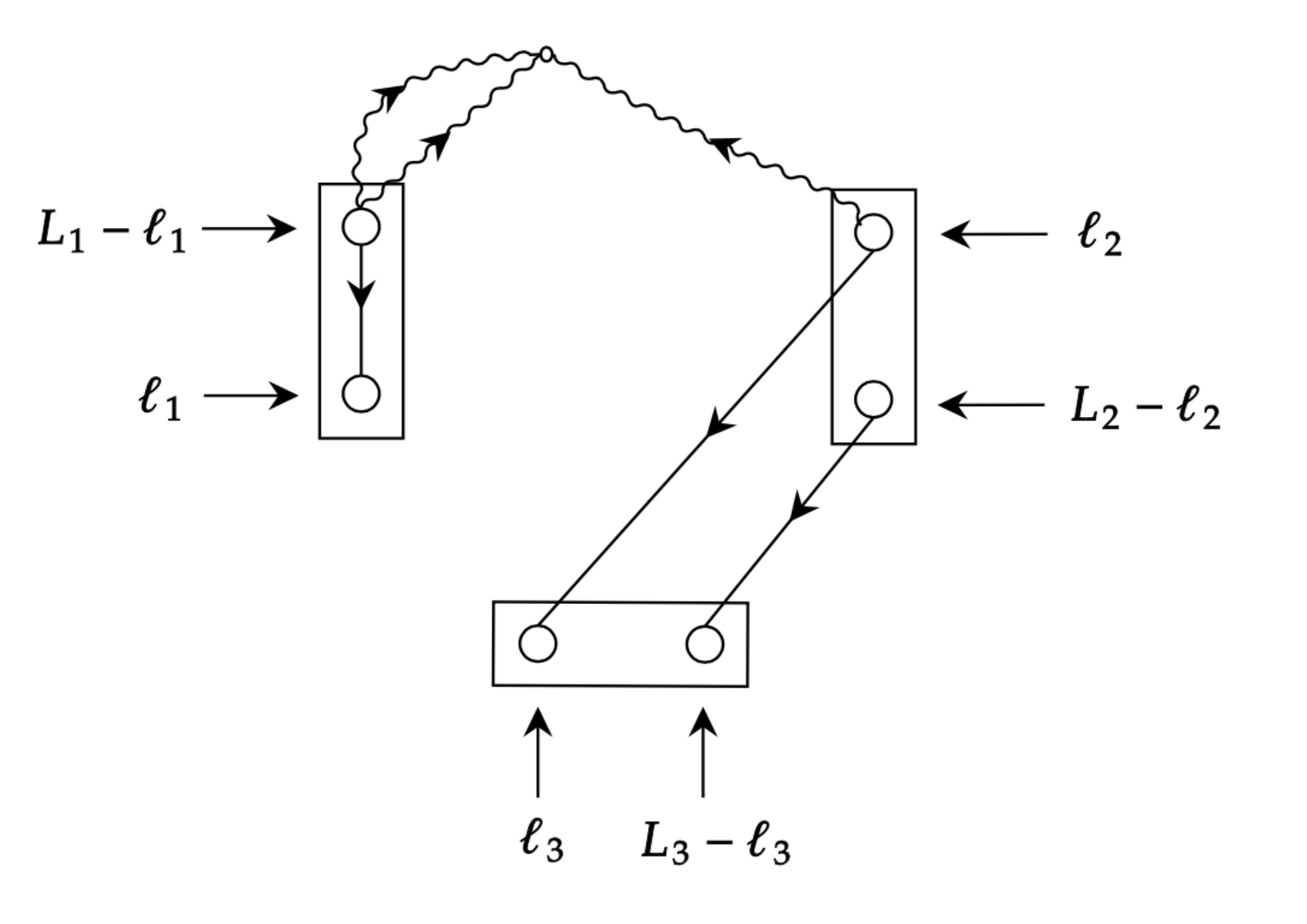} 
        \caption{$N^{(3/2)}_g$} \label{fig:N32_g}
    \end{subfigure}
     \hspace{1cm}
    \begin{subfigure}{0.35\textwidth}
        \centering
        \includegraphics[width=\linewidth]{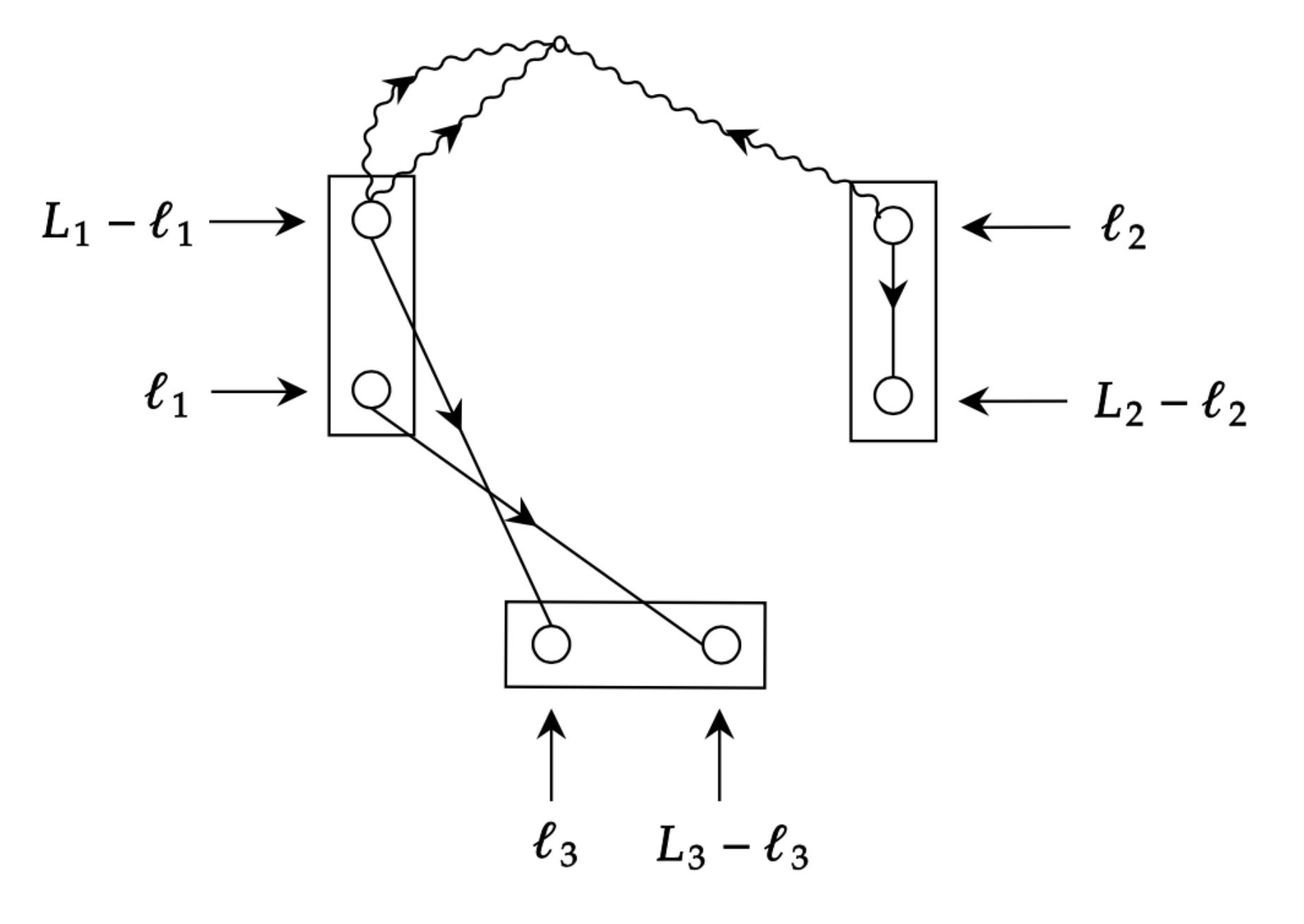} 
        \caption{$N^{(3/2)}_h$} \label{fig:N32_h}
    \end{subfigure}    
    \caption{Independent Feynman diagrams contributing to the $N_\mathrm{B}^{(3/2)}$ noise-bias of $\langle \hat{\kappa} \hat{\kappa}\hat{\kappa}\rangle$ with at least one lensed temperature vertex expanded beyond the linear order \eqref{eq:delta_T}.} \label{fig:N32kkk}
\end{figure} 

The estimation of $N_\mathrm{B}^{(3/2)}$ includes also non-zero contributions from the $\mathcal O(\phi^2)$ and $\mathcal O(\phi^3)$ terms in \eqref{eq:delta_T2}. In particular, the independent Feynman diagrams providing these additional contributions are depicted in Fig.~\ref{fig:N32kkk} and correspond to
\begin{equation}
    \begin{split}
        \text{(a):} &\quad\langle\delta^3 T T\rangle_C \,\langle \widetilde{T}\widetilde{T}\rangle_C \,  \langle \widetilde{T}\widetilde{T}\rangle_C\,,\\
        \text{(d):} &\quad\langle\delta^2 T \delta T\rangle_C \,\langle \widetilde{T}\widetilde{T}\rangle_C \,  \langle \widetilde{T}\widetilde{T}\rangle_C\, ,\\ 
        \text{(b, c, e, f, g, h):}&\quad \langle\delta^2 T \delta T T  T\rangle_C \,  \langle \widetilde{T}\widetilde{T}\rangle_C\,.
    \end{split}
\end{equation}
In the following, by exploiting the Feynman rules in Sec.~\ref{subsec:Feynman}, we give their analytical expression including symmetry factors and permutations over the capital momenta $\bsL_i$'s.
\begin{align} \label{eq:N32a}
N^{(3/2)}_a = 8  \int'_{\bsl_1,\bsl_2,\bsl_3}& \, F(\bsl_1+\bsL_1,-\bsl_1) \, F(\bsl_1,\bsL_2 - \bsl_1) \,  F(\bsl_1-\bsL_2,\bsL_1-\bsl_1)  \nonumber \\
& \times\Big[(\bsl_2 +\bsl_3) \cdot \bsl_1\Big] (\bsl_1 \cdot \bsl_2) \, (\bsl_3 \cdot \bsl_1) \,  B_\phi(|\bsl_2 +\bsl_3|, \ell_2, \ell_3) \nonumber\\
&\times C^{\widetilde{T}\widetilde{T}}_{|\bsL_1 + \bsl_1|} \, C^{\widetilde{T}\widetilde{T}}_{\ell_1} \, C^{\widetilde{T}\widetilde{T}}_{|\bsL_2 - \bsl_1|}  + \mbox{perms($\bsL_i$)} \, .
\end{align}
By definition our noise-bias has to be symmetric under the exchange of the $\bsL_i$'s. By making the change of variable $\bsl_1 = -\bsl_1'$ in the previous term, we easily show that
\begin{equation}
N^{(3/2)}_a  = - N^{(3/2)}_a = 0 \, .    
\end{equation}
The result is that Feynman diagrams with three wiggly lines coming out from a single vertex sum up to 0 and do not contribute to the $N_B^{(3/2)}$ noise-bias. Therefore, only Feynman diagrams with two wiggly lines coming out from a given vertex contribute to higher orders in the lensing expansion. Their contribution is given in the following
\begin{align} \label{eq:N32b}
N^{(3/2)}_b = 8 \int'_{\bsl_1,\bsl_2,\bsl_3} &\, F(\bsl_1,\bsL_1 - \bsl_1) \, F(\bsl_2,\bsL_2 - \bsl_2) \, F(\bsL_2-\bsl_2,\bsL_1+\bsl_2)  \nonumber\\
&\times \Big[(\bsl_2 + \bsL_1) \cdot (\bsl_2-\bsl_1 + \bsL_1)\Big] \Big[\bsl_2 \cdot \bsl_3 \Big]  \Big[\bsl_2 \cdot (\bsl_2-\bsl_1-\bsl_3+\bsL_1)\Big]  \nonumber\\
& \times B_\phi(|\bsl_2-\bsl_1-\bsl_3+\bsL_1|, \ell_3, |\bsl_2-\bsl_1 + \bsL_1|)  \nonumber\\
&\times C^{\widetilde{T}\widetilde{T}}_{|\bsL_1 + \bsl_2|} \, C^{\widetilde{T}\widetilde{T}}_{\ell_2} \, C^{\widetilde{T}\widetilde{T}}_{|\bsL_2 - \bsl_2|}  + \mbox{perms($\bsL_i$)} \, ,
\end{align}
\begin{align} \label{eq:N32d}
N^{(3/2)}_d = 8 \int'_{\bsl_1,\bsl_2,\bsl_3}& \, F(-\bsl_1-\bsL_3, \bsl_1 - \bsL_2) \, F(\bsl_1,\bsL_2 - \bsl_1) \,  F(-\bsl_1,\bsL_3+\bsl_1) \nonumber\\
&\times \Big[\bsl_3 \cdot (\bsL_2 - \bsl_1 - \bsl_3)\Big] \Big[\bsl_3 \cdot \bsl_2 \Big] \Big[\bsl_3 \cdot (\bsl_1+\bsl_3-\bsl_2-\bsL_2)\Big]  \nonumber \\
& \times B_\phi(|\bsl_1+\bsl_3-\bsl_2-\bsL_2|, \ell_2, |\bsL_2 - \bsl_1 - \bsl_3|) \nonumber\\
&\times C^{\widetilde{T}\widetilde{T}}_{|\bsL_3 + \bsl_1|} \,  C^{\widetilde{T}\widetilde{T}}_{\ell_1} \, C^{\widetilde{T}\widetilde{T}}_{\ell_3} + \mbox{perms($\bsL_i$)} \, ,
\end{align}
\begin{align} \label{eq:N32e}
N^{(3/2)}_e = 8 \int'_{\bsl_1,\bsl_2,\bsl_3} & \, F(\bsl_1, \bsL_1 - \bsl_1) \, F(\bsL_3 - \bsl_3 ,  \bsl_3 + \bsL_1) \, F(\bsl_3,\bsL_3-\bsl_3) \nonumber\\
&\times \Big[\bsl_1 \cdot (\bsl_1 - \bsl_3 - \bsL_1)\Big] \Big[\bsl_3 \cdot \bsl_2 \Big]  \Big[\bsl_3 \cdot (\bsL_1 -\bsl_1+\bsl_3-\bsl_2)\Big]  \nonumber\\
& \times B_\phi(|\bsL_1 -\bsl_1+\bsl_3-\bsl_2|, \ell_2, |\bsl_1 - \bsl_3 - \bsL_1|) \nonumber\\
&\times C^{\widetilde{T}\widetilde{T}}_{|\bsL_3 - \bsl_3|} \, C^{\widetilde{T}\widetilde{T}}_{\ell_1} \, C^{\widetilde{T}\widetilde{T}}_{\ell_3} + \mbox{perms($\bsL_i$)} \, ,
\end{align}
\begin{align} \label{eq:N32f}
N^{(3/2)}_f =  8 \int'_{\bsl_1,\bsl_2,\bsl_3}&\, F(\bsl_1, \bsL_1 - \bsl_1) \, F(-\bsl_1 , \bsl_1 - \bsL_1 - \bsL_3) \, F(\bsl_3,\bsL_3-\bsl_3) \nonumber\\
&\times \Big[(\bsL_3-\bsl_3) \cdot (\bsl_1 - \bsl_3 - \bsL_1)\Big] \Big[\bsl_3 \cdot \bsl_2 \Big]  \Big[\bsl_3 \cdot (\bsL_1 -\bsl_1+\bsl_3-\bsl_2)\Big]  \nonumber\\
& \times B_\phi(|\bsL_1 -\bsl_1+\bsl_3-\bsl_2|, \ell_2, |\bsl_1 - \bsl_3 - \bsL_1|) \nonumber \\
&\times C^{\widetilde{T}\widetilde{T}}_{|\bsL_3 - \bsl_3|} \, C^{\widetilde{T}\widetilde{T}}_{\ell_1} \, C^{\widetilde{T}\widetilde{T}}_{\ell_3} + \mbox{perms($\bsL_i$)} \, ,
\end{align}
\begin{align} \label{eq:N32g}
N^{(3/2)}_g = 8 \int'_{\bsl_1,\bsl_2,\bsl_3} & \, F(\bsl_1, \bsL_1 - \bsl_1) \, F(\bsL_3 - \bsl_3, \bsL_1 + \bsl_3) \, F(\bsl_3,\bsL_3-\bsl_3)\nonumber\\
&\times \Big[\bsl_1 \cdot (\bsl_2 - \bsL_1)\Big] \Big[\bsl_1 \cdot \bsl_2 \Big] \Big[\bsl_3 \cdot \bsL_1 \Big]  B_\phi(|\bsL_1 - \bsl_2|, \ell_2, |- \bsL_1|) \nonumber \\
&\times C^{\widetilde{T}\widetilde{T}}_{|\bsL_3 - \bsl_3|} \, C^{\widetilde{T}\widetilde{T}}_{\ell_1} \, C^{\widetilde{T}\widetilde{T}}_{\ell_3} + \mbox{perms($\bsL_i$)} \, ,
\end{align}
\begin{align} \label{eq:N32h}
N^{(3/2)}_h = 8 \int'_{\bsl_1,\bsl_2,\bsl_3} & \, F(\bsL_3 - \bsl_3, \bsL_2 + \bsl_3) \, F(\bsl_1, \bsL_2 - \bsl_1) \, F(\bsl_3,\bsL_3-\bsl_3)  \nonumber\\
&\times \Big[\bsl_3 \cdot \bsl_2 \Big] \Big[\bsl_3 \cdot (-\bsL_2 - \bsl_2)\Big] \Big[\bsl_1 \cdot \bsL_2 \Big] B_\phi(|-\bsL_2 - \bsl_2|, \ell_2, \ell_2) \nonumber \\
&\times C^{\widetilde{T}\widetilde{T}}_{|\bsL_3 - \bsl_3|} \, C^{\widetilde{T}\widetilde{T}}_{\ell_1} \, C^{\widetilde{T}\widetilde{T}}_{\ell_3} + \mbox{perms($\bsL_i$)} \, .
\end{align}
The contribution from $N^{(3/2)}_c$ turns out to be zero. By summing all these terms to those in the previous subsections we get the full $N^{(3/2)}_B$ noise-bias of Eq.~\eqref{eq:kkk_noise_biase_and_signal}. 

\section{Alternative (paired) simulations}
\label{app:paired_sims}

In this appendix, we describe an alternative form of simulations that allow us to estimate the $N^{(0)}_{\text{B}}$ noise-bias and the $4_C \times 2_C$ bias term in the cumulant expansion~\eqref{eq:decomp_6points} of the reconstructed lensing bispectrum. These simulations allow us to target individual terms in the expansion separately, and even particular Wick contractions of the unlensed CMB that contribute to each term if required. The simulations here use exclusively non-Gaussisan lensed CMB maps, which has the advantage over the method used in Sec.~\ref{sec:validation} that there is no reliance on the power spectra of the lensed simulations matching the theory lensed power spectrum to very high accuracy on all scales.

As we shall need to identify explicitly the two temperature fields that enter the quadratic estimator\footnote{We shall refer to these as the two ``legs'' of the quadratic estimator.} for $\kappa(\bsL)$, we adopt the notation $\hat{\kappa}_{\bsL}(i,j)$ for the quadratic estimator applied to temperature fields $i$ and $j$. We use a symmetrised form of the quadratic estimator, as in Eq.~\eqref{eq:lensing_convergence_estimator}, so that $\hat{\kappa}_{\bsL}(i,j) = \hat{\kappa}_{\bsL}(j,i)$ for any temperature fields $i$ and $j$.

In this appendix, the codes to generate the lensed temperature maps and to compute $\hat{\kappa}$ and the binned bispectrum are the same as those described in Sec.~\ref{sec:comp:N}. 
Note that we include instrumental noise in the CMB maps here. The noise model used is that relevant for the Simons Observatory large-aperture telescope with a white noise level of $10 \, \mu \text{K-arcmin}$ and a beam size (full-width at half-maximum) of $1.4\,\text{arcmin}$~\cite{so:forecast}.

\subsection{$N_\mathrm{B}^{(0)}$ bias}
\label{app:N0}

The $N_\mathrm{B}^{(0)}$ bias originates from the $2_C\times 2_C \times 2_C$ term in Eq.~\eqref{eq:decomp_6points}. For the $\bsL_i \neq 0$, the only contractions that contribute couple fields across different quadratic estimators, giving eight possible permutations. These are all equal because of the symmetry of the quadratic estimator, so we have, symbolically,
\begin{equation}
N^{(0)}_\text{B} \sim 8 \langle \hat{\kappa}_{\bsL_1}(I,J) \hat{\kappa}_{\bsL_2}(K,I) \hat{\kappa}_{\bsL_3}(J,K) \rangle_{IJK} \,  ,
\label{eq:appbnzero}
\end{equation}
where the expectation value is over independent realizations $I$, $J$, $K$ of the noisy, lensed CMB temperature anisotropies. Note how the independence of $I$, $J$ and $K$ ensures that only 2-pt correlations arise, despite the fields being non-Gaussian.

\begin{figure}[t]
\centering
    \includegraphics[width=0.8\textwidth]{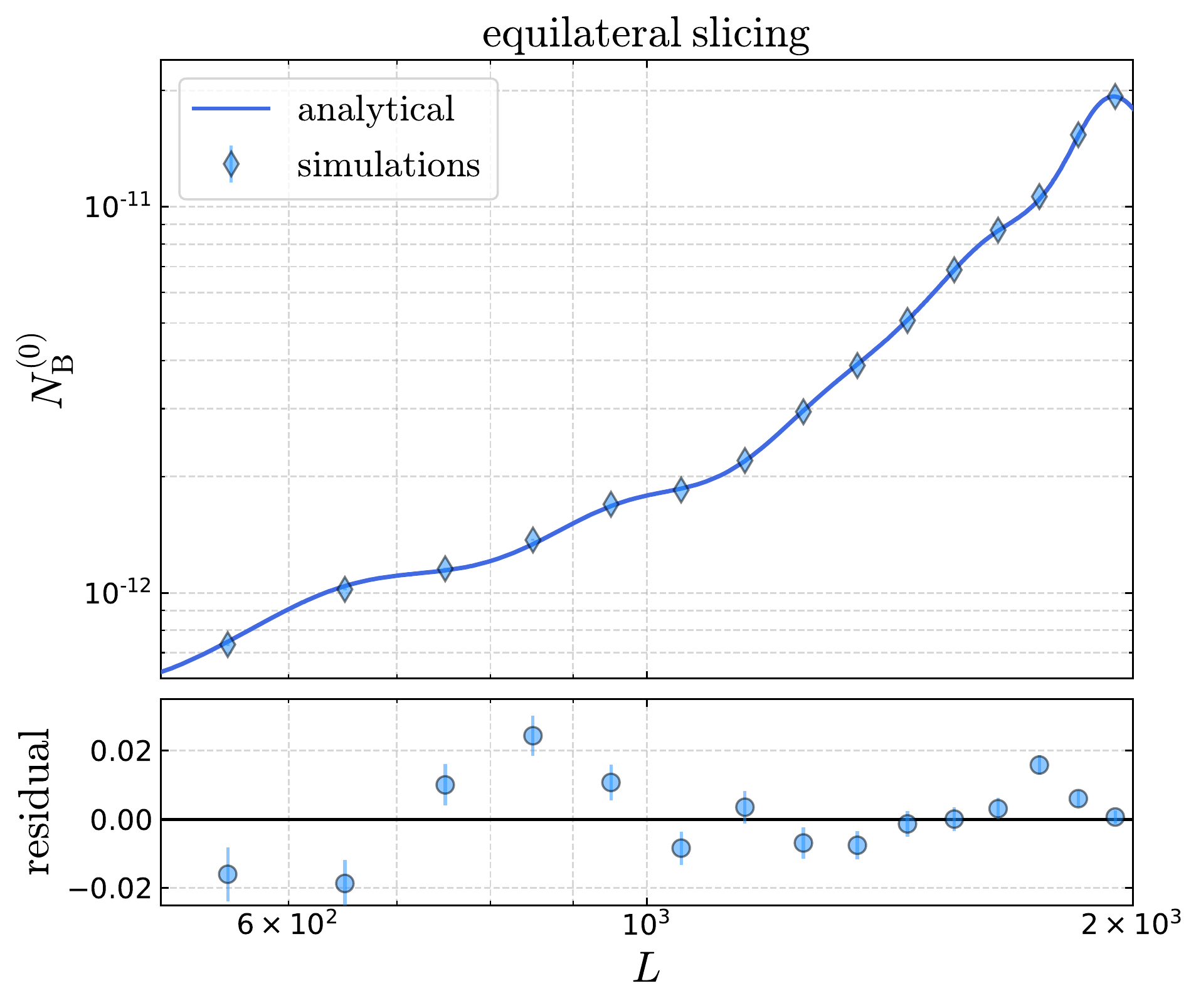}
    \caption{Comparison of $N_\mathrm{B}^{(0)}$ for the equilateral configuration calculated numerically from Eq.~\eqref{eq:N0_wick} and estimated from \eqref{eq:appbnzero}  with 624 noisy and lensed simulations.}
    \label{fig:N0_paired_sims_compare}
\end{figure}

We estimate Eq.~\eqref{eq:appbnzero} by averaging over sets of three independent simulations of the noisy and lensed CMB. Specifically, we use 624 simulations and take $J= I+1$ and $K=I+2$ (cyclically).
Fig.~\ref{fig:N0_paired_sims_compare} shows the comparison between this method, using \eqref{eq:appbnzero}, and the numerical evaluation of $N_\mathrm{B}^{(0)}$ using Eq.~\eqref{eq:N0_wick}, for the equilateral configuration. The two are in good agreement. As seen in Sec.~\ref{subsec:valid_n0}, the percent-level differences in Fig.~\ref{fig:N0_paired_sims_compare} could be due to the flat-sky approximation used in the analytical results. 

\subsection{Bias from $4_C\times 2_C$ terms}
\label{app:N1}

We now estimate the bispectrum noise-bias from the $4_C\times 2_C$ term in Eq.~\eqref{eq:decomp_6points}. This term contributes to $N^{(1)}_\text{B}$, but also to $N^{(2)}_\text{B}$ (and higher-order noise-biases) for the case of Gaussian lenses $\phi$. In Sec.~\ref{subsec:valid_n1}, we compared our leading-order analytic prediction for the $4_C\times 2_C$ bias (computed with the non-perturbative response functions), Eq.~\eqref{eq:N1_final}, with the lensing bispectrum computed directly from lensed CMB simulations with appropriate subtraction of $N^{(0)}_\text{B}$. That comparison includes higher-order contributions to $4_C\times 2_C$, along with those from the $6_C$ term. Here, we instead use suitably paired simulations to isolate the $4_C\times 2_C$ term (at all orders in $\phi$), removing contamination from the $6_C$ term in our comparison with the leading-order anayltic result.

There are three independent permutations contributing to $4_C\times 2_C$, which we label $A$, $B$ and $C$, respectively. These correspond to whether the $2_C$ contraction is between fields with one appearing in $\hat{\kappa}_{\bsL_1}$ and the other in $\hat{\kappa}_{\bsL_2}$, or the $\bsL_1$--$\bsL_3$ or $\bsL_2$--$\bsL_3$ pairings. Symbolically, we have
\begin{equation}
4_C \times 2_C \sim 4 \underbrace{\langle\widetilde{T}_{11}\widetilde{T}_{22}\widetilde{T}_3\widetilde{T}_{33}\rangle_C \langle\widetilde{T}_1\widetilde{T}_2\rangle_C}_{\text{perm. $A$}} + 4\underbrace{\langle\widetilde{T}_{11}\widetilde{T}_{22}\widetilde{T}_2\widetilde{T}_{33}\rangle_C \langle\widetilde{T}_1\widetilde{T}_3\rangle}_{\text{perm. $B$}} + 4\underbrace{\langle\widetilde{T}_{1}\widetilde{T}_{11}\widetilde{T}_{22}\widetilde{T}_{33}\rangle_C \langle\widetilde{T}_2\widetilde{T}_3\rangle_C}_{\text{perm. $C$}} \, .
\end{equation}
The factors of $4$ account for whether the first and first, first and second, second and first, or second and second legs of each quadratic estimator are contracted across (they are all equal by symmetry of the quadratic estimator).
For each of the permutations $A$, $B$ and $C$, there are three possible Wick contractions over the unlensed CMB in the connected 4-pt function. For example, for $C$ we have 
\begin{equation}
\langle \widetilde{T}_{1}\widetilde{T}_{11}\widetilde{T}_{22}\widetilde{T}_{33}\rangle_C = 
    \wick[offset=1.5em]{\langle  \c1 {\widetilde{T}}_1\c1 {\widetilde{T}}_{11} \c1 {\widetilde{T}}_{22} \c1 {\widetilde{T}}_{33}\rangle_{\phi,C}} +
    \wick[offset=1.5em]{\langle  \c1 {\widetilde{T}}_1\c2 {\widetilde{T}}_{11} \c1 {\widetilde{T}}_{22} \c2 {\widetilde{T}}_{33}\rangle_{\phi,C}} +
        \wick[offset=1.5em]{\langle  \c1 {\widetilde{T}}_1\c2 {\widetilde{T}}_{11} \c2 {\widetilde{T}}_{22} \c1 {\widetilde{T}}_{33}\rangle_{\phi,C}} \, ,
        \label{eq:appbunlenscontract}
\end{equation}
where the contractions over the unlensed CMB fields that appear (linearly) in each $\tilde{T}$ are shown explicitly. The second and third terms are equal because of the symmetry of $\hat{\kappa}_{\bsL_1}$.

To extract the different contractions of the unlensed CMB fields in Eq.~\eqref{eq:appbunlenscontract},
we need simulations with independent realizations of the unlensed temperature field but the same realization of the lensing potential. This ensures that the unlensed temperature fields are correctly paired, but the same $\phi$ fields are being averaged in all lensed CMB fields.
We must, however, remove any contractions of $\phi$ that would give only disconnected contributions to the 4-pt function. We therefore estimate, for example, the contribution of the first term in Eq.~\eqref{eq:appbunlenscontract} to permutation $C$ as
\begin{multline}
\int'_{\bsl_1,\bsl_2,\bsl_3} F(\bsl_1,\bar{\bsL}_{11}) F(\bsl_2,\bar{\bsL}_{22})F(\bsl_3,\bar{\bsL}_{33})
\wick[offset=1.5em]{\langle  \c1 {\widetilde{T}}_1\c1 {\widetilde{T}}_{11} \c1 {\widetilde{T}}_{22} \c1 {\widetilde{T}}_{33}\rangle_{\phi,C}} \langle\widetilde{T}_2\widetilde{T}_3\rangle_C \\ = \langle \hat{\kappa}_{\bsL_1}(ii,ii) \hat{\kappa}_{\bsL_2}(Jj,ki) \hat{\kappa}_{\bsL_3}(Jj,ki)\rangle_{Jj,ii,ki} -
\langle \hat{\kappa}_{\bsL_1}(ii,ii) \hat{\kappa}_{\bsL_2}(Jj,kk) \hat{\kappa}_{\bsL_3}(Jj,kk) \rangle_{Jj,ii,kk} \, .
\label{eq:appbterm1permc}
\end{multline}
Here, the first index in a field pair labels the realization of the unlensed temperature field entering $\widetilde{T}$ and the second the realization of the lensing potential. Capital indices indicate that instrumental noise is added to the lensed field. Hence, the first leg of the quadratic estimator $\hat{\kappa}_{\bsL_2}(Jj,ki)$ is a noisy lensed temperature field and the second leg a noiseless lensed temperature field constructed from an independent unlensed temperature realization and lensed by an independent realization of the lensing potential. Since the noise model is Gaussian, the simulations within the connected 4-pt correlation function can be noiseless. Note that in the case where there is no lensing, the right-hand side of Eq.~\eqref{eq:appbterm1permc} vanishes in every realization (not just in the mean), reducing the scatter in our simulation-based estimates. The final expression for the noise-bias to the lensing convergence bispectrum originating from the $4_C\times 2_C$ term is, symbolically,
\begin{align}
4_C\times 2_C &\sim 4 \langle \hat{\kappa}_{\bsL_1}(ii,ii) \hat{\kappa}_{\bsL_2}(Jj,ki) \hat{\kappa}_{\bsL_3}(Jj,ki)\rangle_{Jj,ii,ki} \nonumber \\ & \mbox{} -
4 \langle \hat{\kappa}_{\bsL_1}(ii,ii) \hat{\kappa}_{\bsL_2}(Jj,kk) \hat{\kappa}_{\bsL_3}(Jj,kk) \rangle_{Jj,ii,kk} \nonumber \\ & \mbox{}+ 8 \hat{\kappa}_{\bsL_1}(ii,ki) \hat{\kappa}_{\bsL_2}(Jj,ii) \hat{\kappa}_{\bsL_3}(Jj,ki)\rangle_{Jj,ii,ki} \nonumber \\ & \mbox{} -
8 \langle \hat{\kappa}_{\bsL_1}(ii,kk) \hat{\kappa}_{\bsL_2}(Jj,ii) \hat{\kappa}_{\bsL_3}(Jj,kk) \rangle_{Jj,ii,kk} + (\bsL_1 \leftrightarrow \bsL_2) + (\bsL_1 \leftrightarrow \bsL_3) \, .
\label{eq:N1_paired}
\end{align}

\begin{figure}[t]
\centering
    \includegraphics[width=0.8\textwidth]{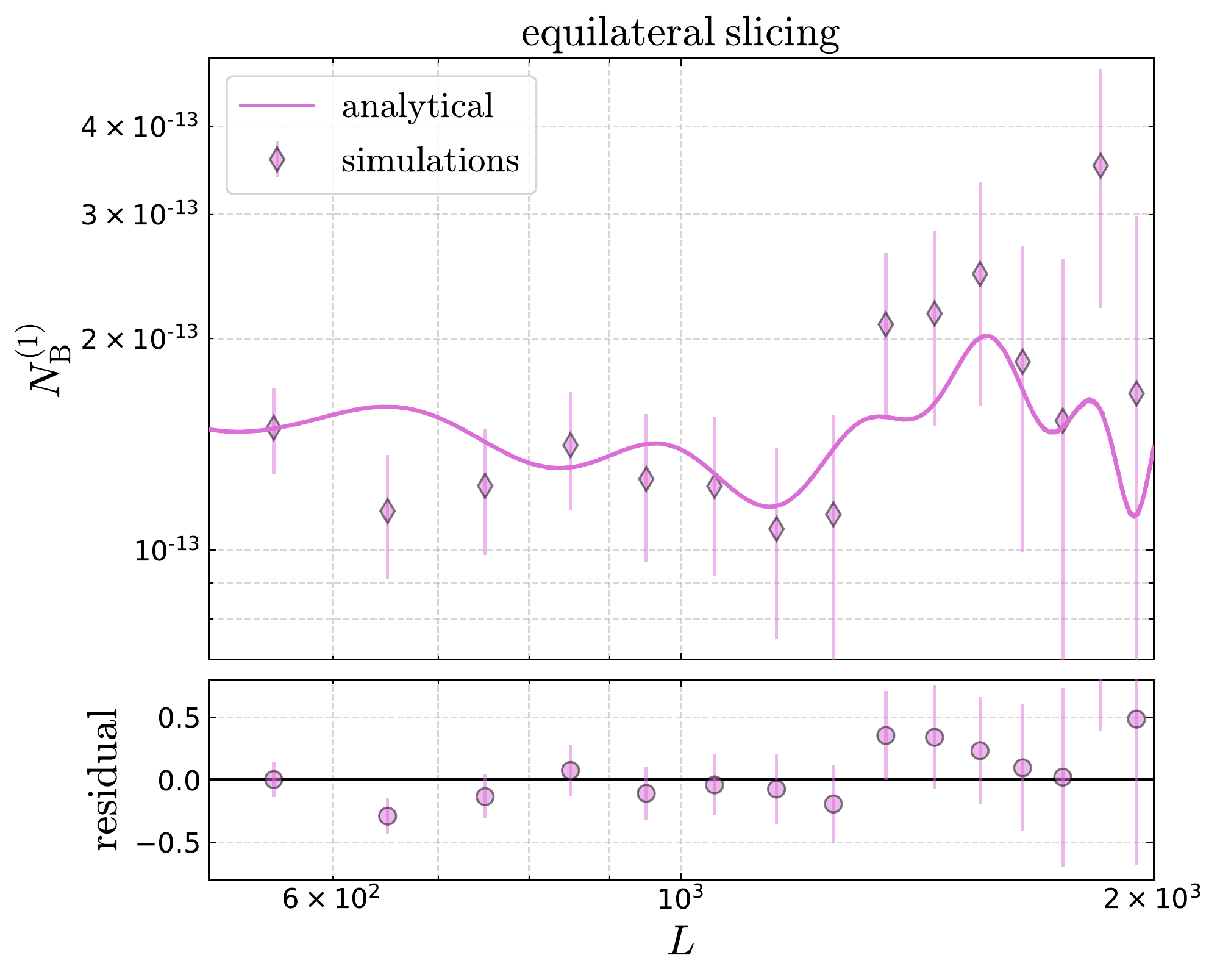}
    \caption{Contribution from $4_C\times 2_C$ terms to the noise-bias on the convergence bispectrum in the equilateral slicing, estimated from Eq.~\eqref{eq:N1_paired} with 624 paired simulations (data points). The solid line shows the flat-sky analytical result for $N^{(1)}_{\text{B}}$, Eq.~\eqref{eq:N1_final}, which agrees well with the data points to within the simulation errors (estimated from the scatter across simulations). The bottom panel shows the fractional residuals.}
    \label{fig:N1_equi_paired_sims_compare}
\end{figure}

\begin{figure}[t]
\centering
    \includegraphics[width=0.8\textwidth]{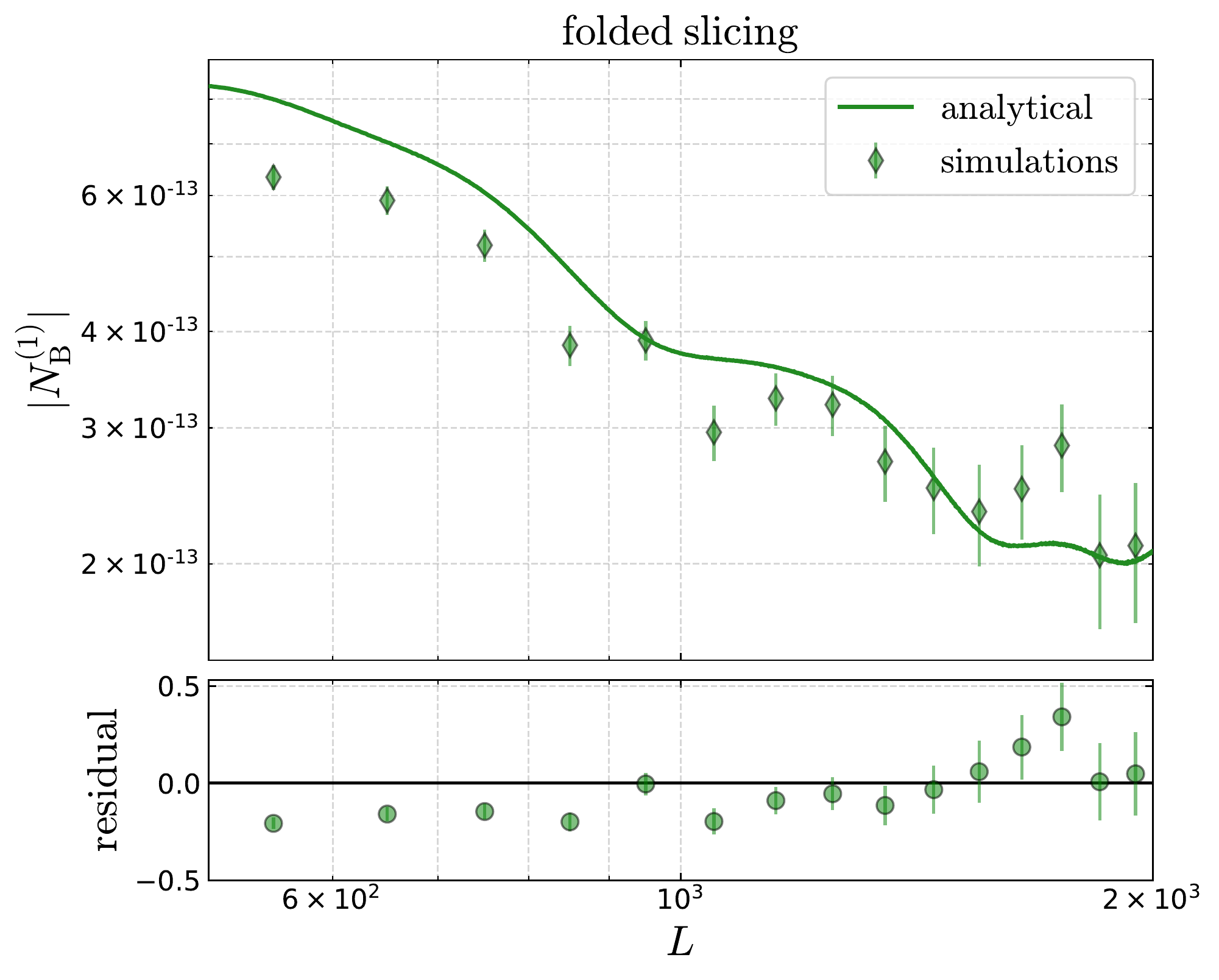}
    \caption{As Fig.~\ref{fig:N1_equi_paired_sims_compare}, but for the folded configuration of the bispectrum.}
    \label{fig:N1_fold_paired_sims_compare}
\end{figure}

In Figs.~\ref{fig:N1_equi_paired_sims_compare} and \ref{fig:N1_fold_paired_sims_compare} we show results of using 624 paired simulations to estimate the contribution from $4_C \times 2_C$ terms to the noise-bias on the bispectrum in the equilateral and folded configurations. These estimates are compared to the flat-sky analytical expression for $N^{(1)}_{\text{B}}$, Eq.~\eqref{eq:N1_final}. The results are in good agreement and support the findings from the alternative simulation approach described in Sec.~\ref{subsec:valid_n1} that the bispectrum noise-bias that originates from $4_C \times 2_C$ terms is well described by the leading-order approximation, $N^{(1)}_{\text{B}}$. Note that the folded slicing suffers from a binning effect as discussed in Sec.~\ref{subsec:valid_n0}, which is expected to become less significant as the number of bins used in the bispectrum estimator is increased.

\subsection{Bias from $6_C$ term}
\label{subapp:n2}

Finally, we estimate the bias from the connected 6-pt function in the case of Gaussian $\phi$. This term contributes to $N^{(2)}_{\text{B}}$ (and higher-order biases), along with the $4_C\times 2_C$ terms. We estimate the $6_C$ noise-bias to the bispectrum with\footnote{An alternative estimator built from the possible Wick contractions of the unlensed CMB fields, analogous to Eq.~\eqref{eq:N1_paired}, can easily be constructed and would likely have less scatter than the simpler form adopted here. However, the latter is adequate for our initial estimate of the relative importance of the bias from $6_C$.}
\begin{equation}
6_C \sim \langle \hat{\kappa}_{\bsL_1}(I,I) \hat{\kappa}_{\bsL_2}(I,I) \hat{\kappa}_{\bsL_3}(I,I) \rangle_I - 4_C \times 2_C - N^{(0)}_{\text{B}} \, . 
\label{eq:N2_estimate}
\end{equation}
We evaluate the $4_C \times 2_C $ term with Eq.~\eqref{eq:N1_paired} and $N^{(0)}_{\text{B}}$ from Eq.~\eqref{eq:appbnzero}. Gaussian instrumental noise does not affect the connected 6-pt function of the noisy CMB so Eq.~\eqref{eq:N2_estimate} could also be evaluated without including noise in any of the simulations used for any of the terms. Since the lensing potential $\phi$ is Gaussian there is no $N_\mathrm{B}^{(3/2)}$ bias. 

\begin{figure}
\centering
    \includegraphics[width=0.8\textwidth]{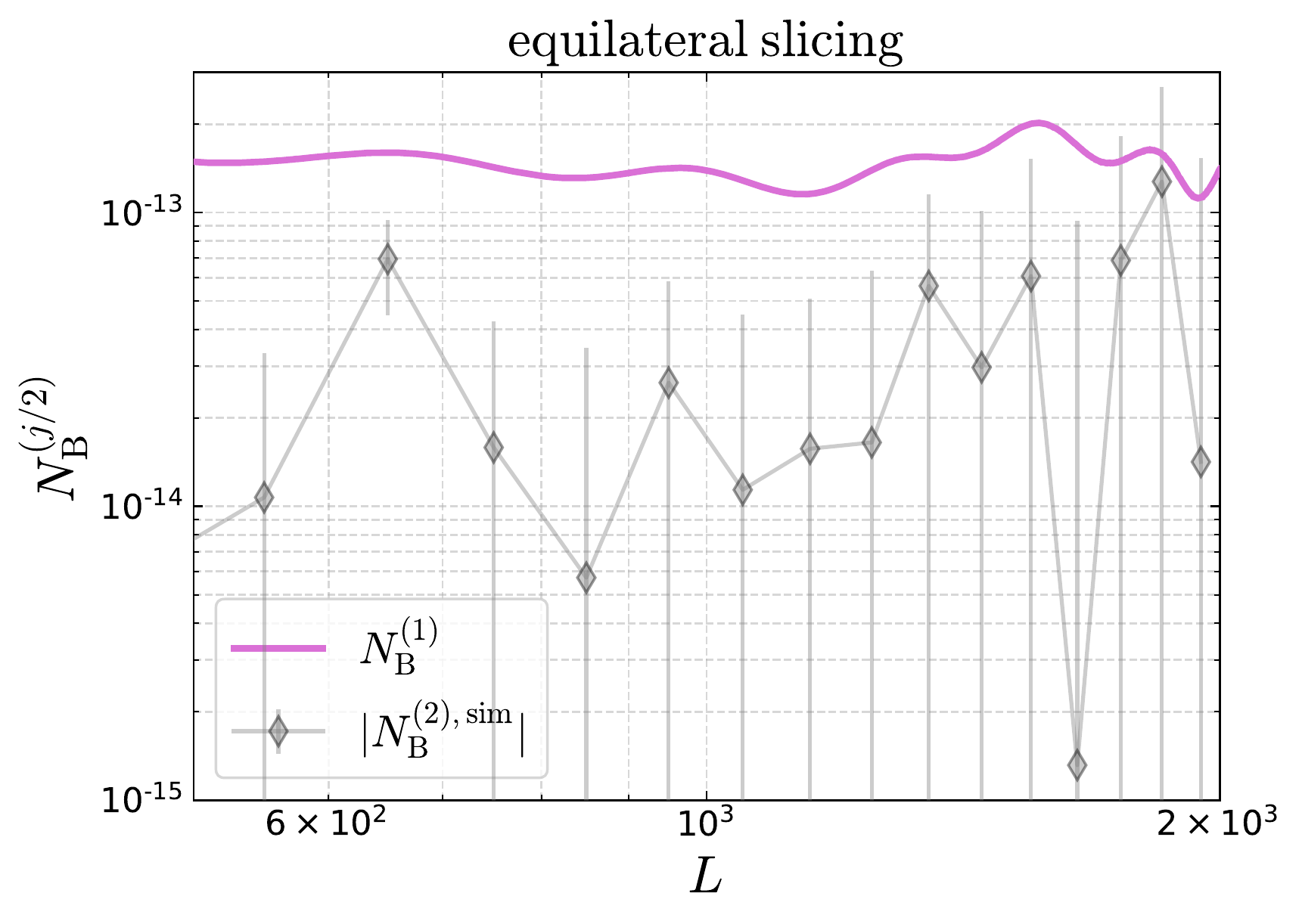}
    \caption{Comparison of the  $N_\mathrm{B}^{(1)}$ bias calculated numerically from Eq.~\eqref{eq:N1_final} and the bias from $6_C$ estimated with Eq.~\eqref{eq:N2_estimate} in the equilateral slicing using 20 multipole bins. The bias from $6_C$ is less significant by around an order of magnitude than the $N_\mathrm{B}^{(1)}$ bias for $L<1000$.}
    \label{fig:N2_est}
\end{figure}

Fig.~\ref{fig:N2_est} compares the estimated $6_C$ term from Eq.~\eqref{eq:N2_estimate} to the size of the $N^{(1)}_{\text{B}}$ bias in the equilateral slicing. Although our estimate is very noisy, it suggests that $6_C$ is around an order of magnitude smaller than $N^{(1)}_{\text{B}}$ for $L < 1000$. However, given the size of $N^{(1)}_{\text{B}}$ compared to the bispectrum signal, we cannot rule out the $6_C$ term being comparable to the signal.

\FloatBarrier
\bibliographystyle{utcaps}
\bibliography{References}

\end{document}